\documentclass{aa}  

\usepackage{graphicx}
\usepackage{subcaption}
\usepackage{titlesec}

\usepackage{txfonts}
\begin{document}

   \title{The POLAR Gamma-Ray Burst Polarization Catalog}
   \titlerunning{The POLAR Gamma-Ray Burst Polarization Catalog}
   \authorrunning{M. Kole et al.}
   \author{M. Kole\inst{1},
          N. De Angelis\inst{1}, F. Berlato\inst{2}, J. M. Burgess\inst{2}, N. Gauvin\inst{3}, J. Greiner\inst{2}, W. Hajdas\inst{4}, H.C. Li\inst{5,6}, Z.H. Li\inst{5,6}, N. Produit\inst{3}, D. Rybka\inst{7}, L.M. Song\inst{5,6}, J.C. Sun\inst{5,6}, J. Szabelski\inst{7}, T. Tymieniecka\inst{7}, Y.H. Wang\inst{5,6}, B.B. Wu\inst{5}, X. Wu\inst{1}, S.L. Xiong\inst{5}, S.N. Zhang\inst{5,6}, Y.J. Zhang \inst{5}
          }

   \institute{Department of Nuclear and Particle Physics, University of Geneva, 24 Quai Ernest-Ansermet, 1205 Geneva, Switzerland\\
              \email{merlin.kole@unige.ch}
         \and
             Max-Planck-Institut fur extraterrestrische Physik, Giessenbachstrasse 1, D-85748 Garching, Germany
         \and
	     University of Geneva, Geneva Observatory, ISDC, 16, Chemin d'Ecogia, 1290 Versoix Switzerland
         \and
	     Paul Scherrer Institut, 5232, Villigen, Switzerland
	 \and
	     Key Laboratory of Particle Astrophysics, Institute of High Energy Physics, Chinese Academy of Sciences, Beijing 100049, China
	 \and
	     University of Chinese Academy of Sciences, Beijing 100049, China
	 \and
	     National Centre for Nuclear Research, ul. A. Soltana 7, 05-400 Otwock, Swierk, Poland
             }

   \date{}

% \abstract{}{}{}{}{} 
% 5 {} token are mandatory
 
  \abstract
 {Despite over 50 years of research, many open questions remain about the origin and nature of gamma-ray bursts (GRBs). Linear polarization measurements of the prompt emission of these extreme phenomena have long been thought to be the key to answering a range of these questions. The POLAR detector was designed to produce the first set of detailed and reliable linear polarization measurements in an energy range of approximately $50-500\,\mathrm{keV}$. During late 2016 and early 2017, POLAR detected a total of 55 GRBs. Analysis results of 5 of these GRBs have been reported in the past. The results were found to be consistent with a low or unpolarized flux. However, previous reports by other collaborations found high levels of linear polarization, including some as high as $90\%$. }{We study the linear polarization for all the 14 GRBs observed by POLAR for which statistically robust inferences are possible. Additionally, time-resolved polarization studies are performed on GRBs with sufficient apparent flux.}{A publicly available polarization analysis tool, developed within the Multi-Mission Maximum Likelihood framework (\texttt{3ML}), was used to produce statistically robust results. The method allows to combine spectral and polarimetric data from POLAR with spectral data from the \textit{Fermi} Gamma-ray Burst Monitor (\textit{Fermi}-GBM) and the {\it Neil Gehrels Swift} Observatory (hereafter {\it Swift}) to jointly model the spectral and polarimetric parameters. }{The time integrated analysis finds all results to be compatible with a low or zero polarization with the caveat that, when time-resolved analysis is possible within individual pulses, we observe moderate linear polarization with a rapidly changing polarization angle. Thus, time-integrated polarization results, while pointing to lower polarization are potentially an artifact of summing over the changing polarization signal and thus, washing out the true moderate polarization. Therefore, we caution against over interpretation of any time-integrated results inferred herein and encourage the community to wait for more detailed polarization measurements from forthcoming missions such as POLAR-2 and LEAP.}{}


   \maketitle
%
%-------------------------------------------------------------------

\section{Introduction}

GRBs are the brightest electromagnetic phenomena in the Universe since the Big Bang. They consist of initial bright bursts of X/gamma-rays, called the prompt phase, which lasts from hundreds of milliseconds up to hours. The prompt emission is followed by a longer lasting afterglow which has been observed from radio wavelengths up to TeV energies. GRBs have historically been divided into two categories \citep{Mazets1981}, short GRBs, which have a prompt phase lasting less than 2 seconds long, and long GRBs, which have a prompt phase of over 2 seconds. In 2017, a GRB was shown to originate from the merger of two neutron stars thanks to the detection of a gravitational wave counterpart to GRB 170817A \citep{LIGO}. As a result of detailed localization measurements with optical instruments, strong evidence exists for long GRBs to be connected to the collapse of massive stars \citep{Woosley}. The spectral and timing properties for both short and long GRBs suggest that the gamma-ray component of the prompt emission is produced in highly relativistic jets \citep{Gehrels2013}. The physical properties of these jets, such as their structure and magnetic topology, remains however poorly understood. This despite the plethora of flux, spectral and timing measurements produced over the last five decades, see for example: \cite{Kienlin2020,Lien2016,Meegan1997,Tsvetkova2017}.

The two remaining key parameters of the gamma-ray emission, the polarization degree (PD) and angle (PA), remain largely unprobed. This despite it being widely believed that precise measurements of these parameters will provide unique information on the nature of the highly relativistic jets in which the prompt emission is produced \citep{Toma2009}. In particular, the linear polarization properties of the gamma-rays are highly dependent on the emission processes at play during the prompt emission as well as the magnetic fields and their structure within the emitting jets. For a detailed overview the reader is referred to, for example, \cite{Toma2009} and \cite{Gill2019}. One can generalize the theoretical predictions as follows: Models predicting the majority of the emission to be the result of synchrotron radiation allow for linear PDs as high as $50\%$ \citep{Luytikov2003}. For emission coming from regions with a highly ordered magnetic field the average linear PD will be $40\%$ within a sample of GRBs \citep{Toma2009}. In dissipative photospheric models, the predicted linear PD is relatively low, at the order of a few percent, although it can be as high as $40\%$ when the jet is seen at large off-axis angles \citep{Lundman2018}. Additionally, a high linear PD (around $50\%$) is possible in such models as well at lower energies (typically below $10\,\mathrm{keV}$), while at higher energies it is close to $0\%$ \citep{Lundman2018}. Finally, the Internal-Collision-Induced Magnetic Reconnection and Turbulence (ICMART) models predict a decaying linear PD during each pulse of the prompt-emission. In these models the maximum value at the beginning of the pulse can be as high as $60\%$ while the minimum value decreases to a few $\%$ as presented in \cite{ICMART}. In order to properly distinguish between such models, one requires both a large sample of GRB measurements as well as measurements capable of determining the evolution of the linear polarization parameters within a single emission pulse \citep{Gill2019}. It should be noted that circular polarization is not predicted in any of these models \citep{Luytikov2003,Lundman2014,ICMART}, therefore for the remainder of this work the term polarization can be assumed to mean linear polarization.

The large resolving power of the polarization parameters has prompted many attempts to perform detailed measurements in the past. The results however often suffer from large statistical uncertainties as a result of the low efficiency of gamma-ray polarimeters, which is typically an order of magnitude less than that of spectrometers. Many of these measurements have additionally been shown to suffer from problems with systematics or mistakes in the analysis \citep{MCCONNELL20171}. Overall, the existing measurements do now show a coherent result on the PD of the prompt emission. An overview can be found in \citep{Covino:2016cuw}. Although some of the reported measurements correspond to low PDs, the majority of the published results concern measurements of high PDs. There are however several reasons for this bias towards larger PDs. The first is that at least several measurements where the PD was found to be compatible with an unpolarized flux were not published. This is for example the result for at least 4 GRBs measured by the GAP collaboration \citep{Yonetoku_slides}. A second and perhaps more important reason is that systematic errors in polarization measurements will almost always result in an over-estimation of the PD. This is a result of both the positive definite nature of the PD, as well as the way polarization is measured. The latter reason will be clarified in the following section in more detail. A significant number of detailed measurements from a dedicated gamma-ray polarimeter are required to allow for a more clear view of the polarization parameters of the prompt emission of GRBs. The POLAR mission, which underwent detailed calibration measurements both on ground \citep{Kole} and in-orbit \citep{Li2018}, thereby reducing systematic errors significantly, measured a total of 55 GRBs during the mission life time. Using this data we aim here to provide the first catalog of detailed polarization measurements capable of resolving some of the open questions regarding GRBs.

\section{The POLAR mission}

\begin{figure}[!ht]
   \centering
     \resizebox{\hsize}{!}{\includegraphics{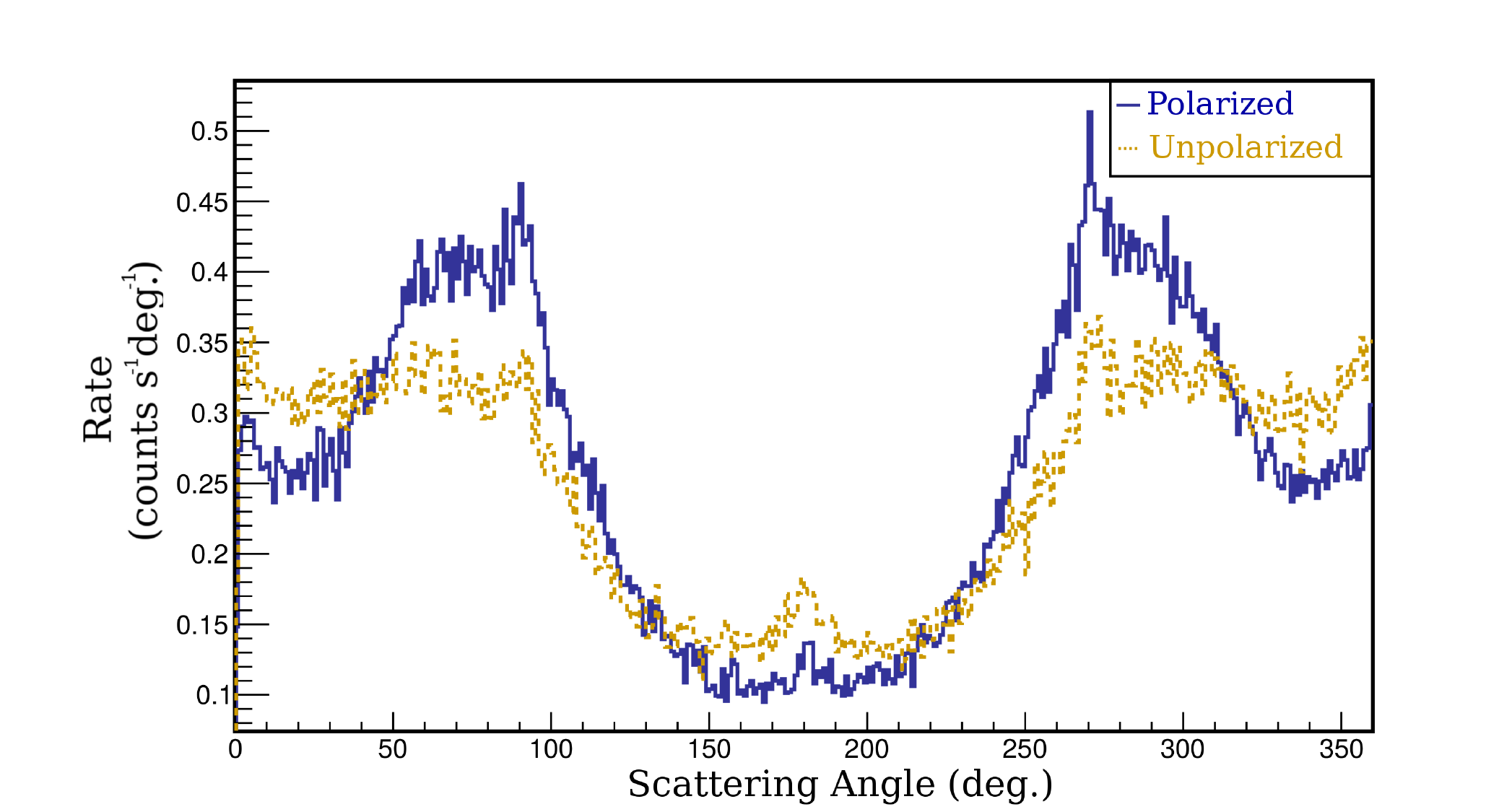}}
   \caption[Example of scattering angle distributions.]
 {An example of a simulated scattering angle distribution of a polarized (blue) and unpolarized (brown dotted) flux as it would be produced by POLAR for an off-axis GRB (170207A is taken as an example here). Both distributions show a $360^\circ$ modulation induced by the off-axis incoming angle of the GRB as well as some instrument induced effects. The polarized distribution additionally shows a $180^\circ$ modulation, as well as an additional small $360^\circ$ modulation induced by the polarization of the incoming flux.}
 \label{scatter_example}
 \end{figure}

POLAR was a dedicated GRB polarimeter launched as part of the Chinese Tiangong-2 (TG-2) Space Lab on September 15th 2016. The instrument, of which a detailed description can be found in \cite{Produit}, made use of a segmented scintillator array of $40\times40$ plastic bars, optimized for performing polarization measurements in the $50-500\,\mathrm{keV}$ energy range. The plastic scintillators were readout in groups of 64 using multi-anode photomutiplier tubes (MAPMTs). The polarization of an incoming gamma-ray flux can be derived from measurements of the scattering angle distribution of the incoming photons. This scattering angle is measured by POLAR using the scintillator array. Useful polarization events consist of photons which scatter at least once in one of the scintillators and subsequently (within a 100 ns time window) interact, either through Compton scattering or through photo-absorption, in a second scintillator. The scattering angle is then deduced from the position of the two scintillators in which an energy deposition was measured. Finally a distribution of the scattering angles, often referred to as a modulation curve in the polarimetry community, can be produced for a full measurement. An example of such distributions, simulated using the POLAR Monte Carlo software \citep{Kole}, for a $100\%$ polarized and unpolarized flux can be seen in figure \ref{scatter_example}.

Using a perfect instrument, the scattering angle distribution from a polarized flux would show only a $180^\circ$ modulation, while that from an unpolarized flux would be flat. For actual instruments the scattering angle distributions look significantly more complex. Firstly, because GRBs typically enter the instrument with a non-zero incoming angle with respect to zenith (the off-axis angle). A $360^\circ$ modulation is induced as photons scattering back towards the incoming direction of the GRB have a larger chance of escaping the instrument. Secondly, instrumental effects are seen in the distribution. For POLAR one clear effect is the presence of peaks at $90^\circ$ intervals induced by the square geometry of the instrument which favours some scattering angles over others. Other, less clearly, visible effects stem from non-uniformity in detector sensitivity, for example, as a result of differences in efficiency for different MAPMT channels. The $180^\circ$ modulation induced by the polarization is added to all these other effects in case of a polarized flux. In addition to a $180^\circ$ modulation the polarization can induce a second $360^\circ$ degree modulation especially for GRBs with a large off-axis incoming angle. It should therefore be noted that the frequently used method of dividing the measured scattering angle distribution by a simulated distribution of an unpolarized flux, which removes all effects not induced by polarization from the distribution, and subsequently fitting this with a $180^\circ$ modulation is therefore not correct. This is especially true for wide field-of-view instruments.

It should furthermore be stressed that instrumental effects can also induce a $180^\circ$ modulation which, if not modeled correctly in the MC, will appear as a polarization signal in the analysis. It is therefore crucial for polarimeters to be well calibrated such that all features in the scattering angle distribution are understood and well modeled in the instrument response. Any discrepancies between the real and modeled instrument response can, in the analysis, result in a fake polarization signal, whereas they are unlikely to result in an under-estimation of the polarization. This results from the effect that polarization adds fluctuations to the scattering angle distribution, which is flat for an unpolarized flux; while model induced errors have a large possibility to induce any fluctuations which make the distribution less flat, it is unlikely for such errors to flatten a distribution. To reduce systematic effects the POLAR instrument was carefully calibrated both on ground, of which a detailed report can be found in \cite{Kole} and subsequently in orbit \citep{Li2018,Xiao2018}. As reported in \cite{Kole}, the polarization measured for an unpolarized flux, which is induced due to uncertainties in the instrument response as well as due to statistical uncertainties, was of the order of $2\%$. This error is taken as a systematic error induced by uncertainties in the instrument response in all the analysis presented in this work. The error is directly taken into account for all the results presented here by adding it to the simulated scattering angle distributions. 

In-orbit calibration of the polarization response of POLAR to verify the systematic error is not possible due to the lack of standard-candle like sources for gamma-ray polarization at the moment. The source closest to being a standard-candle is the Crab as several measurements exist which largely agree as detailed in for example \cite{PoGO} and references therein. As POLAR is a large field-of-view instrument, polarization measurements of the nebula are not possible as it is a continuous source. Measurements of the pulsar are possible, albeit challenging, using time-resolved analysis. The Crab pulsar was previously used to calibrate the timing precision of POLAR \citep{Li2017} and the energy response of POLAR \citep{Li2019}. Analysis of the polarization is currently ongoing \citep{Li2020}. Although the preliminary results agree with previous measurements, like those in \cite{PoGO}, the precision of the POLAR measurements is not sufficient, due to the significant difficulty in such measurements with a wide field-of-view instrument, to use this as a calibration source. Although the careful in-orbit calibration has significantly improved our understanding of the detector response, and therefore likely the systematic error, it is therefore currently not possible to quantify this improvement. The systematic error of $2\%$ is therefore kept despite likely being an over-estimation of this error.

\section{Analysis}

As the analysis method used in this catalog differs from that conventionally used in polarization analysis, as well as that used previously by the POLAR collaboration, we provide some details on the analysis method here. It should be noted that the method used here was previously used to produce the time-resolved analysis results of GRB 170114A detected by POLAR \citep{170114A_BALROG}. While here we give an overview of the analysis and a justification for using it, the reader is referred to \cite{170114A_BALROG} for details.

\subsection{Original POLAR data analysis}

The first polarization results from GRBs detected by POLAR were presented in \cite{Zhang+Kole}. The analysis method applied there, which was used to extract the polarization parameters of 5 GRBs, consists of the following steps:

\begin{itemize}
    \item Selecting a signal and background interval for the GRB. The background interval is taken to be significantly longer than the GRB in order to minimize statistical errors. Additionally, special care is taken for the background to not contain any low-energy afterglow or other forms of contamination.
    \item The data from both intervals are processed using the data pipeline described in \cite{Li2018} in order to select polarization events and to calculate the scattering angles of these events.
    \item The scattering angle distributions of the background is subtracted from that of the signal region while taking into account the difference in live time of the two intervals.
    \item A set of 61 simulations for the particular GRB is performed using the MC software described in \cite{Kole}. The simulations are performed using an energy spectrum provided by other instruments, such as \textit{Fermi}-GBM or Konus-Wind and a location of the GRB, again provided by other instruments. The simulations are performed for both an unpolarized flux as well as for $100\%$ polarized fluxes with 60 different PA's. Additionally, the temperatures as measured throughout the POLAR detector, which slightly modify the electronics behaviour \citep{Li2018}, are taken into account when producing the response.
    \item The 61 simulated results are processed using the same analysis pipeline used for the measurement data.
    \item Scattering angle distributions are produced for 101 different PDs and 60 different PAs (giving a total of 6060 different parameter configurations) by combining the scattering angle distributions of the 61 simulations in the right ratio.
    \item The scattering angle distribution of the measured signal is directly fitted to all the 6060 simulated distributions. The $\chi^2$ values of the 6060 fits are used to produce a $\Delta\chi^2$ distribution in PD and PA space. Finally, the polarization of the measured flux along with confidence intervals are deduced from this distribution.
\end{itemize}

A very similar method was used previously for the analysis of the GAP data \citep{GAP}. Several improvements to this analysis method are, however, possible. Firstly, the background subtraction method is, although frequently used for such types of analysis, not the proper way of statistically handling data with Poisson errors. As a result of this procedure, statistical information is lost and the error propagation is not accurate. As described in more detail in \cite{170114A_BALROG} a better way to correct for the background is by fitting the background and subsequently taking it into account by modeling it in the analysis. This method ensures both a correct way of handling the statistics while also getting a more accurate estimation of the background and its fluctuations during the signal region.

A second issue is that the simulations are performed for one specific spectrum. The first problem with this is that in order to study the effects of a different input spectrum on the polarization measurement, the simulation process has to be fully repeated. Secondly, errors on the spectral parameters used are not naturally carried over into the polarization results. In order to propagate the spectral errors to the polarization parameters a range of simulations was performed in \cite{Zhang+Kole}. In these simulations the spectral parameters were varied within their uncertainty followed by the full reanalysis of the polarization for each new spectrum. This process was repeated 10'000 times. The distribution of the polarization parameters produced with these simulations was then used to estimate the systematic error. As this process was highly time consuming it was performed for only one GRB and the errors were assumed to be representative for all 5 GRBs studied in that work. Although this method is satisfactory, it is clearly not the optimum method to take the spectral uncertainties into account. 

Finally, the method fully relies on spectral information which is acquired independently from the polarization analysis. For example, when using spectral data from \textit{Fermi}-GBM, the spectral fit results as provided by that collaboration in their Gamma-ray Coordination Network notices (GCN)s are used as fixed input parameters in the analysis. The polarization analysis is then performed fully independently. The spectral information from POLAR remains fully unused despite the possibility for this data to improve the spectral analysis and thereby to reduce systematic errors in the polarization analysis.

\subsection{Improved analysis using \texttt{3ML}}

An analysis method which tackles all the above issues was developed and for the first time implemented in the time-resolved analysis of GRB 170114A described in \cite{170114A_BALROG}. There, instead of performing simulations with a fixed spectrum, the polarization response for 150 discrete energy ranges (from $5$ to $755\,\mathrm{keV}$ in $5\,\mathrm{keV}$ steps) were simulated separately. By subsequently adding the produced responses in the correct ratio, the polarization response for each possible spectrum can be produced by simply folding the spectrum through this response. While the initial simulation is time consuming, it allows to study the effect of the spectral parameters on the results of the polarization analysis without the need for starting simulations for a specific new spectrum. More importantly, the spectral response of POLAR was simulated, allowing the polarization parameters to be fitted at the same time as the spectrum using POLAR data only. This makes it possible to analyse GRBs which were only observed by POLAR.

All this analysis can be performed directly using the \texttt{3ML} framework \citep{3ML} which contains both spectral analysis and a polarization analysis tool. Although the latter was designed for POLAR specifically, it was designed to be easily adapted to work with data from any other gamma-ray polarimeter. An additional advantage of using the \texttt{3ML} framework is that it allows to perform joint fits using data from different instruments. Therefore the spectral information of POLAR can be combined with that of for example \textit{Fermi}-GBM, in order to perform a joint fit of the spectrum (using POLAR and \textit{Fermi}-GBM data) and the polarization (using POLAR data only). This method therefore not only allows to study GRBs which were previously not analyzable due to the lack of spectral information, it also allows to get more precise measurements thanks to joint fitting of the spectrum and the polarization. The background is furthermore fitted and subsequently modeled, thereby improving both the precision and allowing for a proper error propagation.

\begin{figure*}[]
   \centering
     \includegraphics[width=17.0 cm]{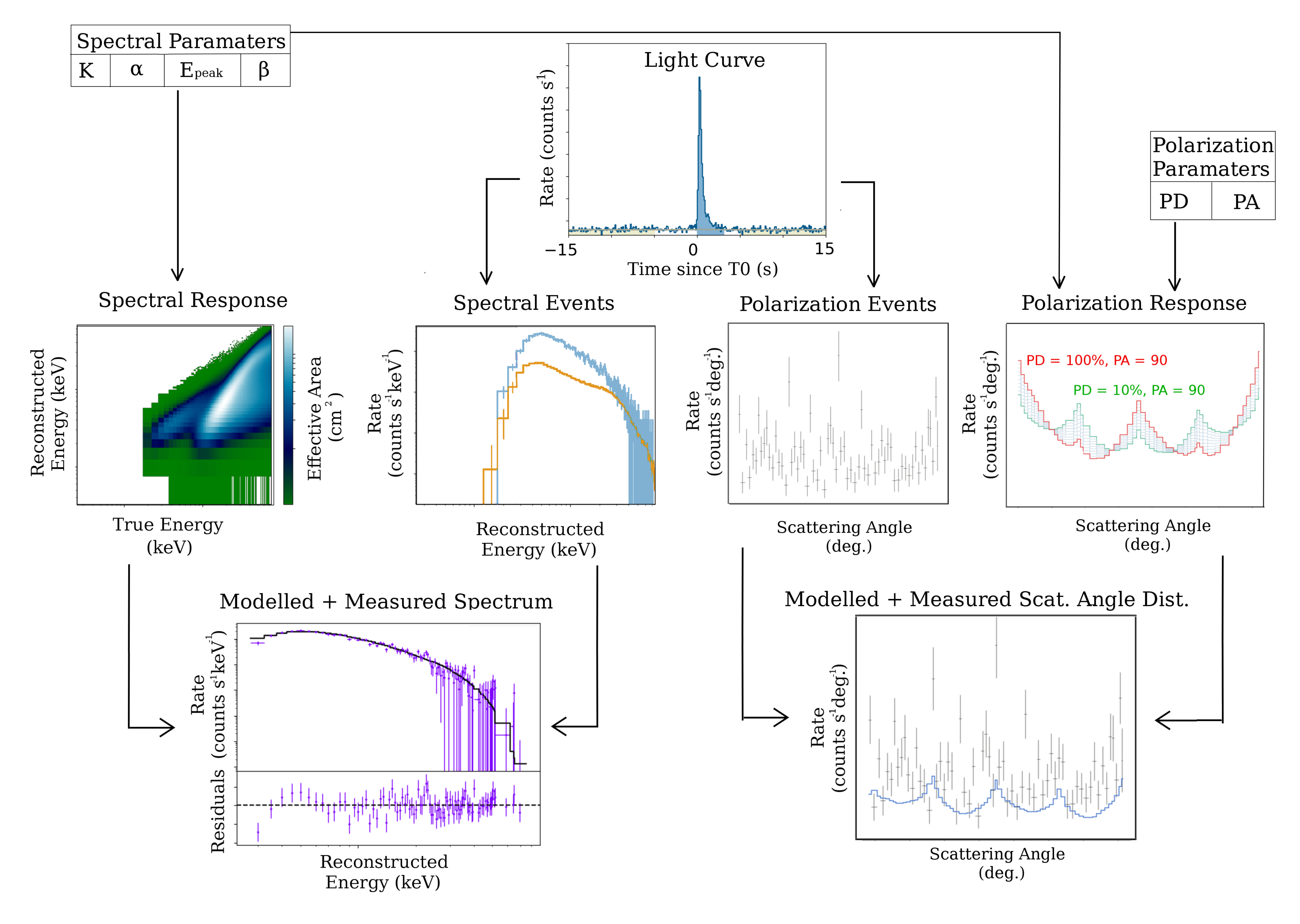}
   \caption[Schematic representation of the analysis.]
 {A schematic representation of the analysis method used here. In step one both the spectral and polarization events are extracted from the data. They are used to produce a spectrum (in count space) and a scattering angle distribution. Spectral parameters are picked from the provided priors and combined with the spectral response to produce a spectrum (again in count space) which is used together with the measured spectrum to calculate a likelihood. In parallel, these spectral parameters are, together with polarization parameters from a prior, provided to the polarization response in order to produce a scattering angle distribution. This is then used to calculate a likelihood using the measured distribution. Going through this process for different parameters in order to minimize the likelihood, a posterior distribution is produced for both the spectral and polarization parameters.}
 \label{fig:flow_chart}
 \end{figure*}

A schematic overview of the analysis procedure can be seen in figure \ref{fig:flow_chart}. First, from the light curve a signal and background period are selected based on $T_{90}$ calculations. Typically one second is added before the start time of the $T_{90}$ period and an equal period is added towards the end to ensure the full emission episode being captured. This period was often corrected by eye when needed, in case of low emission at the start of the GRB for example, in which case it is clarified in the upcoming analysis section of the specific GRB. For the background selection, a period both before and after the GRB are selected while excluding potential late emission or other features in the light curve. It should be noted that previous extensive studies presented in \cite{Zhang+Kole} have shown that the background region selection has no significant influence on the polarization results. 

Based on the selection, both a spectrum (in count space) and scattering angle distribution (also in count space) can be produced for the signal and the background region. The instrument response for the spectrum, which has been produced for one specific incoming angle of the GRB, can subsequently be used, with spectral parameters picked from a prior distribution, to produce a range of modeled spectra (also in count space). The modeled and the measured spectra can then be used to calculate the likelihood. In parallel, the same spectral parameters are, together with polarization parameters (also picked from their respective flat prior distributions), used to produce the scattering angle distributions corresponding to the specific spectral shape described by the spectral parameters. Each of these modeled scattering angle distributions is therefore for one specific set of spectral parameters and polarization parameters, and can be used together with the measured distribution to calculate the likelihood. The method differs here from all previous analyses where the spectral fit is performed first followed by performing polarization analysis using the results from the spectral analysis as input. Here the two are performed in parallel.

It is worth noting here that the likelihood for the scattering angle distribution is calculated using distributions with 360 bins for each GRB even if, when presenting the scattering angle distribution in upcoming figures, the number of bins is often reduced for illustrative purposes. A joint likelihood can then be calculated for the set of spectral and polarization parameters which is subsequently minimized. The final posterior parameter distributions are obtained using MultiNest (version PyMultiNest 2.9) \citep{MULTI1, MULTI2}. It should finally be noted that, opposed to analysis in for example \cite{Zhang+Kole}, \cite{GAP} and \cite{AstroSAT}, the measured and modeled scattering angle distributions are not normalized to ensure an equal number of total counts in both. Rather, the number of events per scattering bin is in units of $counts\,s^{-1}$, both in the measured and modeled distributions. This was one here to ensure that potential issues with the effective area become apparent in the analysis. Such issues are ignored in the traditional analysis through the normalization process.

\subsection{Catalog analysis}

The production of the simulation responses, as well as the Bayesian analysis used for joint fitting the spectral and polarization parameters, is described in detail in \cite{170114A_BALROG}. Essentially the same analysis procedure is followed for the work presented here. However, it should be noted that in \cite{170114A_BALROG} the spectral fits were performed using a synchrotron model. For the production of the catalog presented here the spectral fits will be performed using the Band function \citep{Band} or a cutoff powerlaw (CPL). The Band function is an exponentially cutoff power law smoothly connected to a power law at high energy. The CPL is defined here as $F(E)=K\,(E/100\,\mathrm{keV})^{-index}\,e^{-E/E_c}$. Here $K$ ($counts\,s^{-1}\,cm^{-2}\,keV^{-1}$) is the normalization constant, \textit{E} is the energy in $keV$, $index$ is the powerlaw index and $E_{c}$ is the cutoff energy. 

The decision to use empirical models, such as the Band function and CPL, was taken to make the results independent on the physical model. Both models have ubiquitous use as a phenomenological fitting function for GRB spectra as they describe the shape of most GRBs adequately. However, recent studies have shown that the Band function is a poor model for the physical emission spectra (e.g. \cite{Burgess2015}, \cite{Burgess2019}, \cite{Zhang2019} ) and should only be used as a qualitative description of the spectra such as the peak energy and flux. However, \cite{Yassine2019} have also shown that the Band function may not adequately model the true peak energy as well. The use of physical models to perform the analysis, as done in \cite{170114A_BALROG}, would be preferable to avoid over-interpretation of the spectral parameters found using empirical models. However, in order to allow for a model independent interpretation of the results presented here the use of empirical models was chosen.

The selection to use either a Band function or a CPL in the final analysis is based on an initial analysis where only the spectrum of each GRB was analyzed. The use of either of these two empirical models in the final analysis was based on the Deviance Information Criterion \citep{DIC} found in this initial analysis.

As we use the Band function or CPL for spectral fitting, we will perform a simultaneous fit for each GRB using the spectral parameters $\alpha\,,E_{peak}\,,\beta$ for Band and $index,\,E_{c}$ for CPL. The parameter $K$ which serves as a normalisation constant for both functions is also fitted along with the two polarization parameters PD and PA. In case spectral data is available from either \textit{Fermi}-GBM or the \textit{Swift}-BAT, such data is used in the analysis. In case data from \textit{Fermi}-GBM is used, the data from at least 3 of its NaI sub-detectors is used in the analysis as well as one of the BGO detectors. These 4 detectors are selected based on the significance of the detection of the GRB as calculated using the standard tools available in \texttt{3ML} \footnote{\url{https://threeml.readthedocs.io}}. The 3 NaI detectors with the highest significance are selected along with the BGO detector with the highest significance. In case data from more than 1 mission is used in the spectral analysis, additional normalization parameters are fitted to allow for the correction of the effective area of either of the instruments. This method is typically applied by the \textit{Fermi} collaboration to correct for unknown inter-calibration issues between its different detectors \citep{catalogI}. In all the analysis performed in this work, the corrections are found to be compatible to be within $25\%$ of 1.0 when using POLAR and  \textit{Fermi}-GBM data, and below $10\%$ when using POLAR and \textit{Swift}-BAT. Such values are similar to those found, for example, in catalog analysis of \textit{Fermi}-GBM for inter-calibration detection for the \textit{Fermi}-LAT and the different NaI (corrections of the order of $5\%$) and BGO detector (corrections of the order of $25\%$) \citep{catalogI}. Here we therefore find that that POLAR, \textit{Swift}-BAT and \textit{Fermi}-GBM are well calibrated with respect to one another. Additionally, for none of the GRBs analyzed here obvious systematic issues in the spectral fits were found, indicating that the POLAR instrument response is modeled well. This includes GRBs with large off-axis incoming angles (those exceeding the limit set in \cite{Zhang+Kole} of $45^\circ$). The analysis of GRBs with large off-axis incoming angles was not performed in previous POLAR publications as there was concern that material in the vicinity of POLAR, such as the surface of Tiangong-2 or its solar panels, would not be modeled correctly. The concern was that incorrectly modeled materials would induce systematic errors in the response. It was previously not possible to properly asses the existence of such systematics whereas now, using the joint spectral fits with other well established spectrometers this is possible. As we do not find obvious systematic issues for off-axis GRBs we can relatively safely assume the polarization response, produced using the same MC software, to also be accurate. Finally it should be clarified that the analysis procedure for time-resolved analyses presented here for certain GRBs is identical to that employed for the analysis of the full GRB. As joint fits are performed for the spectrum and polarization, spectral evolution during the GRB, which could otherwise induce significant systematic errors in the polarization results, is automatically taken into account using the method employed here.

After launch in September 2016, POLAR detected a total of at least 55 GRBs \citep{Xiong2017}. Operation of POLAR ended in April 2017 after the instrument suffered from a problem with the high voltage power supply. It should be noted that all 55 GRBs reported in \cite{Xiong2017} have been detected by other instruments. Additional GRBs, not detected by other instruments, could be present in the POLAR data, however, no significant candidates have been found in preliminary studies. We therefore only use the 55 GRBs reported in \citep{Xiong2017} in this study. Initial spectral analysis (without fitting the polarization parameters) was performed using \texttt{3ML} on all GRBs detected while POLAR was taking data in its science mode (data was taken in calibration mode as well for several weeks) and for which the incoming angle of the GRB was found to be $90^\circ$ or smaller. The spectral results for all these GRBs, 38 in total, are presented in table \ref{tab:spectra}. Additionally the $T_{90}$ for all these GRBs was calculated for all such GRBs using the POLAR data and is reported in table \ref{tab:spectra} as well. The $T_{90}$ was calculated based on counts in the $10$ to $750\,\mathrm{keV}$ energy range. Table \ref{tab:spectra} additionally contains a fit result for the late emission of GRB 170127C using both POLAR and \textit{Fermi}-GBM data, as well as a fit result for a specific time period of GRB 170131A. This last information was added as this is the only GRB observed by all three instruments, meaning POLAR, \textit{Fermi}-GBM and \textit{Swift}-BAT, although \textit{Swift}-BAT data is not available for the full GRB. Despite a lack of data for he full GRB a fit of a specific time interval was still performed as it allowed to study the inter-calibration of the three instruments together for the first time. The analysis found effective area corrections below $10\%$ for all instruments, indicating that the three instruments are well calibrated against one another.

\begin{table*}[!ht]
  \scriptsize
  \centering
 \caption{The spectral properties of the 38 GRBs detected by POLAR with an off-axis angle below $90^\circ$ while the instrument was in science mode. The $T_{90}$ parameter is additionally presented (*in units of erg/cm$^2$ in 10-1000 keV based on $T_{90}$ period, **afterglow time interval only (from $T0$+2.0 s to 20.0 s), ***GRB fitted using only the time interval where \textit{Swift}-BAT has data (from $T0$+16.6 s to 26.5 s).}\label{tab:spectra}
\hspace*{-2.5cm}\begin{tabular}{|c||c|c|c|c|c|c||}
  \hline
  GRB & Band or CPL & $\mathrm{T_{90}}$ (s) & Fluence* & $\alpha$/index & $\mathrm{E_{peak}}$/$\mathrm{E_c}$ (keV) & $\beta$  \\[5pt] \hline\hline
  161203A & Band & $4.1\pm{0.1}$ &  $(5.0\substack{+0.7 \\ -0.8})\times10^{-6}$ & $0.13\substack{+0.27 \\ -0.25}$ & $344\substack{+19 \\ -12}$ & $-3.41\substack{+0.39 \\ -0.46}$  \\[5pt] \hline
  161205A & Band & $15.8\pm{3.4}$ &   $(1.4\substack{+0.1 \\ -0.1})\times10^{-6}$ & $-1.39\substack{+0.17 \\ -0.09}$ & $119\substack{+28 \\ -21}$ & $-3.08\substack{+0.48 \\ -0.58}$  \\[5pt] \hline
  161207A & Band & $33.3\pm{1.4}$ &  $(6.1\substack{+0.6 \\ -0.2})\times10^{-6}$ & $-0.44\substack{+0.17 \\ -0.13}$ & $197\substack{+29 \\ -26}$ & $-2.30\substack{+0.27 \\ -0.40}$  \\[5pt] \hline
  161207B & Band & $5.1\pm{0.3}$ &  $(4.1\substack{+2.2 \\ -1.0})\times10^{-7}$ & $-0.98\substack{+0.36 \\ -0.28}$ & $1680\substack{+1350 \\ -790}$ & $-3.17\substack{+0.61 \\ -0.50}$  \\[5pt] \hline
  161210A & CPL & $3.14\pm{1.26}$ &   $(5.4\substack{+7.6 \\ -4.4})\times10^{-7}$ & $-0.89\substack{+0.18 \\ -0.18}$ & $830\substack{+310 \\ -310}$ & N.A.  \\[5pt] \hline
  161212A & CPL & $1.44\pm{0.16}$ &  $(3.3\substack{+4.3 \\ -3.2})\times10^{-7}$ & $-1.20\substack{+0.16 \\ -0.16}$ & $1040\substack{+290 \\ -290}$ & N.A.  \\[5pt] \hline
  161217B & CPL & $2.66\pm{0.2}$ &  $(1.7\substack{+2.9 \\ -0.9})\times10^{-7}$ & $-1.32\substack{+0.25 \\ -0.15}$ & $300\substack{+650 \\ -180}$ & N.A.  \\[5pt] \hline
  161217C & Band & $6.33\pm{0.25}$ &  $(9.0\substack{+2.7 \\ -1.9})\times10^{-6}$ & $-1.08\substack{+0.43 \\ -0.25}$ & $143\substack{+32 \\ -34}$ & $-2.76\substack{+0.36 \\ -0.61}$  \\[5pt] \hline
  161218A & Band & $11.48\pm{0.14}$ &   $(2.0\substack{+0.3 \\ -0.2})\times10^{-5}$ & $-0.54\substack{+0.07 \\ -0.06}$ & $144\substack{+12 \\ -11}$ & $-2.51\substack{+0.14 \\ -0.15}$  \\[5pt] \hline
  161228A & Band & $67.4\pm{2.58}$ &  $(3.8\substack{+1.3 \\ -0.7})\times10^{-6}$ & $-0.96\substack{+0.22 \\ -0.19}$ & $182\substack{+56 \\ -46}$ & $-3.05\substack{+0.58 \\ -0.56}$  \\[5pt] \hline
  161228B & Band & $9.75\pm{1.55}$ &  $(9.6\substack{+1.6 \\ -1.0})\times10^{-7}$ & $-1.22\substack{+0.12 \\ -0.09}$ & $171\substack{+37 \\ -27}$ & $-2.98\substack{+0.48 \\ -0.60}$  \\[5pt] \hline
  161228C & Band & $21.76\pm{1.52}$ &  $(3.4\substack{+0.8 \\ -0.4})\times10^{-6}$ & $-0.57\substack{+0.14 \\ -0.13}$ & $232\substack{+30 \\ -27}$ & $-2.96\substack{+0.53 \\ -0.58}$  \\[5pt] \hline
  161229A & Band & $31.26\pm{0.44}$ &  $(3.1\substack{+0.2 \\ -0.1})\times10^{-5}$ & $-0.64\substack{+0.03 \\ -0.03}$ & $339\substack{+12 \\ -14}$ & $-3.07\substack{+0.72 \\ -1.49}$  \\[5pt] \hline
  161230A & CPL & $3.92\pm{0.13}$ &  $(4.4\substack{+6.0 \\ -2.5})\times10^{-7}$ & $-1.24\substack{+0.12 \\ -0.10}$ & $650\substack{+360 \\ -260}$ & N.A.  \\[5pt] \hline
  170101A & CPL & $2.02\pm{0.11}$ &  $(6.4\substack{+1.6 \\ -1.3})\times10^{-6}$ & $-1.55\substack{+0.06 \\ -0.03}$ & $323\substack{+34 \\ -30}$ & N.A.  \\[5pt] \hline
  170101B & Band & $11.14\pm{0.38}$ &  $(8.0\substack{+0.7 \\ -0.5})\times10^{-6}$ & $-0.59\substack{+0.05 \\ -0.05}$ & $232\substack{+10 \\ -10}$ & $-3.28\substack{+0.46 \\ -0.49}$  \\[5pt] \hline
  170105A & Band & $3.23\pm{1.6}$ &  $(4.3\substack{+0.6 \\ -1.1})\times10^{-6}$ & $-1.22\substack{+0.46 \\ -0.2}$ & $50\substack{+14 \\ -13}$ & $-3.16\substack{+0.30 \\ -0.43}$  \\[5pt] \hline
  170109A & Band & $258.4\pm{1.7}$ &  $(8.9\substack{+0.7 \\ -0.4})\times10^{-5}$ & $-0.88\substack{+0.04 \\ -0.04}$ & $869\substack{+108 \\ -83}$ & $-3.14\substack{+0.41 \\ -0.48}$  \\[5pt] \hline
  170114A & Band & $10.48\pm{0.16}$ &  $(2.0\substack{+0.3 \\ -0.2})\times10^{-5}$ & $-0.68\substack{+0.09 \\ -0.09}$ & $211\substack{+31 \\ -25}$ & $-1.87\substack{+0.04 \\ -0.05}$  \\[5pt] \hline
  170120A & Band & $8.55\pm{2.72}$ &  $(8.7\substack{+2.2 \\ -1.0})\times10^{-7}$ & $-0.19\substack{+0.21 \\ -0.22}$ & $126\substack{+14 \\ -11}$ & $-3.23\substack{+0.37 \\ -0.46}$  \\[5pt] \hline
  170121A & Band & $6.12\pm{0.19}$ &  $(3.3\substack{+0.7 \\ -0.4})\times10^{-6}$ & $-0.32\substack{+0.15 \\ -0.12}$ & $567\substack{+57 \\ -60}$ & $-1.88\substack{+0.18 \\ -0.24}$  \\[5pt] \hline
  170127C & CPL & $0.14\pm{0.01}$ &  $(3.4\substack{+5.3 \\ -1.4})\times10^{-6}$ & $0.25\substack{+0.12 \\ -0.11}$ & $358\substack{+31 \\ -28}$ & N.A.  \\[5pt] \hline
  170127C** & Band & N.A. &  $(5.8\substack{+0.8 \\ -1.1})\times10^{-6}$ & $-1.14\substack{+0.22 \\ -0.21}$ & $1500\substack{+800 \\ -900}$ & $-3.1\substack{+0.6 \\ -0.6}$  \\[5pt] \hline
  170130A & Band & $31.08\pm{0.69}$ &  $(3.1\substack{+0.8 \\ -0.5})\times10^{-6}$ & $-0.94\substack{+0.16 \\ -0.14}$ & $202\substack{+43 \\ -41}$ & $-2.58\substack{+0.40 \\ -0.76}$  \\[5pt] \hline
  170131A & Band & $30.4\pm{9.4}$ &  $(4.6\substack{+0.6 \\ -0.4})\times10^{-6}$ & $-1.39\substack{+0.07 \\ -0.07}$ & $158\substack{+29 \\ -22}$ & $-3.05\substack{+0.48 \\ -0.55}$  \\[5pt] \hline
  170131A*** & Band & N.A. &  N.A. & $-1.28\substack{+0.09 \\ -0.08}$ & $196\substack{+41 \\ -30}$ & $-3.06\substack{+0.45 \\ -0.59}$  \\[5pt] \hline
  170206A & Band & $1.26\pm{0.01}$ &  $(6.5\substack{+0.4 \\ -0.3})\times10^{-6}$ & $-0.49\substack{+0.04 \\ -0.03}$ & $344\substack{+13 \\ -12}$ & $-2.68\substack{+0.14 \\ -0.19}$  \\[5pt] \hline
  170206C & CPL & $22.48\pm{0.31}$ &  $(1.4\substack{+1.3 \\ -0.8})\times10^{-5}$ & $-1.8\substack{+0.1 \\ -0.1}$ & $880\substack{+300 \\ -300}$ & N.A.  \\[5pt] \hline
  170207A & Band & $38.76\pm{0.26}$ &  $(6.7\substack{+0.2 \\ -0.2})\times10^{-5}$ & $-0.87\substack{+0.02 \\ -0.02}$ & $475\substack{+19 \\ -20}$ & $-3.37\substack{+0.74 \\ -1.26}$  \\[5pt] \hline
  170208C & Band & $45.93\pm{0.23}$ &   $(1.4\substack{+0.3 \\ -0.2})\times10^{-5}$ & $-0.75\substack{+0.11 \\ -0.11}$ & $176\substack{+25 \\ -21}$ & $-2.01\substack{+0.09 \\ -0.16}$  \\[5pt] \hline
  170210A & Band & $47.63\pm{2.51}$ &  $(6.6\substack{+0.3 \\ -0.2})\times10^{-5}$  & $-0.96\substack{+0.02 \\ -0.02}$ & $462\substack{+22 \\ -22}$ & $-2.72\substack{+0.39 \\ -0.49}$  \\[5pt] \hline
  170219A & Band & $0.09\pm{0.02}$ &  $(3.9\substack{+1.9 \\ -0.8})\times10^{-7}$ & $-0.40\substack{+0.20 \\ -0.16}$ & $510\substack{+150 \\ -110}$ & $-2.10\substack{+0.29 \\ -0.44}$  \\[5pt] \hline
  170228B & CPL & $26.99\pm{0.78}$ &  $(4.9\substack{+1.9 \\ -1.7})\times10^{-6}$ & $-1.36\substack{+0.05 \\ -0.05}$ & $347\substack{+63 \\ -54}$ & N.A.  \\[5pt] \hline
  170305A & Band & $0.45\pm{0.01}$ &  $(1.6\substack{+0.2 \\ -0.2})\times10^{-6}$  & $-0.35\substack{+0.09 \\ -0.09}$ & $253\substack{+17 \\ -16}$ & $-3.20\substack{+0.41 \\ -0.49}$  \\[5pt] \hline
  170306B & Band & $19.88\pm{0.22}$ &  $(1.3\substack{+0.2 \\ -0.1})\times10^{-5}$ & $-0.62\substack{+0.04 \\ -0.04}$ & $273\substack{+12 \\ -12}$ & $-2.96\substack{+0.43 \\ -0.49}$  \\[5pt] \hline
  170309A & Band & $2.03\pm{0.03}$ &  $(2.9\substack{+0.5 \\ -0.2})\times10^{-6}$ & $-1.42\substack{+0.04 \\ -0.03}$ & $303\substack{+126 \\ -90}$ & $-3.02\substack{+0.64 \\ -0.60}$  \\[5pt] \hline
  170315A & Band & $22.70\pm{5.68}$ &  $(1.1\substack{+0.2 \\ -0.2})\times10^{-6}$ & $-1.06\substack{+0.24 \\ -0.20}$ & $52.8\substack{+6.3 \\ -5.8}$ & $-3.28\substack{+0.44 \\ -0.45}$  \\[5pt] \hline
  170317A & Band & $13.74\pm{0.82}$ &  $(3.2\substack{+2.2 \\ -1.3})\times10^{-6}$ & $-1.21\substack{+0.40 \\ -0.22}$ & $60\substack{+39 \\ -29}$ & $-2.78\substack{+0.34 \\ -0.51}$  \\[5pt] \hline
  170320A & Band & $6.83\pm{0.09}$ &  $(6.7\substack{+0.8 \\ -0.6})\times10^{-6}$ & $-0.24\substack{+0.13 \\ -0.17}$ & $228\substack{+13 \\ -15}$ & $-2.32\substack{+0.16 \\ -0.21}$  \\[5pt] \hline
  170325B & Band & $7.82\pm{0.76}$ &  $(3.5\substack{+0.5 \\ -0.2})\times10^{-6}$  & $-0.62\substack{+0.37 \\ -0.36}$ & $125\substack{+12 \\ -12}$ & $-3.41\substack{+0.40 \\ -0.42}$  \\[5pt] \hline

\end{tabular}
\end{table*}

For the results presented previously in \cite{Zhang+Kole} the selection criteria to perform polarization analysis on a GRB were as follows:

\begin{itemize}
 \item The GRB has been observed by detectors other than POLAR and measurements of both the spectrum and location are provided by other instruments.
 \item The fluence of the GRB, as provided by other instruments in the $10-1000\,\mathrm{keV}$ energy range, exceeds $5\times10^{-6}\mathrm{erg/cm^2}$.
 \item The incoming angle with respect to the POLAR instrument zenith, $\theta$, is below $45^\circ$.
\end{itemize}

For the study here the first criterion is dropped as, thanks to fitting the spectrum and polarization at the same time, no external spectrum is required to perform the analysis. Although no external spectrum is required, relying on the POLAR data for the spectral fit will typically result in a larger systematic error. These, however, are automatically propagated to the posterior distribution.  In case \textit{Fermi}-GBM or \textit{Swift}-BAT data is available this data is used, together with the POLAR data, to perform a joint fit. For Konus-Wind observations this is currently not possible as the data is not publicly available in a format compatible with the \texttt{3ML} software. Therefore, in the analysis performed here the Konus-Wind spectral parameters, when available, are used as the priors for the spectral parameters in the fit.

For the 3rd criterion listed above, the accepted off-axis angle, is increased to $90^\circ$ in the selection procedure as no significant problems were found in the instrument response for large off-axis angles using joint spectral studies.

The main selection criterion remaining is therefore based on the fluence of the GRB. As such fluence measurements are not reported for all GRBs by other instruments, fluence measurements based on the POLAR data are used instead. An average flux, as measured over a 1 second bin is used as the selection criterion here. The average flux was used here instead of the fluence as it is more representative for the signal to background counts in the data. The lower limit was set to $1\times10^{-6}\mathrm{erg/cm^2/s}$ measured in the $10-1000\,\mathrm{keV}$ energy range.

Finally, the POLAR instrument response is simulated based on the best available localization measurement of the GRB. Typically this is either a location from \textit{Swift}, the localization provided by \textit{Fermi}-LAT, or calculated using InterPlentary Network (IPN) triangulation. When none of these are available, the \textit{Fermi}-GBM data is used for localization using the BAyesian Location Reconstruction Of GRBs (BALROG) method described in \citep{BALROG}. If additionally no \textit{Fermi}-GBM data is available the location is calculated using POLAR data using the method described in \cite{Yuanhao}. 

The uncertainty on the location will induce a systematic error in the polarization results. This effect was previously studied for the POLAR mission and is presented in \cite{Yuanhao} and in \cite{Zhang+Kole}. The systematic error resulting from the location is taken into account here in the same way as it was done in \cite{Zhang+Kole}. This method consists of adding the uncertainty to the polarization model by adding errors to the data points in the modeled scattering angle distribution. The size of these errors are based on the work performed in \cite{Zhang+Kole}, where it was found that the uncertainty on the data point in the model is approximately linear with the error in the uncertainty, where a $1\sigma$ error of $1^\circ$ on the location results in a $1\%$ uncertainty on the data points in the modeled scattering angle distribution. While this method results in systematic errors of approximately right size, it does not allow to properly take into account asymmetric errors on the location (such as those typically seen in IPN locations). For asymmetric location errors the average location error with a single digit precision is used to get a good approximation of the systematic error. A method to overcome this short coming in our analysis is to perform joint fits for both the spectrum, location and polarization. Although this is theoretically possible such a method would require very large CPU time and will require development and careful studies which forms a future project of the authors.

Finally, the above criteria would pass a total of 15 GRBs. However, GRB 161129A is dropped from the analysis as it occurred on the tail of a large solar flare \citep{RHESSI}.  As a result, potential systematics from a quickly changing background could exist making the measurement unreliable.

In total 14 different GRBs \footnote{The data products used for the analysis for these 14 GRBs analysis will be placed on \url{https://www.astro.unige.ch/polar/} and the collaboration strongly encourages additional analysis by external groups.} will be discussed here. Out of these two are clear single pulse GRBs, 170101A and 170114A, while the rest consist of several overlapping pulses. Below we present the details for the analysis of GRB 170101A, a single pulse GRB where \textit{Swift}-BAT data was used in the analysis to produce more constraining measurements than those previously published in \cite{Zhang+Kole} for this GRB. Additionally, we present the analysis of 170207A, a bright GRB previously not considered in the analysis due to its large off-axis angle. Details on the analysis of the remaining 12 GRBs can be found the appendices of this work, while a summary of all the results can be found in section \ref{sec:conclusions}.

 \clearpage

\section{GRB Analysis}

\subsection{170101A}

\begin{figure}[h]
\begin{subfigure}{.5\textwidth}
  \centering
  % include first image
  \includegraphics[width=.95\linewidth]{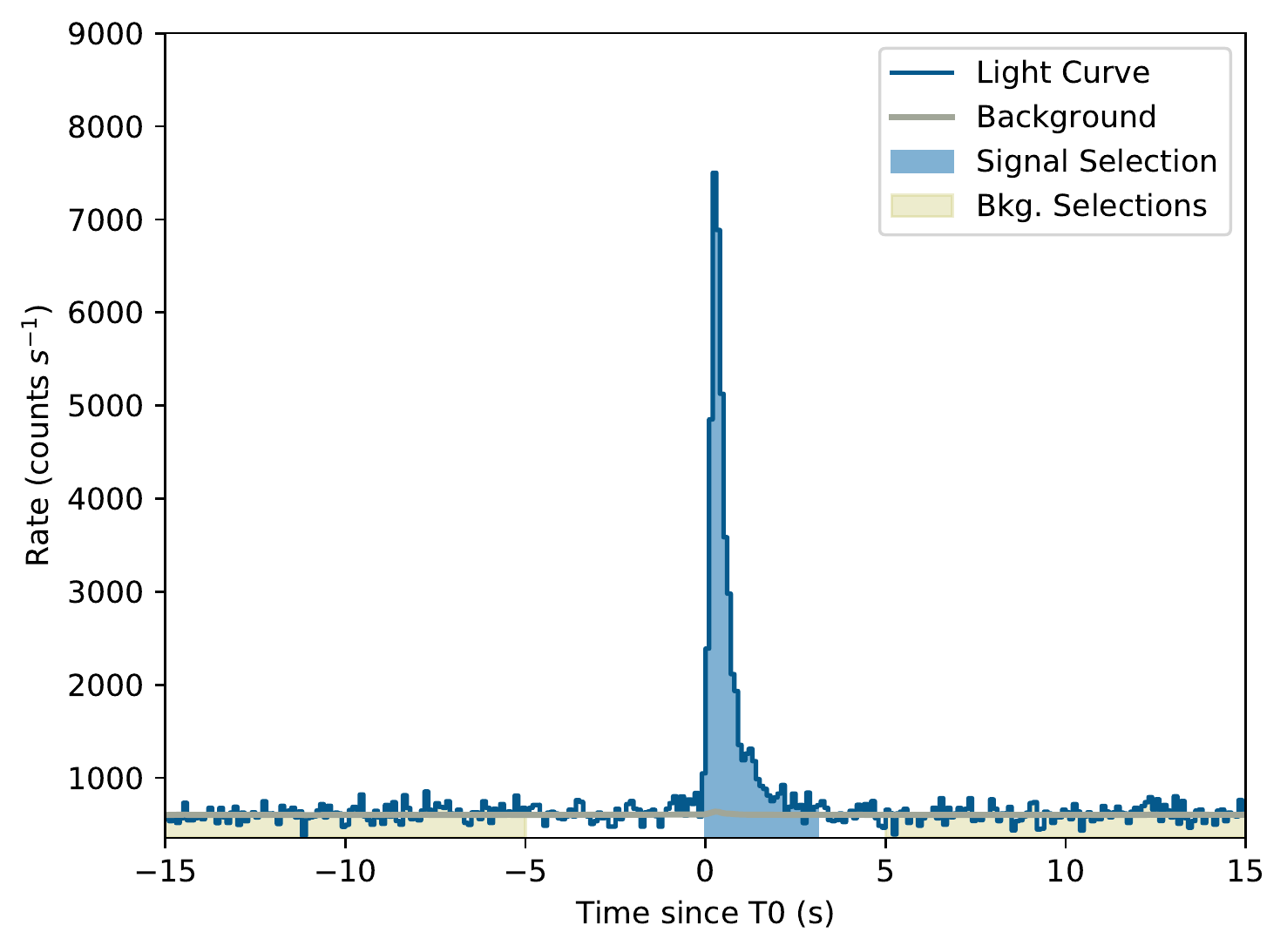}  
  \caption{The light curve of GRB 170101A as measured by POLAR where $T=0\,\mathrm{s}$ is defined as the $T0$ defined by \textit{Swift}-BAT for this GRB.}
  \label{fig:170101A_lc}
\end{subfigure}
\newline
\begin{subfigure}{.5\textwidth}
  \centering
  % include second image
  \includegraphics[width=.95\linewidth]{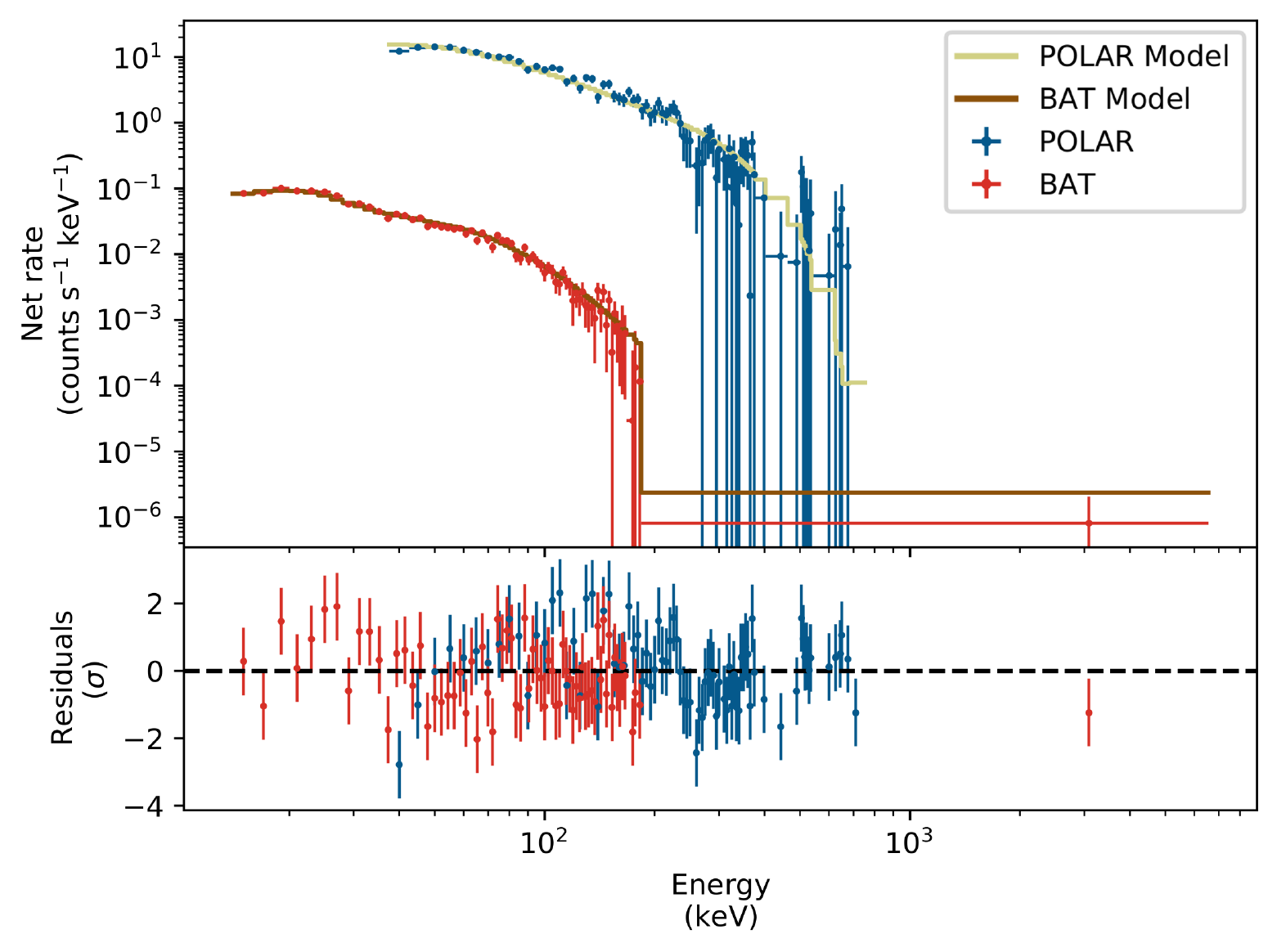}  
  \caption{The joint spectral fit result for 170101A. The number of counts as detected by both POLAR (blue) and \textit{Swift}-BAT (red) are shown along with the best fitting spectrum folded through the instrument responses in yellow for POLAR data and in brown for \textit{Swift}-BAT data. The residuals for both data sets are shown in the bottom of the figure.}
  \label{fig:170101A_cs}
\end{subfigure}
\caption{The light curve as measred by POLAR for GRB 170101A (a) along with the joint spectral fit results of POLAR and \textit{Swift}-BAT for the signal region indicated in yellow in  figure (a).}
\label{fig:170101A_lc_cs}
\end{figure}

\begin{figure}[!ht]
   \centering
     \resizebox{\hsize}{!}{\includegraphics{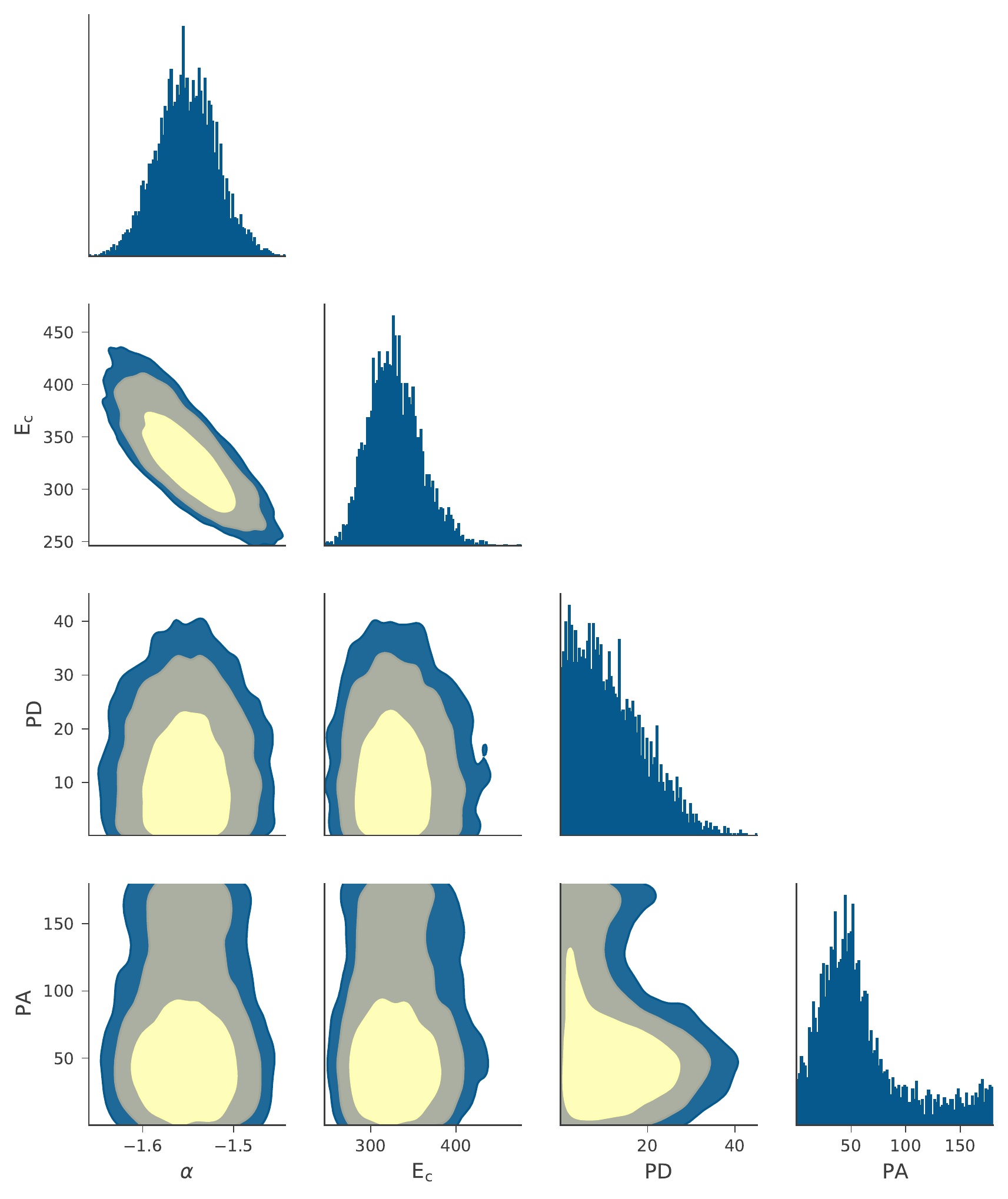}}
   \caption[Spectral and polarization posterior distributions.]
 {The spectral and polarization posterior distributions for GRB 170101A. The 1 and 2 $\sigma$ credibility intervals as well as that corresponding to $99\%$ are indicated. The PA shown here is in the POLAR coordinate system, a rotation in the positive direction of 56 degrees transforms this to the coordinate system as defined by the IAU.}
 \label{fig:post_170101A}
 \end{figure}

\begin{figure}[ht]
\begin{subfigure}{.5\textwidth}
  \centering
  % include first image
  \includegraphics[width=.85\linewidth]{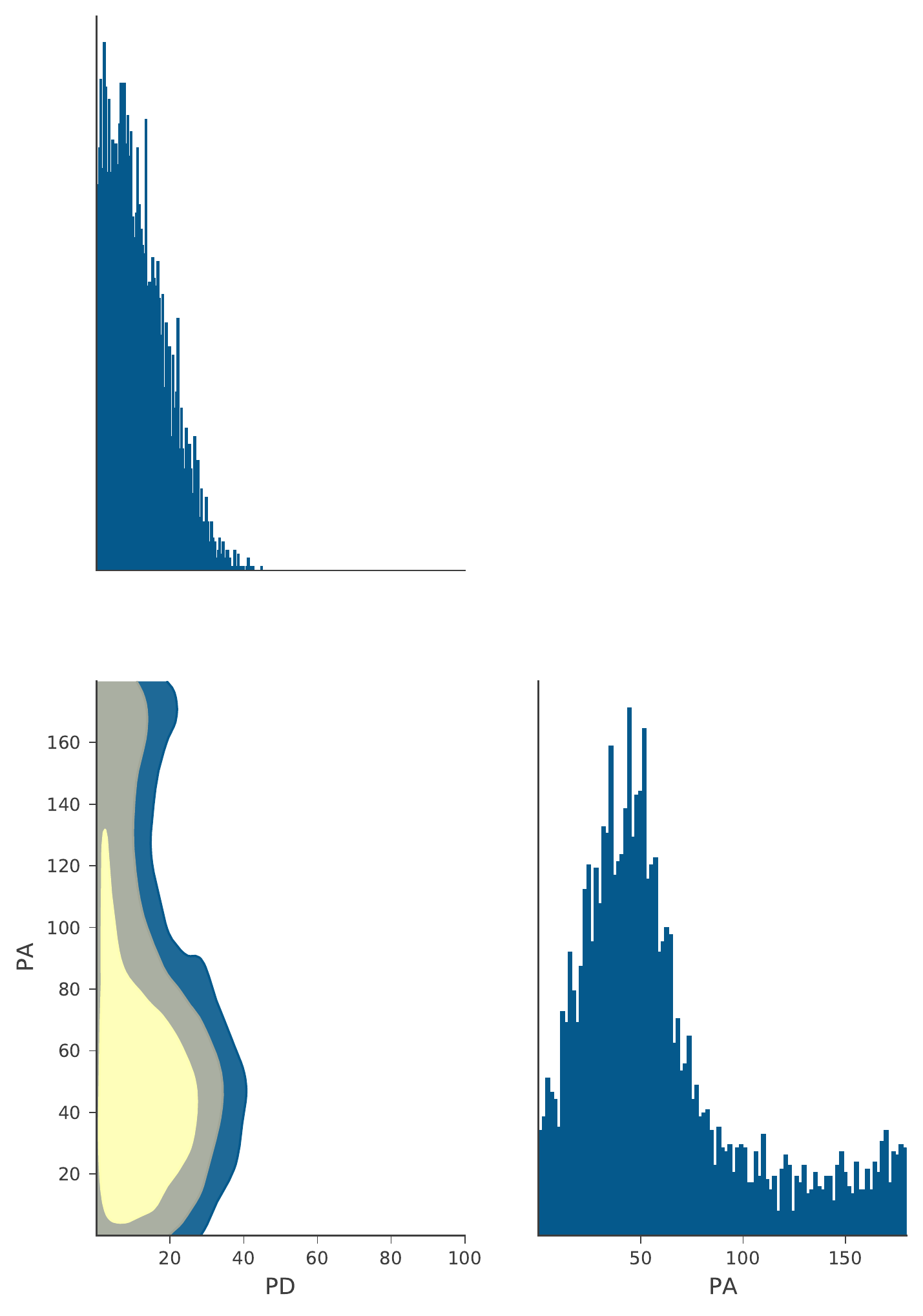}  
  \caption{The polarization posterior distributions for GRB 170101A with the 1 and 2 $\sigma$ credibility intervals as well as that corresponding to $99\%$. The PA shown here is in the POLAR coordinate system, a rotation in the positive direction of 56 degrees transforms this to the coordinate system as defined by the IAU.}
  \label{fig:170101A_PD}
\end{subfigure}
\newline
\begin{subfigure}{.5\textwidth}
  \centering
  % include second image
  \includegraphics[width=.85\linewidth]{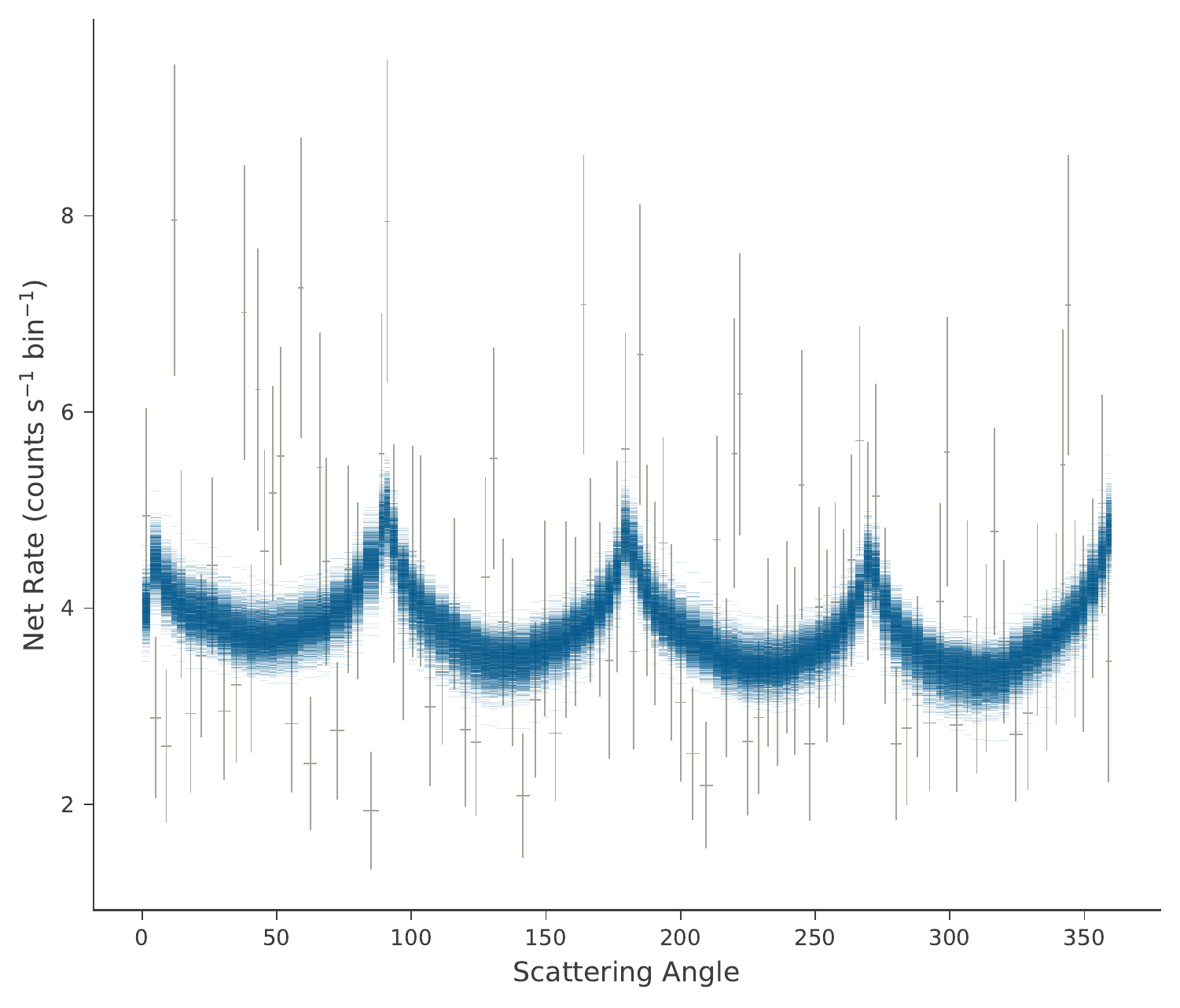}  
  \caption{The measured scattering angle distribution (gray data points with 3 degree bin size) superimposed by the posterior model predictions (blue). The errors on the data points are the Poisson errors corrected for the background.}
  \label{fig:170101A_sd}
\end{subfigure}
\caption{The posterior distribution of the polarization parameters (a) together with the scattering angle distribution of GRB 170101A.}
\label{fig:170101A_PD_sd}
\end{figure}

GRB 170101A was detected by POLAR and by \textit{Swift}-BAT. The latter defines $T0$ as 2017-01-01 at 02:26:00.679 (UT) which, for convenience, will also be used as $T0$ in the analysis presented here. A $T_{90}$ of $(2.02\pm0.11)\,\mathrm{s}$ was measured using POLAR data for this GRB. The light curve, including the signal region (blue) and part of the background region (yellow) can be seen in figure \ref{fig:170101A_lc}. Spectral and polarization analysis was performed using both POLAR and \textit{Swift}-BAT data. The response of POLAR was produced using the refined location provided by \textit{Swift}-BAT: RA (J2000) = $267.089^\circ$, Dec (J2000) = $11.642^\circ$ \citep{GCN_SWIFT_170101A}. The spectral results of the fit can be seen in figure \ref{fig:170101A_cs}. The effective area correction (applied to the POLAR data) found in the analysis was $1.07\pm0.04$. The posterior distributions of the spectral and polarization parameters are shown in figure \ref{fig:post_170101A}. The posterior distribution of the polarization parameters is shown, together with the measured scattering angle distribution superimposed by the posterior model predictions (blue), in figure \ref{fig:170101A_PD_sd}. A PD of $6.3\substack{+10.8 \\ -6.3}\%$ (errors correspond to the $1\sigma$ uncertainty bounds here and throughout the remainder of this  paper unless otherwise specified) is found. This is compatible with that reported in \cite{Zhang+Kole}, it should be noted that the PA used in \cite{Zhang+Kole} is measured in the coordinate system as defined by the International Astronomical Union (IAU), see \cite{IAU} for details, while here we present it in the POLAR coordinate system. The PA as measured following the IAU convention is presented in figure \ref{fig:PA_catalog} towards the end of this paper. A $99\%$ credibility upper limit of $35.1\%$ is found.

\begin{figure}[!ht]
   \centering
     \resizebox{\hsize}{!}{\includegraphics{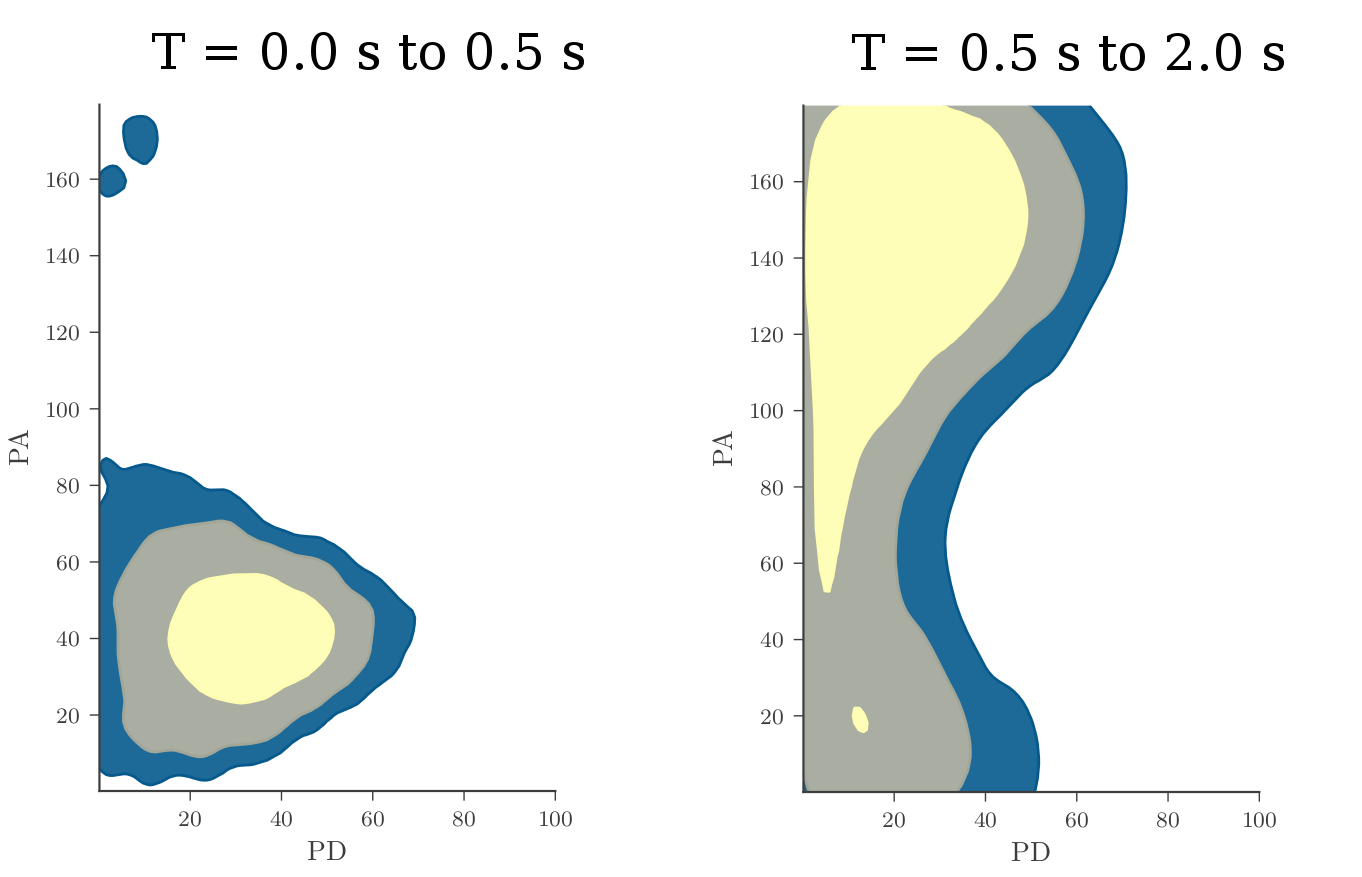}}
   \caption[Light curve of GRB 170101A as measured by POLAR.]
{The polarization posterior distributions for the 2 time bins of GRB 170101A are shown with the 1 and 2 $\sigma$ credibility intervals as well as that corresponding to $99\%$ credibility. On the left the posteriors are for the first time bin ranging from $T=0.0\,\mathrm{s}$ to $T=0.5\,\mathrm{s}$, on the right it is for the second time bin ranging from $T=0.5\,\mathrm{s}$  to $T=2.0\,\mathrm{s}$ . The polarization angle shown here is in the POLAR coordinate system, a rotation in the positive direction of 56 degrees transforms this to the coordinate system as defined by the IAU.}
 \label{time_res_170101A}
 \end{figure}

As GRB 170101A has a sufficiently high signal to background counts ratio to perform basic time-resolved analysis, the GRB was divided into two time bins with similar statistics. The first time bin ranged from $T=0.0\,\mathrm{s}$ to $T=0.5\,\mathrm{s}$, the second from $T=0.5\,\mathrm{s}$ to $T=2.0\,\mathrm{s}$ seconds. The polarization parameter posteriors for these time bins can be seen in figure \ref{time_res_170101A}. The results are consistent with an unpolarized flux for both time bins within 99\% credibility. However, for the shorter first time bin a PD of $(32\pm12)\%$ is found which gives a hint of polarization. For the longer time bin a PD of $13\substack{+18 \\ -13}\%$ and therefore consistent with a PD of $0\%$ is found. The results, with a PD of $32\%$ in the first time bin and a potentially smeared out polarization signal in the longer second time bin, hint that with a finer time binning a significant polarization with an evolving polarization angle could be found. The signal to background count ratio in the POLAR data from this GRB do not allow for such an analysis.

\clearpage

\subsection{170207A}

\begin{figure}[ht]
\begin{subfigure}{.5\textwidth}
  \centering
  % include first image
  \includegraphics[width=.95\linewidth]{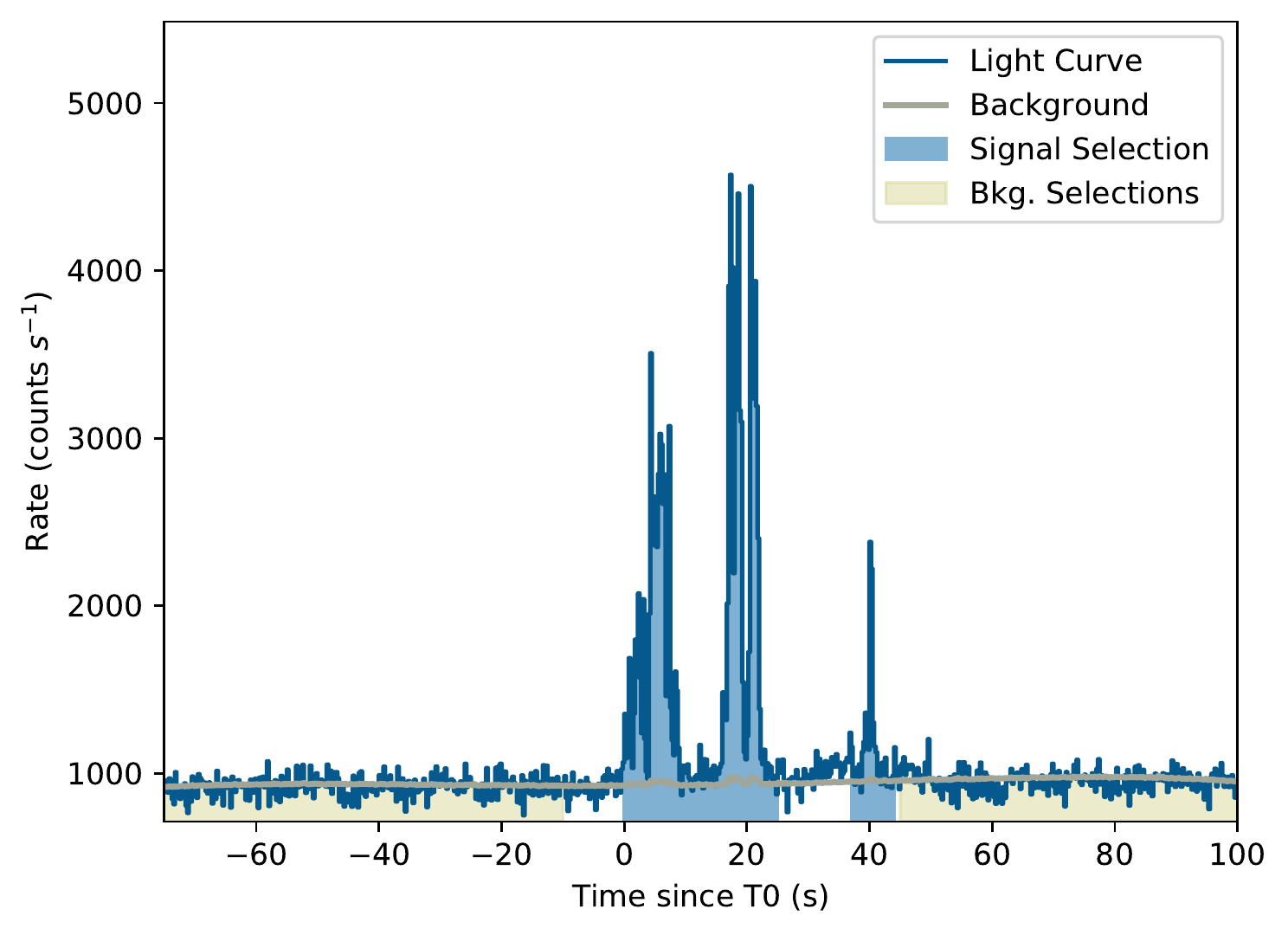}  
  \caption{The light curve of GRB 170207A as measured by POLAR, where $T=0\,\mathrm{s}$ is defined as the $T0$ employed in the \textit{Fermi}-GBM data products \citep{GCN_170207A_Fermi}.}
  \label{fig:170207A_lc}
\end{subfigure}
\newline
\begin{subfigure}{.5\textwidth}
  \centering
  % include second image
  \includegraphics[width=.95\linewidth]{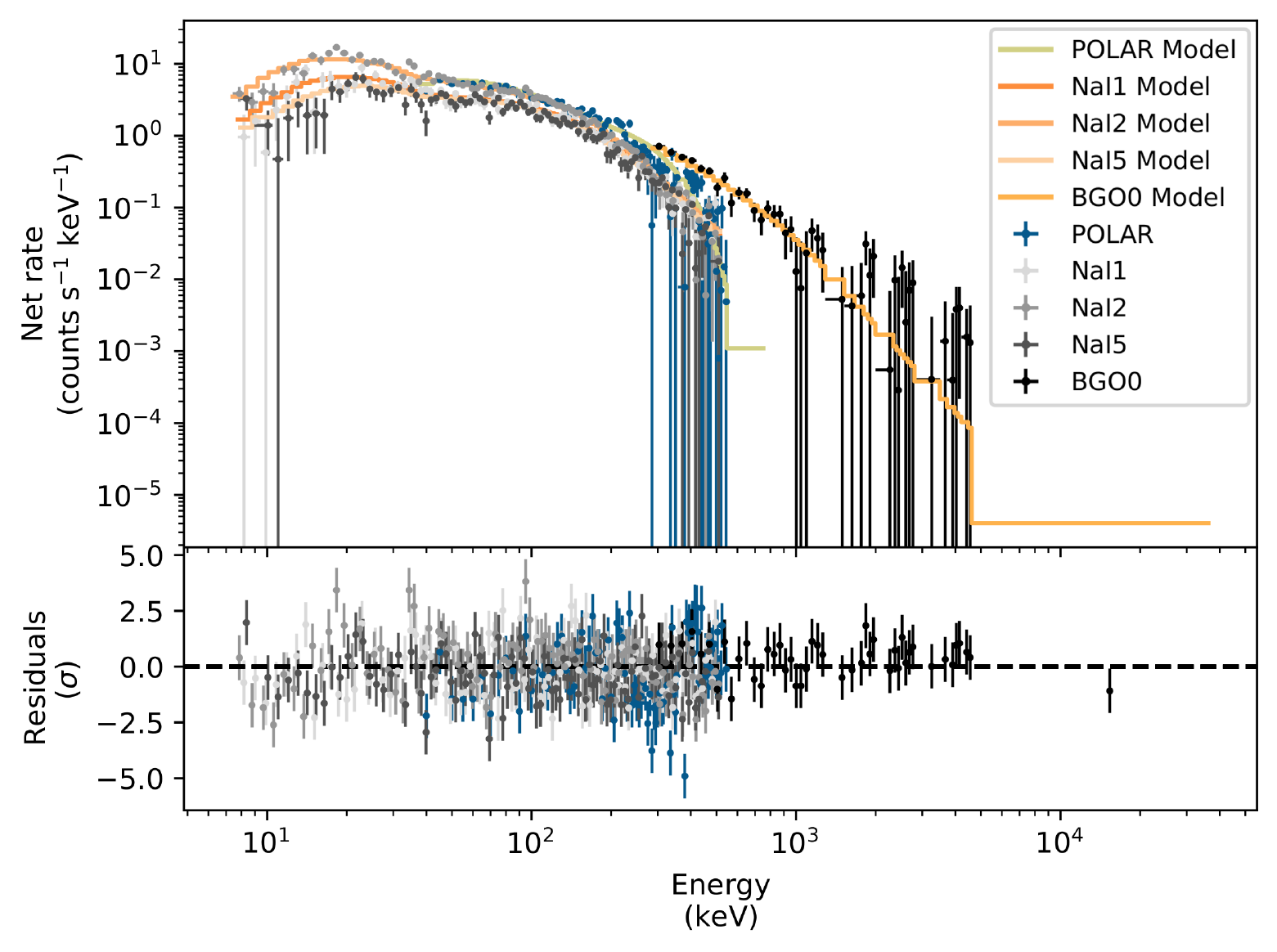}  
  \caption{The joint spectral fit result for 170207A. The number of counts as detected by both POLAR (blue) and the different NaI and BGO detectors of \textit{Fermi}-GBM (gray tints) are shown along with the best fitting spectrum folded through the instrument responses in yellow for POLAR data and in orange tints for the \textit{Fermi}-GBM data. The residuals for both data sets are shown in the bottom of the figure.}
  \label{fig:170207A_cs}
\end{subfigure}
\caption{The light curve as measured by POLAR for GRB 170207A (a) along with the joint spectral fit results of POLAR and \textit{Fermi}-GBM for the signal region indicated in yellow in  figure (a).}
\label{fig:170207A_lc_cs}
\end{figure}

GRB 170207A, a long GRB with 3 clearly separated active periods, each of which consists of overlapping pulses, was detected by both POLAR and \textit{Fermi}-GBM. The latter reported a $T0$ of 2017-02-07 at 21:45:03.67 (UT) \citep{GCN_170207A_Fermi}, which, for convenience will be taken as $T0$ for the analysis presented here as well. A $T_{90}$ of $(38.76\pm0.26)\,\mathrm{s}$ was measured using POLAR data. The light curve, including the signal region (blue) and part of the background region (yellow) can be seen in figure \ref{fig:170207A_lc}. The GRB was detected by \textit{Fermi}-GBM \citep{GCN_170207A_Fermi} the data from 4 of its detectors was therefore used for a combined fit with the POLAR data. The results of the spectral fit can be seen in figure \ref{fig:170207A_cs}. The effective area correction (applied to the POLAR data) found in the analysis was $0.94\pm0.02$. The polarization response of POLAR was produced using the location calculated using the BALROG method, detailed in \cite{BALROG} using \textit{Fermi}-GBM data for this GRB: RA (J2000) = $316.9 ^\circ$, Dec (J2000) = $59.1^\circ$, a localization error of $2^\circ$ was assumed in the response. The location of this GRB implies a significant off-axis incoming angle for the GRB of $67.2^\circ$. 

\begin{figure}[!ht]
   \centering
     \resizebox{\hsize}{!}{\includegraphics[width=15 cm]{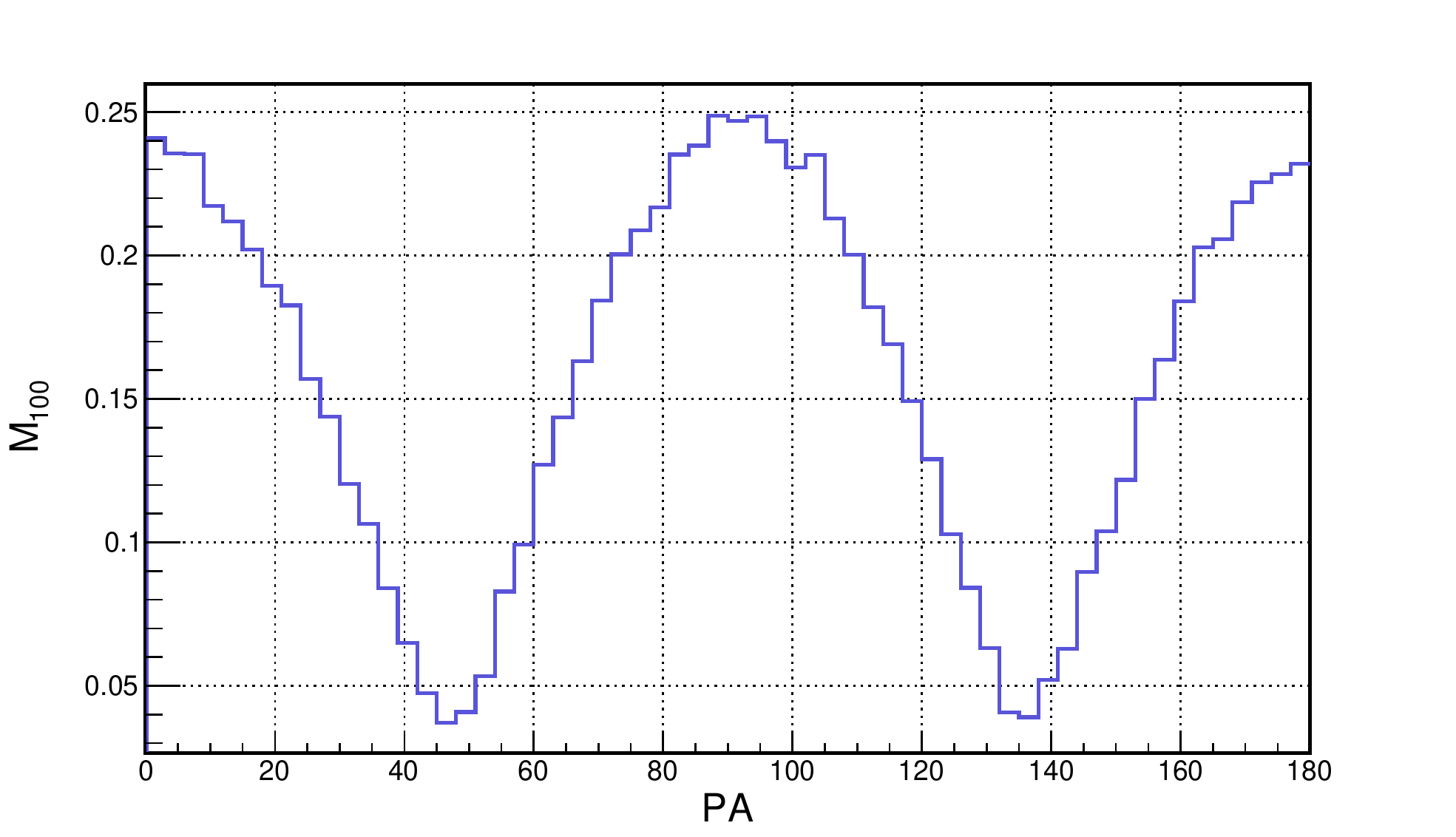}}
   \caption[Light curve of GRB 161203A as measured by POLAR.]
 {Simulated $M_{100}$ value (as defined in the text of this section) for a GRB entering POLAR with an off-axis angle of $85^\circ$ as a function of the PA (measured in the POLAR coordinate system). It can be seen that the sensitivity for polarization measurements are almost 0 for PAs of $45^\circ$ and $135^\circ$, this is a result of the large off-axis angle of such a GRB.}
 \label{fig:161203A_mod}
 \end{figure}

As a result of the large off-axis incoming angle of the GRB, the sensitivity of POLAR is reduced for specific PAs. This is a result of the design of POLAR which uses scintillator bars to measure the interaction locations of the photons. As there is no information of the interaction location within a bar only a 2d location, with a precision of the bar cross section of $6\times6\,\mathrm{mm^2}$, inside of the detector is provided for each interaction. For GRBs which occur on-axis this 2d plane is perpendicular to the incoming direction of the photons and therefore all polarization angles can be measured with the same precision. For GRBs which enter POLAR from the side this is no longer the case. The effect is illustrated in in figure \ref{fig:161203A_mod} which shows the $M_{100}$  as a function of the polarization angle for a GRB entering POLAR at $85^\circ$ off-axis. $M_{100}$ is a figure of merit which is defined as the amplitude of the $180^\circ$ modulation for a $100\%$ polarized incoming flux. It is generally acquired by dividing a scattering angle distribution produced using a $100\%$ polarized beam by the simulated scattering angle distribution for an unpolarized beam. The result is a scattering angle distribution which is a near perfect harmonic function \footnote{Significant $360^\circ$ modulation can also be found in this corrected scattering angle distribution, especially for an off-axis beam} with a $180^\circ$ modulation of which the amplitude is taken to be the $M_{100}$. It can be seen that for polarization angles of $45^\circ$ and $135^\circ$ (corresponding to polarization angles of $45^\circ$ degrees with respect to zenith direction) the sensitivity is almost 0. As all the plots in this paper give the PA as measured in the POLAR coordinate system these minima are in the same location for all the GRBs presented in this work. This effect becomes only significant for off-axis incoming angles, larger than approximately $65^\circ$.

\begin{figure}[!ht]
   \centering
     \resizebox{\hsize}{!}{\includegraphics[width=11 cm]{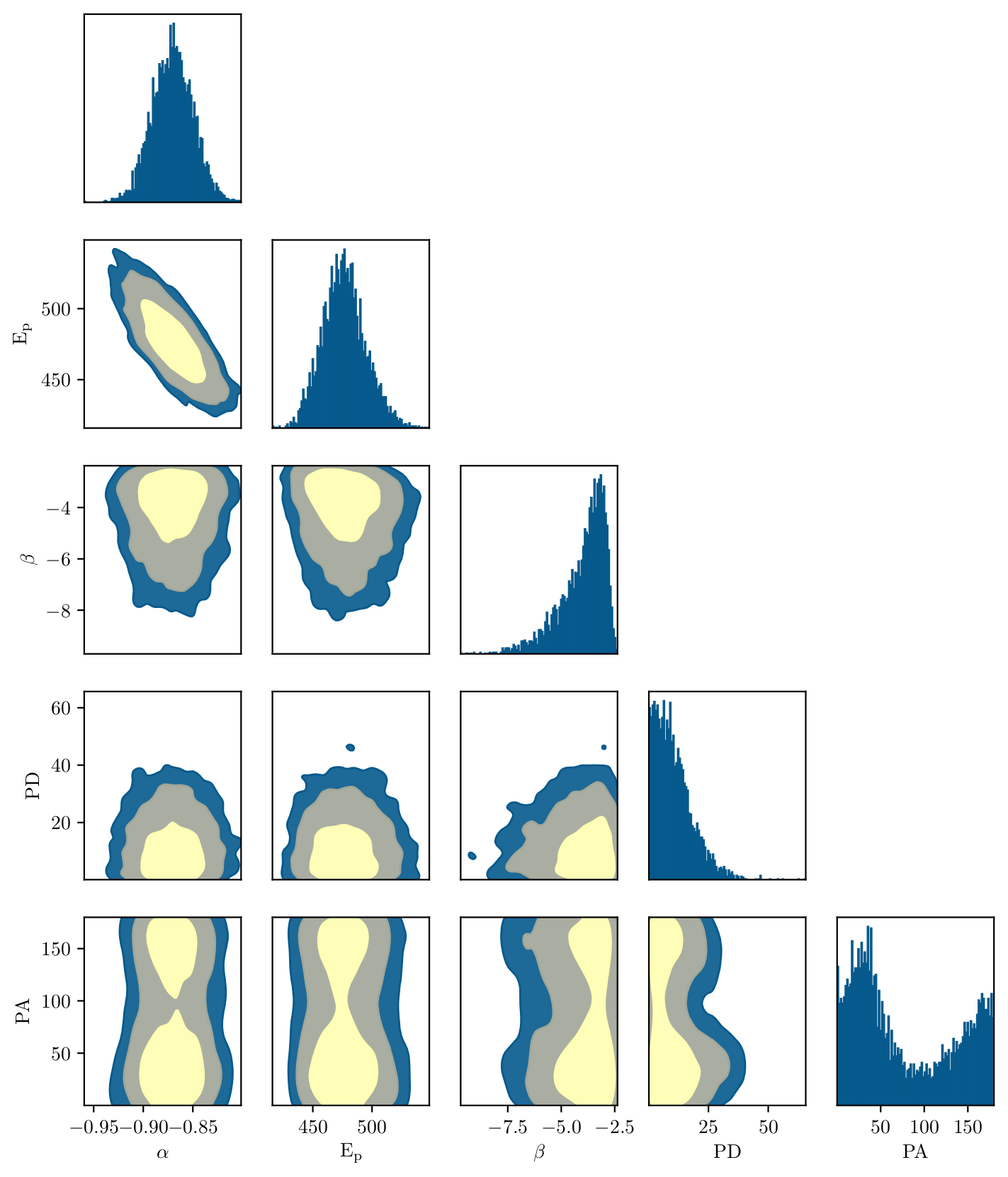}}
   \caption[Spectral and polarization posterior distributions.]
 {The spectral and polarization posterior distributions for GRB 170207A. The 1 and 2 $\sigma$ credibility intervals as well as that corresponding to $99\%$ are indicated. The polarization angle shown here is in the POLAR coordinate system, a rotation in the positive direction of 176 degrees transforms this to the coordinate system as defined by the IAU.}
 \label{fig:post_170207A}
 \end{figure}

\begin{figure}[ht]
\begin{subfigure}{.5\textwidth}
  \centering
  % include first image
  \includegraphics[width=.85\linewidth]{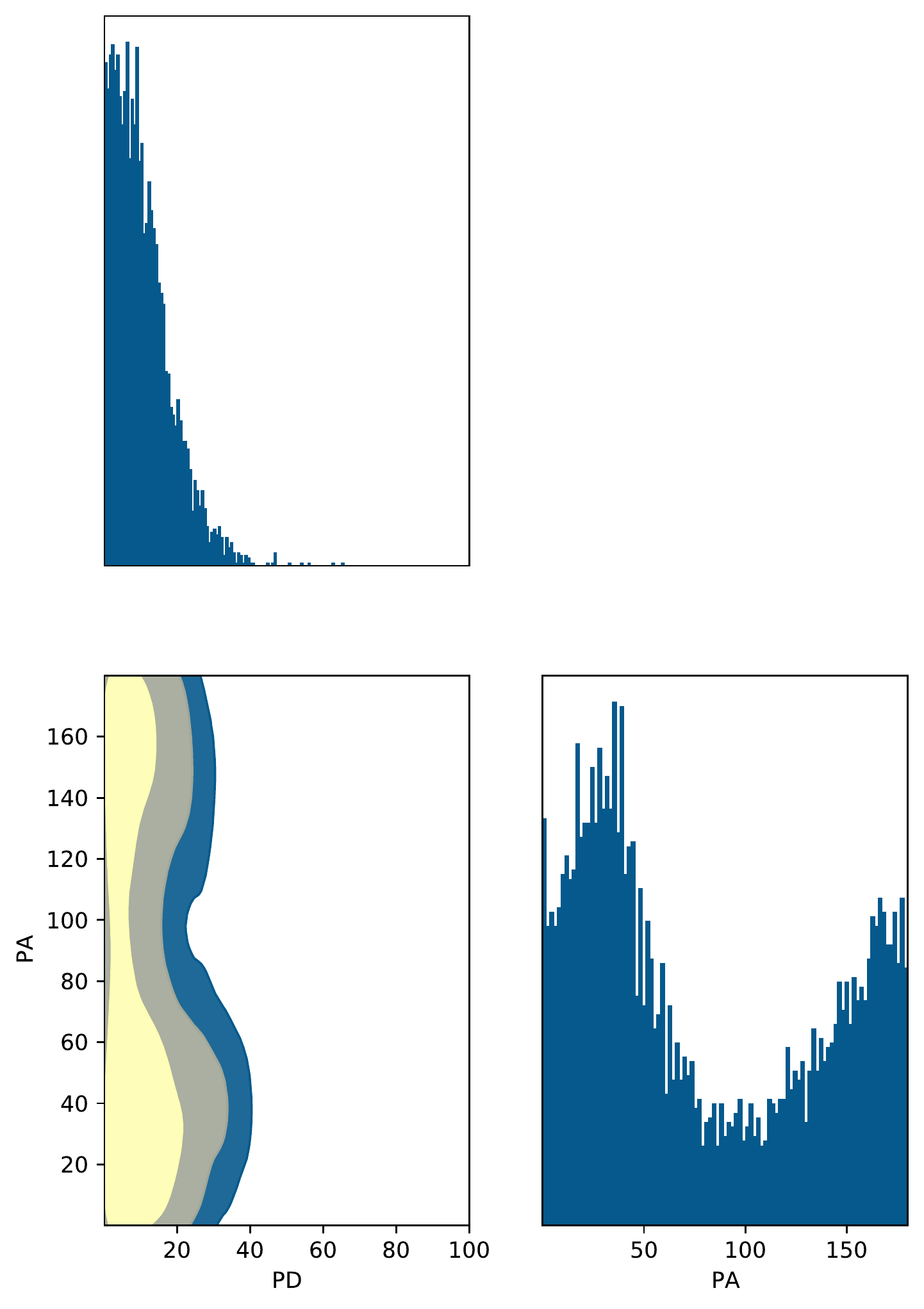}  
  \caption{The polarization posterior distributions for GRB 170207A with the 1 and 2 $\sigma$ credibility intervals as well as that corresponding to $99\%$ credibility. The polarization angle shown here is in the POLAR coordinate system, a rotation in the positive direction of 176 degrees transforms this to the coordinate system as defined by the IAU. }
  \label{fig:170207A_PD}
\end{subfigure}
\newline
\begin{subfigure}{.5\textwidth}
  \centering
  % include second image
  \includegraphics[width=.85\linewidth]{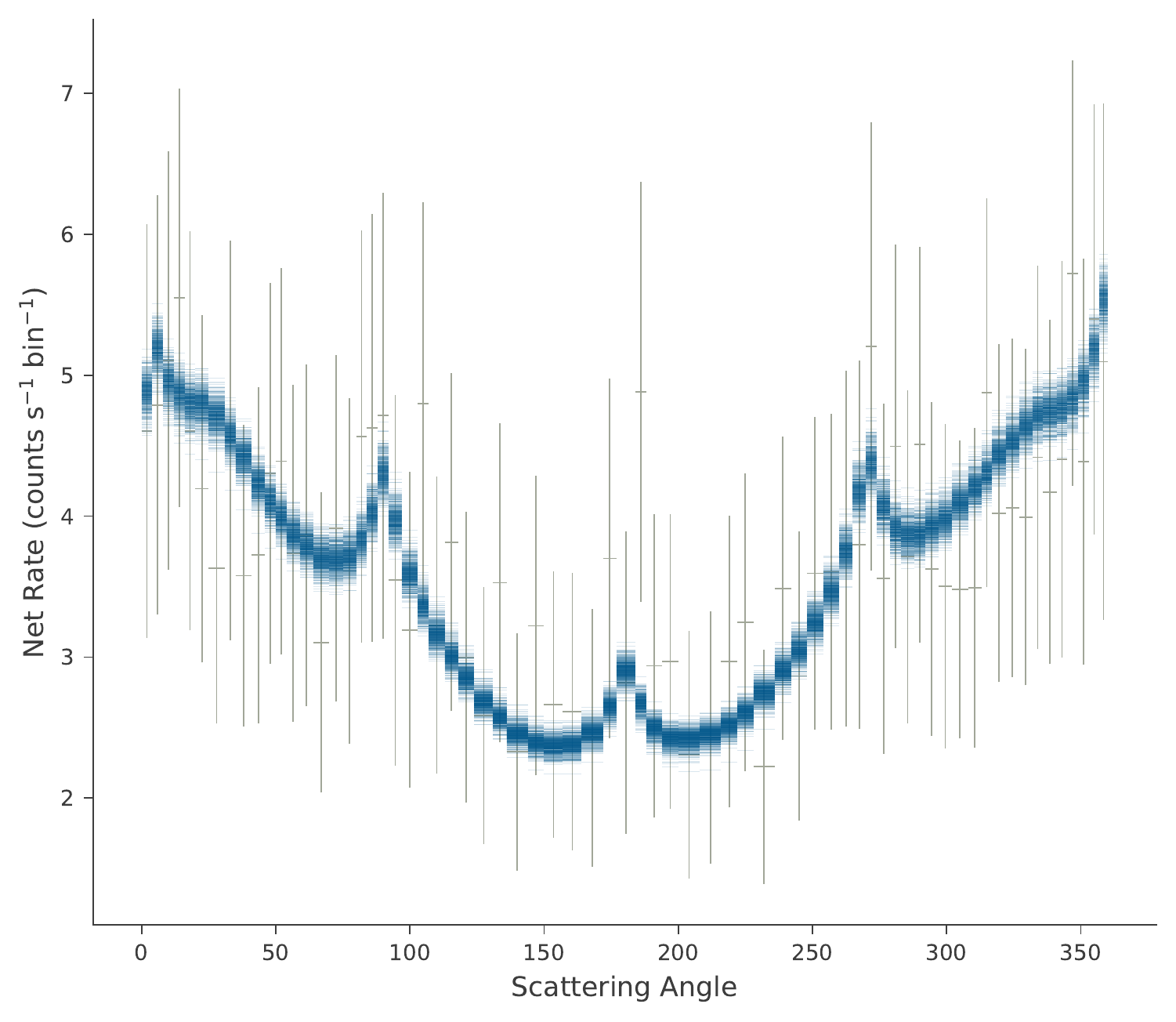}  
  \caption{The measured scattering angle distribution (gray data points with a 5 degree bin size) superimposed by the posterior model predictions (red) is shown.  The errors on the data points are the Poisson errors corrected for the background. }
  \label{fig:170207A_sd}
\end{subfigure}
\caption{The posterior distribution of the polarization parameters (a) together with the scattering angle distribution of GRB 170207A.}
\label{fig:170207A_PD_sd}
\end{figure}

The posterior distributions of the spectral and polarization parameters are shown in figure \ref{fig:post_170207A}. Finally the posterior distribution of the polarization parameters is shown together with measured scattering angle distribution superimposed by the posterior model predictions (blue) in figure \ref{fig:170207A_PD_sd}. The scattering angle distribution indicates clearly the large off-axis angle resulting in the $360^\circ$ modulation in the distribution. A PD of $5.9\substack{+9.6 \\ -4.8}\%$ was found along with $99\%$ credibility upper limit for PD of $35.9\%$.

\begin{figure}[!ht]
   \centering
     \resizebox{\hsize}{!}{\includegraphics[width=14 cm]{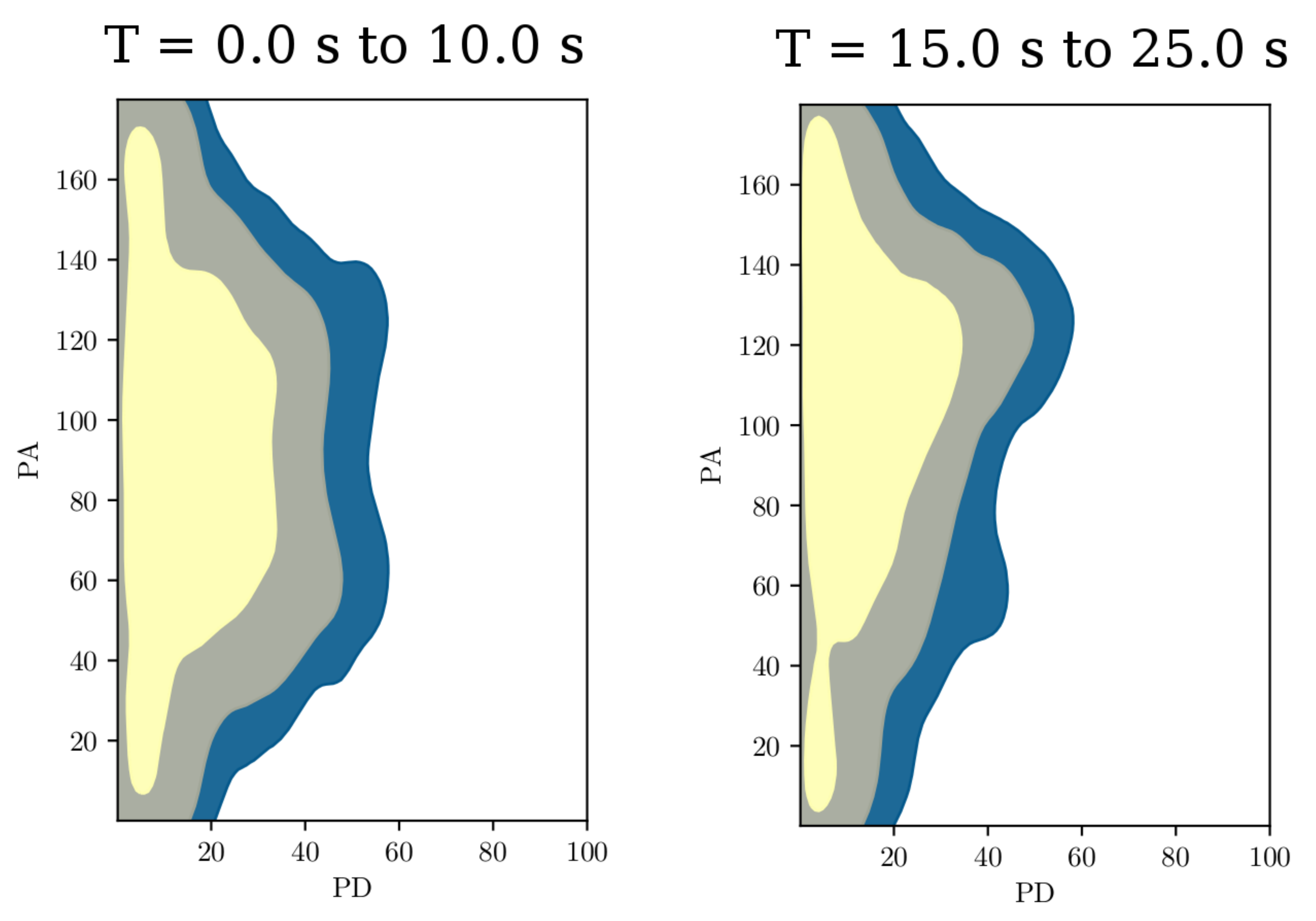}}
   \caption[Light curve of GRB 170207A as measured by POLAR.]
 {The polarization posterior distributions for the 2 time intervals studied for GRB 170207A are shown with the 1 and 2 $\sigma$ credibility intervals as well as that corresponding to $99\%$ credibility. On the left the posteriors are for the first selected time interval ranging from $T=0.0\,\mathrm{s}$ to $T=10.0\,\mathrm{s}$  (corresponding to the first emission period), on the right it is for the second time bin ranging from $T=15.0\,\mathrm{s}$  to $T=25.0\,\mathrm{s}$  (corresponding to the second emission period). The polarization angle shown here is in the POLAR coordinate system, a rotation in the positive direction of 176 degrees transforms this to the coordinate system as defined by the IAU.}
 \label{time_res_170207A}
 \end{figure}

GRB 170207A has a sufficiently high signal to background counts ratio to perform basic time-resolved analysis. The 3 separate emission episodes were therefore analyzed individually. While for the first two periods constraining measurements were possible for the last period it was not due to a lack of signal counts. Therefore, only the first 2 periods are discussed here. The first time bin ranged from T=0.0 to 10.0 seconds, the second from 15.0 to 25.0 seconds. The polarization parameter posteriors can be seen in figure \ref{time_res_170207A}. The results indicate PDs of $9.1\substack{+16.3 \\ -7.8}\%$ and $7.4\substack{+14.3 \\ -6.1}\%$ for time interval 1 and 2 respectively and are therefore consistent with an unpolarized flux. Upper limits with $99\%$ credibility for PD of $53.1\%$ and $50.9\%$ are found for these two time intervals. It should be noted here that despite being clearly separate time emission episodes within the GRB, both episodes consist of several overlapping pulses.

 \clearpage

\section{Conclusion}
\label{sec:conclusions}

\begin{figure}[!ht]
   \centering
     \resizebox{\hsize}{!}{\includegraphics{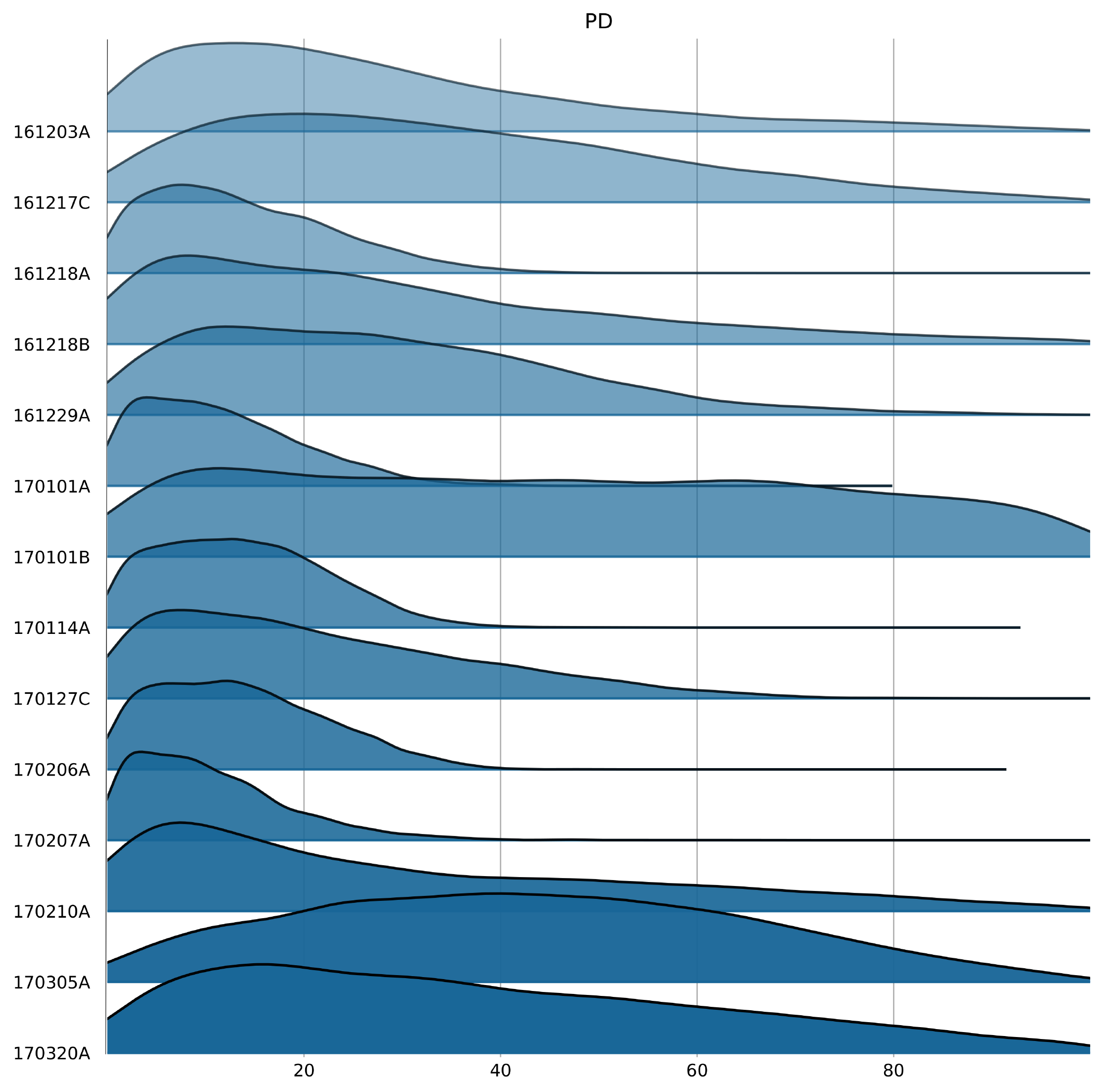}}
   \caption[The polarization results for the POLAR catalog.]
 {The posterior distributions of the polarization degree of the 14 GRBs studied in this work.}
 \label{fig:post_sum}
 \end{figure}

\begin{figure}[!ht]
   \centering
     \resizebox{\hsize}{!}{\includegraphics{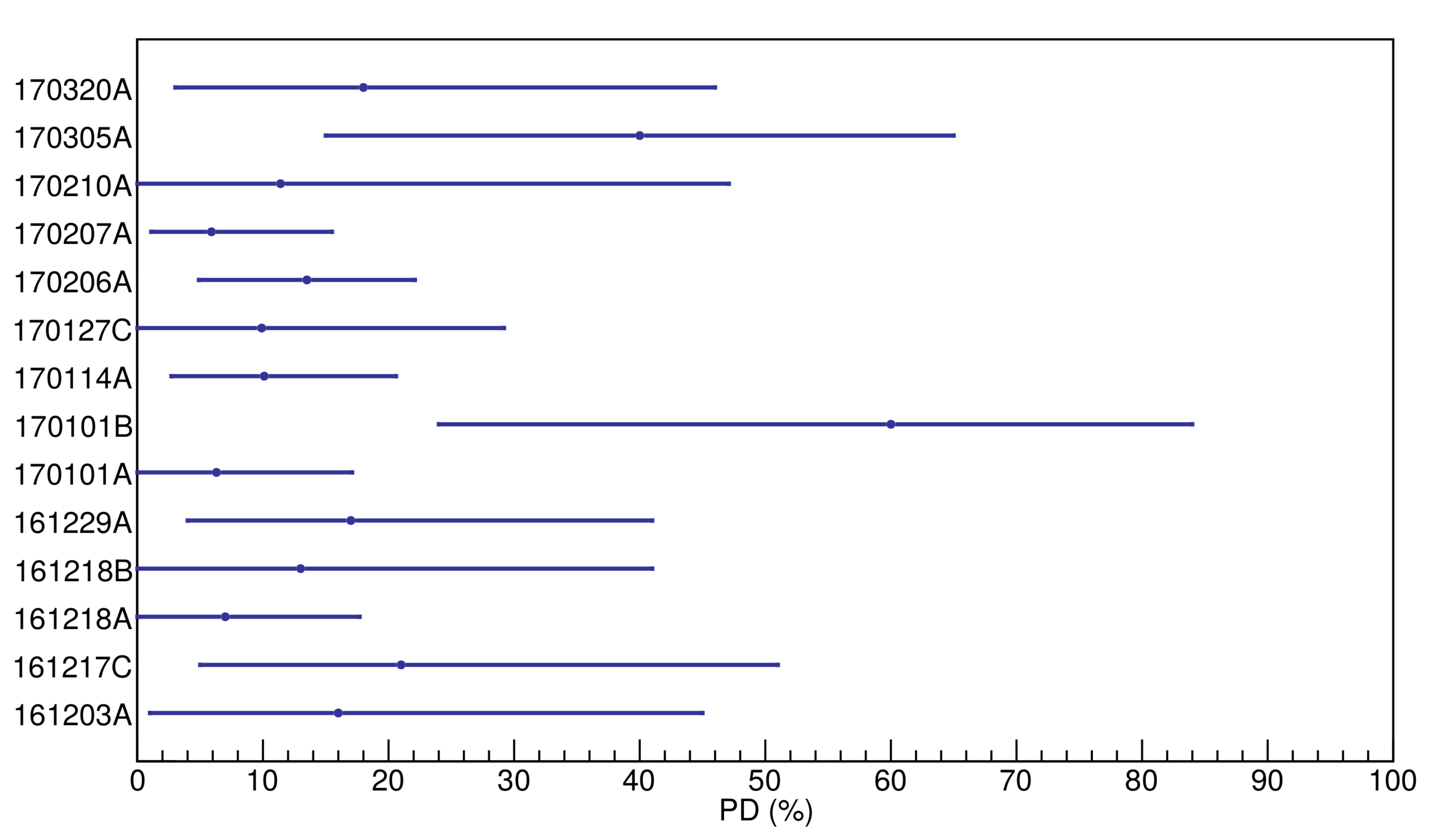}}
   \caption[The polarization results for the POLAR catalog.]
 {Overview of the polarization degrees found for the 14 GRBs analyzed here. The errors shown are those corresponding to a $68\%$ credibility.}
 \label{fig:catalog}
 \end{figure}

\begin{figure}[!ht]
   \centering
     \resizebox{\hsize}{!}{\includegraphics{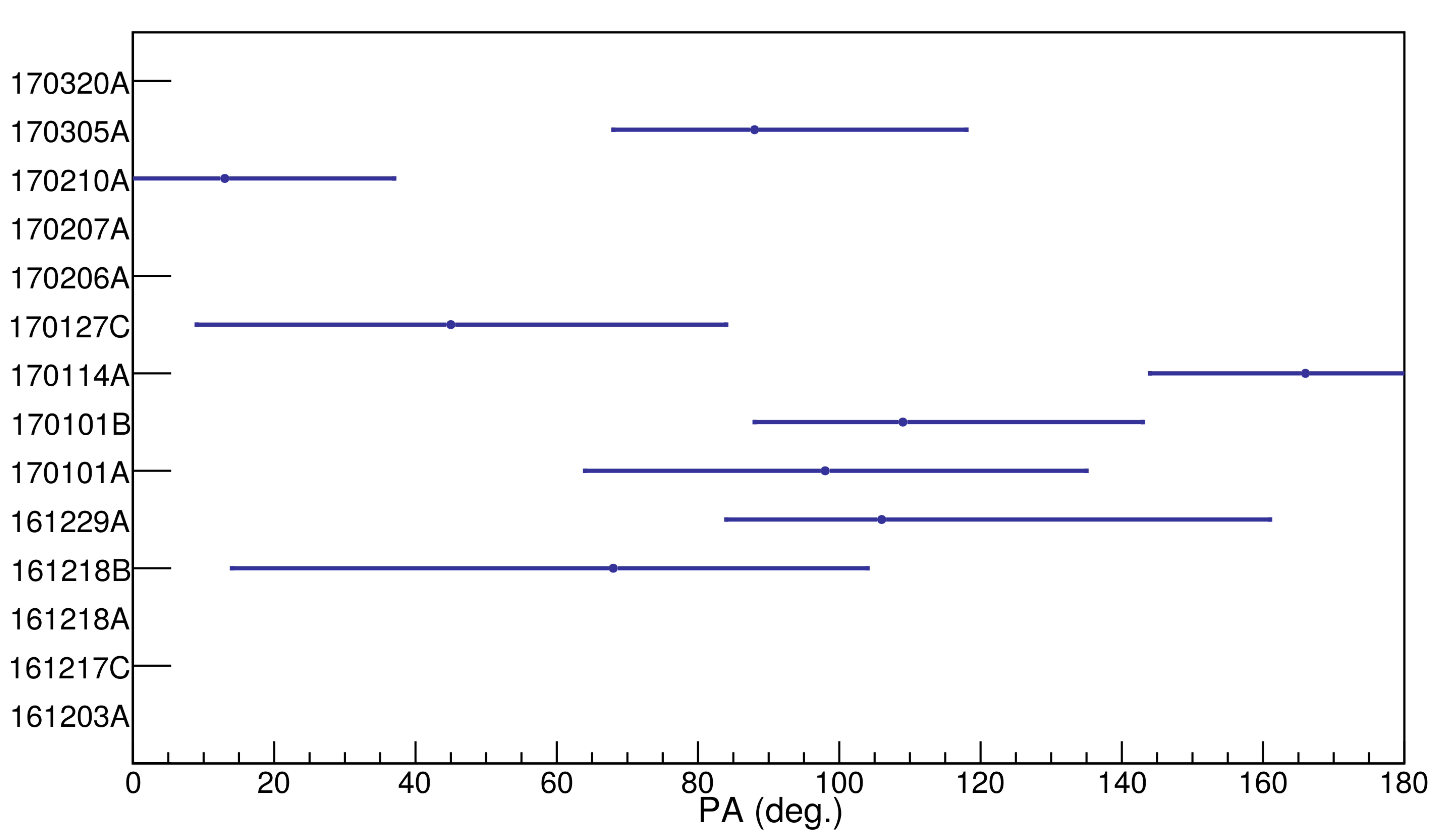}}
   \caption[The polarization results for the POLAR catalog.]
 {Overview of the polarization angles, with an angle definition following the IAU convention, found for the 14 GRBs analyzed here. The errors shown are those corresponding to a $68\%$ credibility. For GRBs for which no constraining PA was found (meaning that the total $1\sigma$ error exceeds $180^\circ$ no value is provided.}
 \label{fig:PA_catalog}
 \end{figure}

A more detailed analysis method, previously used in \cite{170114A_BALROG}, was employed to analyse the polarization of 14 GRBs detected by POLAR. The summary of the posteriors of the PD for the 14 GRBs can be seen in figure \ref{fig:post_sum}, while a simplified figure showing the PDs together with the $1\,\sigma$ error bars, is shown in figure \ref{fig:catalog}. The corresponding PAs are shown in Fig. \ref{fig:PA_catalog} where the angles are defined following the IAU convention. Finally, the spectral and polarization parameters (including the fluence) found in this analysis for all 14 GRBs are summarized in table \ref{tab:sum}. Looking at the results we can firstly conclude that the new results are compatible with previously published polarization results of POLAR for 5 of these GRBs \citep{Zhang+Kole,170114A_BALROG}. We typically find low levels of polarization with the exception of GRB 170101B for which a PD of $60\%$ is found. However it is within a credibility of $90\%$ to come from an unpolarized flux. The combination of the 14 posteriors shown in figure \ref{fig:post_sum} is, despite the high PD found for that GRB, compatible with what one would expect for unpolarized or lowly polarized emission. While for several GRBs no tight constraints on the polarization degree are possible due to a lack of sensitivity for specific PAs, it remains possible for these GRBs to exclude high PDs for all polarization angles with the exception of those for which there is poor sensitivity. This is specifically the case for GRBs 170210A and 161218A and in a lesser extent for GRBs 161203A, 161229A and 170320A. 

Despite an agreement for the time integrated polarization results with a low or unpolarized GRB flux, we again find strong hints for intra-pulse evolution of the PA in single pulse GRBs. The emission within single pulses appears to be polarized at around $30\%$. However, the PD is washed out due to a quickly evolving PA. We confirm the strong hints previously presented for GRB 170114A in both \cite{Zhang+Kole} and \cite{170114A_BALROG}. Here we find additionally that GRB 170101A shows hints of this behavior. More precise measurements, such as those expected to come from POLAR-2 \citep{POLAR-2}, scheduled for launch in 2024, or LEAP \citep{LEAP}, currently proposed with a potential launch in 2025, are important to fully probe the time evolution of the polarization parameters. Finally, we only see signs of an evolving PA inside of single pulses. We do not see any signs of evolving polarization angles outside of individual pulses or in overlapping pulses, such as those reported for example by GAP for GRB 100826A \citep{GAP} or by the AstroSat CZT-Imager collaboration for GRB 160821A and 160325A \citep{Astro_160821A, Astro_160325A}.

\begin{table*}[!ht]
  \scriptsize
  \centering
 \caption{The length, fluence, location and polarization properties of the 14 GRBs resulting from the analysis presented in this work. The $T_{90}$ parameter as calculated using POLAR data is provided along with  the off-axis incoming angle ($\theta$) and the RA (J2000) and Dec (J2000) positions used in the analysis. For the RA (J2000) and Dec (J2000) parameters we report those used to produce the instrument response in the analysis. In case PA, which is given according to the IAU convention, is not constrained (meaning that the total $1\sigma$ error exceeds $180^\circ$ an x is added in this table. (*in units of erg/cm$^2$ in 10-1000 keV).}\label{tab:sum}
\hspace*{-2.1cm}\begin{tabular}{|c|c|c|c|c|c|c|c|}
  \hline
  GRB & T90 (s) & Fluence* & PD (\%) & PA (deg.) & $\theta$ & RA (deg.) & Dec (deg.) \\[5pt] \hline\hline
  161203A & $4.1\pm{+0.1}$ & $(7.84\pm1.05)\times10^{-6}$& $16\substack{+29 \\ -15}$ & x &$85^\circ$ & $13.5^\circ$ & $-146^\circ$ \\[5pt] \hline
   161217C & $6.3\pm{+0.3}$ & $(4.37\pm1.05)\times10^{-6}$ & $21\substack{+30 \\ -16}$ & x &  $35^\circ$ & $34.6^\circ$ & $-4.0^\circ$ \\[5pt] \hline
   161218A & $11.5\pm{+0.1}$ & $(8.72\pm1.44)\times10^{-6}$ & $7.0\substack{+10.7 \\ -7.0}$ & x & $24.3^\circ$ & $245.3^\circ$ & $-4.1^\circ$ \\[5pt] \hline
    161218B & $25.1\pm{+0.2}$ & $(8.55\pm0.30)\times10^{-5}$ & $13\substack{+28 \\ -13}$& $68\substack{+36 \\ -54}$ & $80.5^\circ$ & $0.9^\circ$ & $-14.7^\circ$ \\[5pt] \hline
    161229A & $31.3\pm{+0.4}$ & $(4.21\pm0.22)\times10^{-5}$ & $17\substack{+24 \\ -13}$ & $106\substack{+55 \\ -22}$ & $87.6^\circ$ & $78.9^\circ$ & $45.7^\circ$ \\[5pt] \hline
     170101A & $2.0\pm{+0.1}$ & $(6.8\pm1.4)\times10^{-6}$ & $6.3\substack{+10.8 \\ -6.3}$& $98\substack{+37 \\ -34}$ & $6.0^\circ$ & $276.1^\circ$ & $11.7^\circ$ \\[5pt] \hline
     170101B &$11.1\pm{+0.4}$ & $(1.2\pm0.12)\times10^{-5}$ & $60\substack{+24 \\ -36}$ & $109\substack{+34 \\ -21}$  & $75.0^\circ$ & $69.6^\circ$ & $-1.0^\circ$ \\[5pt] \hline
     170114A &$10.4\pm{+0.2}$ & $(1.84\pm0.30)\times10^{-5}$ & $10.1\substack{+10.5 \\ -7.4}$ & $166\substack{+25 \\ -22}$ & $26.4^\circ$ & $13.1^\circ$ & $-13.0^\circ$ \\[5pt] \hline
     170127C & $0.14\pm{+0.01}$ & $(5.4\pm2.4)\times10^{-6}$ & $9.9\substack{+19.3 \\ -8.4}$ & $45\substack{+39 \\ -36}$ & $41.8^\circ$ & $339.3^\circ$ & $-63.9^\circ$ \\[5pt] \hline
     170206A &$1.26\pm{+0.01}$ & $(1.04\pm0.06)\times10^{-5}$& $13.5\substack{+7.4 \\ -8.6}$ & x & $19.5^\circ$ & $212.8^\circ$ & $14.5^\circ$ \\[5pt] \hline
     170207A & $38.8\pm{+0.3}$ & $(6.80\pm0.47)\times10^{-5}$ & $5.9\substack{+9.6 \\ -5.9}$& x & $67.2^\circ$ & $316.9^\circ$ & $59.1^\circ$ \\[5pt] \hline
     170210A & $47.6\pm{+2.5}$ & $(7.21\pm0.28)\times10^{-5}$ & $11.4\substack{+35.7 \\ -9.7}$& $13\substack{+24 \\ -21}$ & $80.6^\circ$ & $226.1^\circ$ & $-65.1^\circ$ \\[5pt] \hline
     170305A & $0.45\pm{+0.01}$ & $(1.08\pm0.12)\times10^{-6}$ & $40\substack{+25 \\ -25}$& $88\substack{+30 \\ -20}$  & $31.4^\circ$ & $39.7^\circ$ & $9.9^\circ$ \\[5pt] \hline
     170320A & $6.8\pm{+0.1}$ & $(9.6\pm4.8)\times10^{-6}$ & $18\substack{+32 \\ -18}$& x & $84.7^\circ$ & $320.1^\circ$ & $55.1^\circ$ \\[5pt] \hline
\end{tabular}
\end{table*}

\section{Discussion}

The time integrated results presented here indicate that the PD of the GRB prompt emission in the energy range of approximately 30 to 750 keV is unpolarized or lowly polarized. This result is found consistently for both short and long GRBs, multi-peak and single peak GRBs. The results are therefore in contradiction to the results reported by the AstroSat CZT-Imager collaboration presented in \cite{AstroSAT} where typically polarization levels around $50\%$ are found and an unpolarized flux is excluded with $3\sigma$ certainty for several GRBs. Future joint polarization analysis for GRBs observed by both instruments, which are possible using the \texttt{3ML} framework, is vital to resolve the origin of these incompatible results. As the AstroSat mission has reported an observation for GRB 161218B \citep{GCN_161218B_Astro} for which we here exclude a high level of polarization for the majority of polarization angles, this GRB forms a prime target for such a joint analysis in the future.

These POLAR results appear to disfavor synchrotron models with globally ordered toroidal magnetic fields \citep{Gill2019}. The time integrated results are however consistent with synchrotron emission with a radial or normal to the radial magnetic field as well as Compton drag models. Although high values for PD can be produced in special cases in such models, polarization measurements for a sample of GRBs would result in a distribution peaking at $0\%$ \citep{Gill2019}. Additionally, the results are compatible with photospheric emission which predicts PD distributions peaking at $0\%$ and not exceeding $40\%$ \citep{Lundman2014}. The time integrated results presented here therefore only exclude synchrotron emission with globally ordered toroidal magnetic fields. 

The results additionally agree with the predictions of the majority of models regarding the evolution of the PD and PA over multiple pulses \citep{Gill2019}, as no significant PD is found during time-resolved analysis where pulses are separated. However, the results presented here do confirm the hints of intra-pulse evolution found for GRB 170114A in previous studies and finds an additional, although not as strong, hint for this for GRB 170101A. For all GRBs for which intra-pulse time-resolved studies were possible hints are found for PD values around $30\%$ with an evolving PA. The authors are not aware of any theoretical work performed on evolution of the PA within single pulse GRBs and therefore encourage the community to provide predictions for this. 

It should finally be noted that, although the work presented here contains the vast majority of all the information which can be extracted from the POLAR mission regarding GRBs, energy resolved studies have not yet been performed. Such studies have the possibility to, for example, test predictions such as those made in \cite{Lundman2018} for photospheric emission. In such models the PD for the low energy emission (10's of keV) can be relatively high, while at higher energies the PD is lost due to comptonization of this emission component. Whilst this analysis appears straightforward, the energy dispersion of gamma-ray detectors requires one to develop dedicated analysis techniques for this. Such studies are therefore expected to be performed with existing POLAR data in the future.
 
 \begin{acknowledgements}
  We gratefully acknowledge the financial support from the National Natural Science Foundation of China (Grant No. 11961141013, 11503028), the Joint Research Fund in Astronomy under the cooperative agreement between the National Natural Science Foundation of China and the Chinese Academy of Sciences (Grant No. U1631242), the Xie Jialin Foundation of the Institute of High Energy Phsyics, Chinese Academy of Sciences (Grant No. 2019IHEPZZBS111),  the National Basic Research Program (973 Program) of China (Grant No. 2014CB845800), the Strategic Priority Research Program of the Chinese Academy of Sciences (Grant No. XDB23040400), the Swiss Space Office of the State Secretariat for Education, Research and Innovation (ESA PRODEX Programme), the National Science Center of Poland (Grant No. 2015/17/N/ST9/03556), and the Youth Innovation Promotion Association of Chinese Academy of Sciences (Grant No. 2014009). J.M. Burgess acknowledges support from the Alexander von Humboldt Foundation. The authors are grateful to the \textit{Fermi}-GBM team and HEASARC for public access to \textit{Fermi} data products and to the \textit{Swift}-BAT team for public access to their data products. Specifically, we would like to thank Dr. Amy Lien for assistance with the analysis of the \textit{Swift}-BAT data. Finally, we are extremely thankful for the anonymous referee who went through the paper in extreme detail and who, through many suggestions, helped to significantly improve this work.
  
 \end{acknowledgements}

\appendix{}

\section{161203A}

\begin{figure}[ht]
\begin{subfigure}{.5\textwidth}
  \centering
  % include first image
  \includegraphics[width=.95\linewidth]{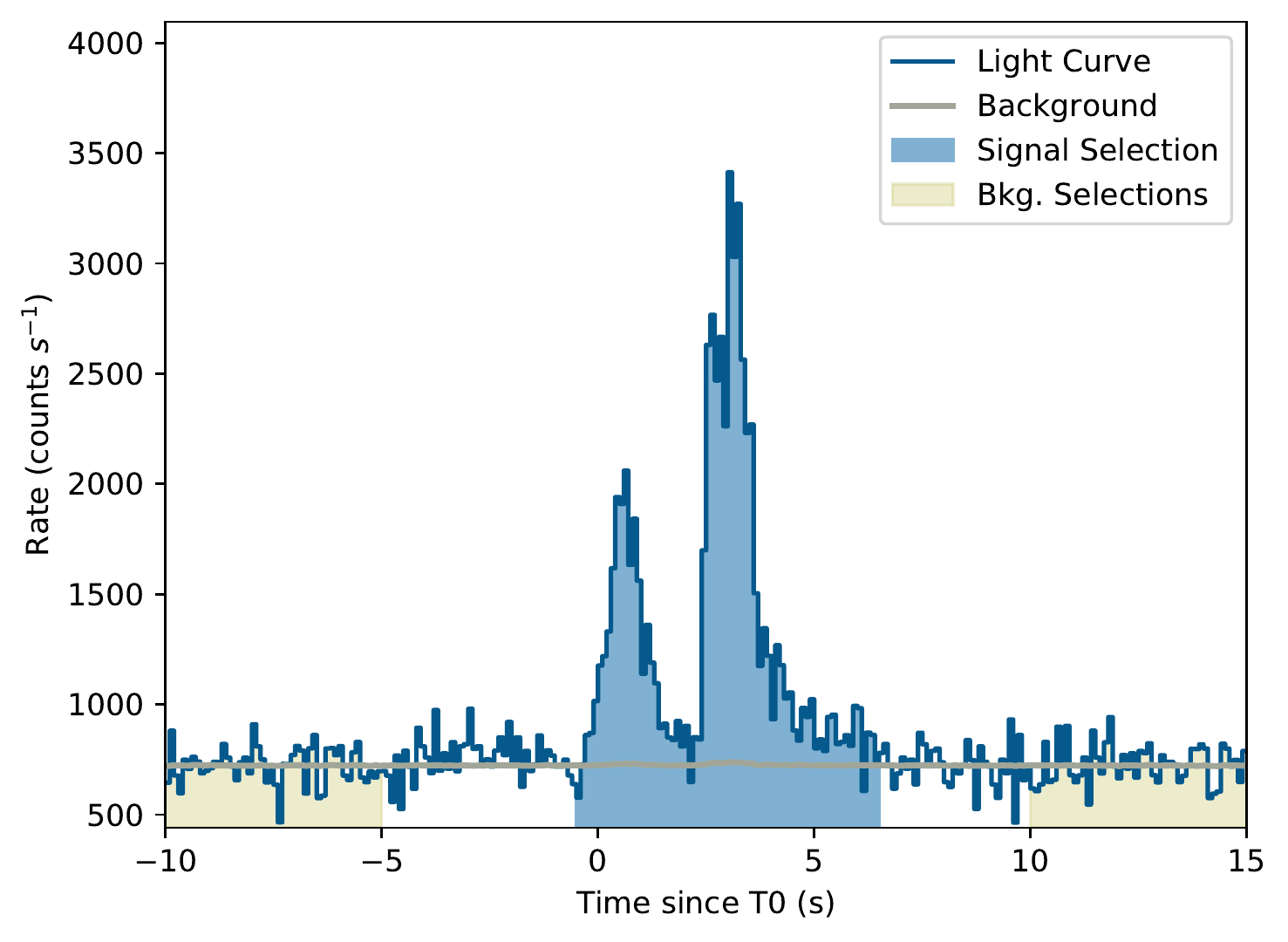}  
  \caption{The light curve of GRB 161203A as measured by POLAR, where $T=0\,\mathrm{s}$ is defined as the detection time measured by POLAR \citep{POLAR_GCN_161203A}}
  \label{fig:161203A_lc}
\end{subfigure}
\newline
\begin{subfigure}{.5\textwidth}
  \centering
  % include second image
  \includegraphics[width=.95\linewidth]{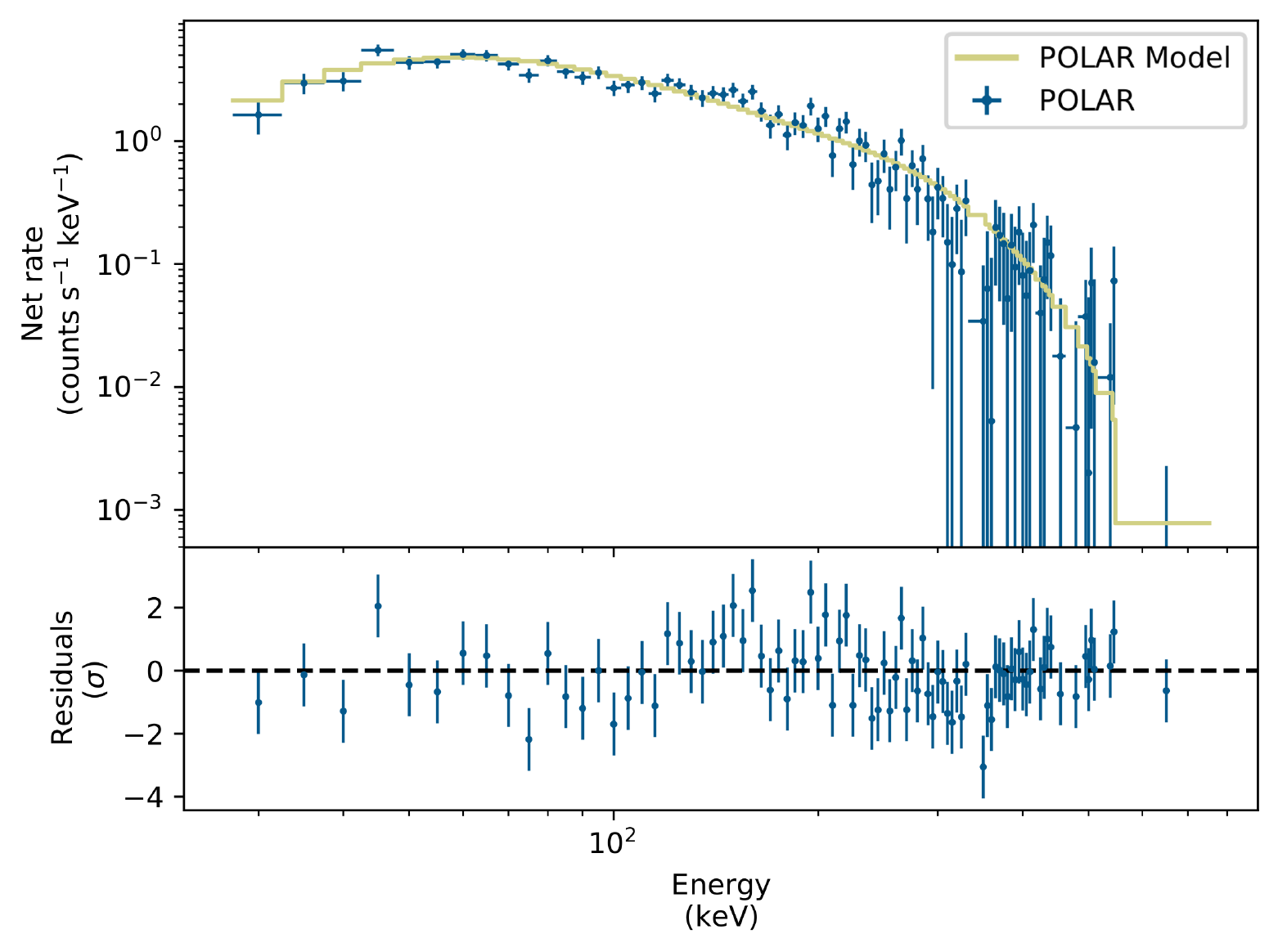}  
  \caption{The joint spectral fit result for 161203A. The number of counts as detected by both POLAR (blue) is shown along with the best fitting spectrum folded through the instrument response in yellow. The residuals are shown in the bottom of the figure.}
  \label{fig:161203A_cs}
\end{subfigure}
\caption{The light curve as measured by POLAR for GRB 161203A (a) along with the joint spectral fit result of POLAR for the signal region indicated in yellow in figure (a).}
\label{fig:161203A_lc_cs}
\end{figure}

GRB 161203A was detected by POLAR on 2016-12-03 at 18:41:07.75 (UT) \citep{POLAR_GCN_161203A}, which is taken to be $T0$ for the analysis presented here. A $T_{90}$ of $(4.1\pm0.1)\,\mathrm{s}$ was measured using POLAR data. The light curve, including the signal region (blue) and part of the background region (yellow) can be seen in figure \ref{fig:161203A_lc}. It was not detected by \textit{Fermi}-GBM or \textit{Swift}-BAT. The GRB was detected by Konus-Wind \citep{KONUS_cat_161203A}, however, no spectral parameters were reported and therefore no priors could be used. Furthermore, no location was reported for this GRB by any instrument. For this purpose the POLAR localization method reported in \cite{Yuanhao} was used to calculate a relative position for POLAR corresponding to a $\theta=85^\circ$ and $\phi=170^\circ$ where $\theta$ is the angle with respect to the zenith of POLAR and $\phi$ is the angle with respect to the x-axis of POLAR (which corresponds to a scattering angle of $0^\circ$). These coordinates correspond to RA (J2000)=$194^\circ$ and Dec (J2000)=$-34^\circ$ with an uncertainty of $7^\circ$. The spectral results from the joint fit can be seen in figure \ref{fig:161203A_cs}.

\begin{figure}[!ht]
   \centering
     \resizebox{\hsize}{!}{\includegraphics{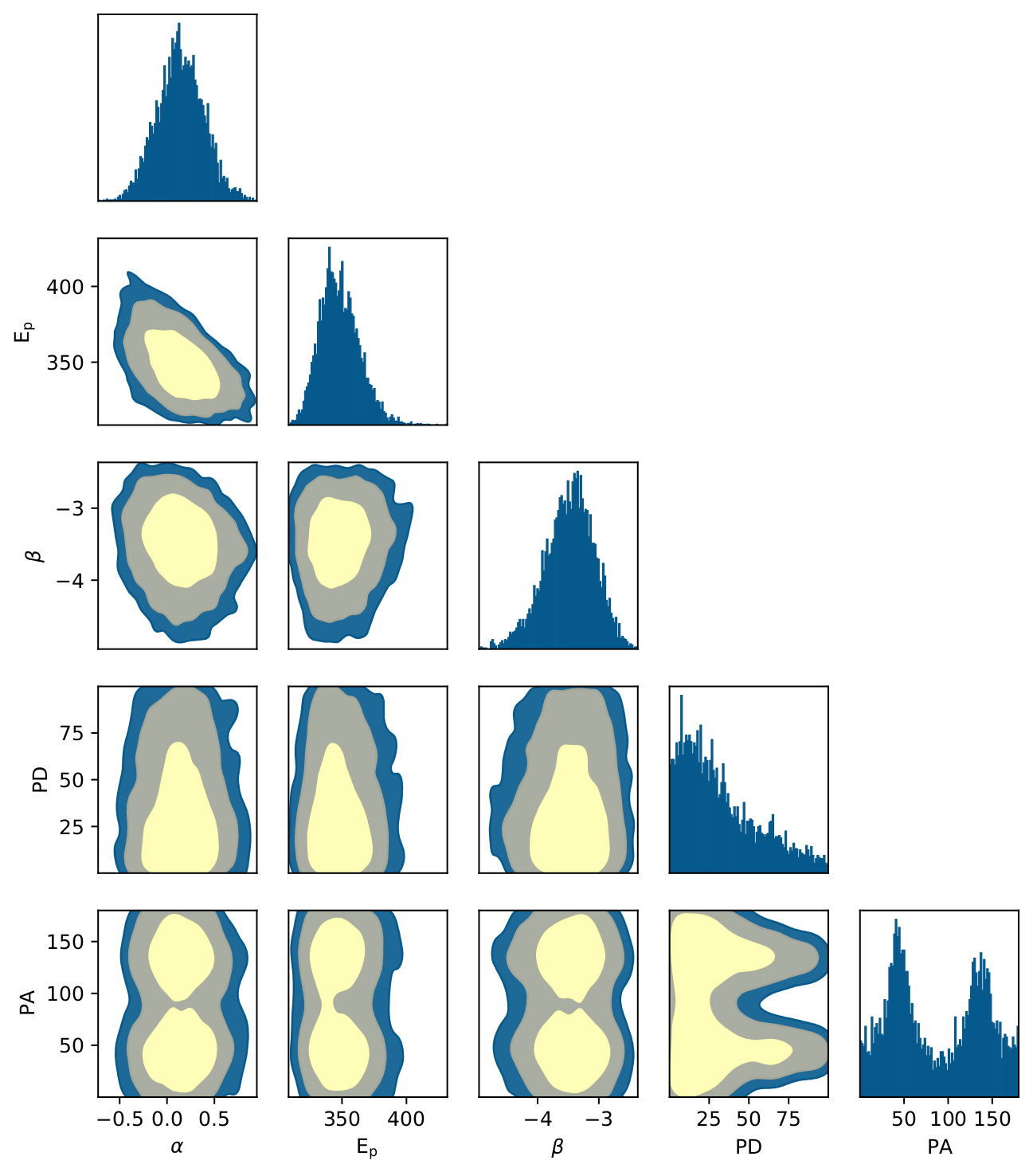}}
   \caption[Spectral and polarization posterior distributions.]
 {The spectral and polarization posterior distributions for GRB 161203A. The 1 and 2 $\sigma$ credibility intervals as well as that corresponding to $99\%$ are indicated. The PA shown here is in the POLAR coordinate system, a rotation in the positive direction of 98 degrees transforms this to the coordinate system as defined by the IAU.}
 \label{fig:post_161203A}
 \end{figure}

\begin{figure}[ht]
\begin{subfigure}{.5\textwidth}
  \centering
  % include first image
  \includegraphics[width=.85\linewidth]{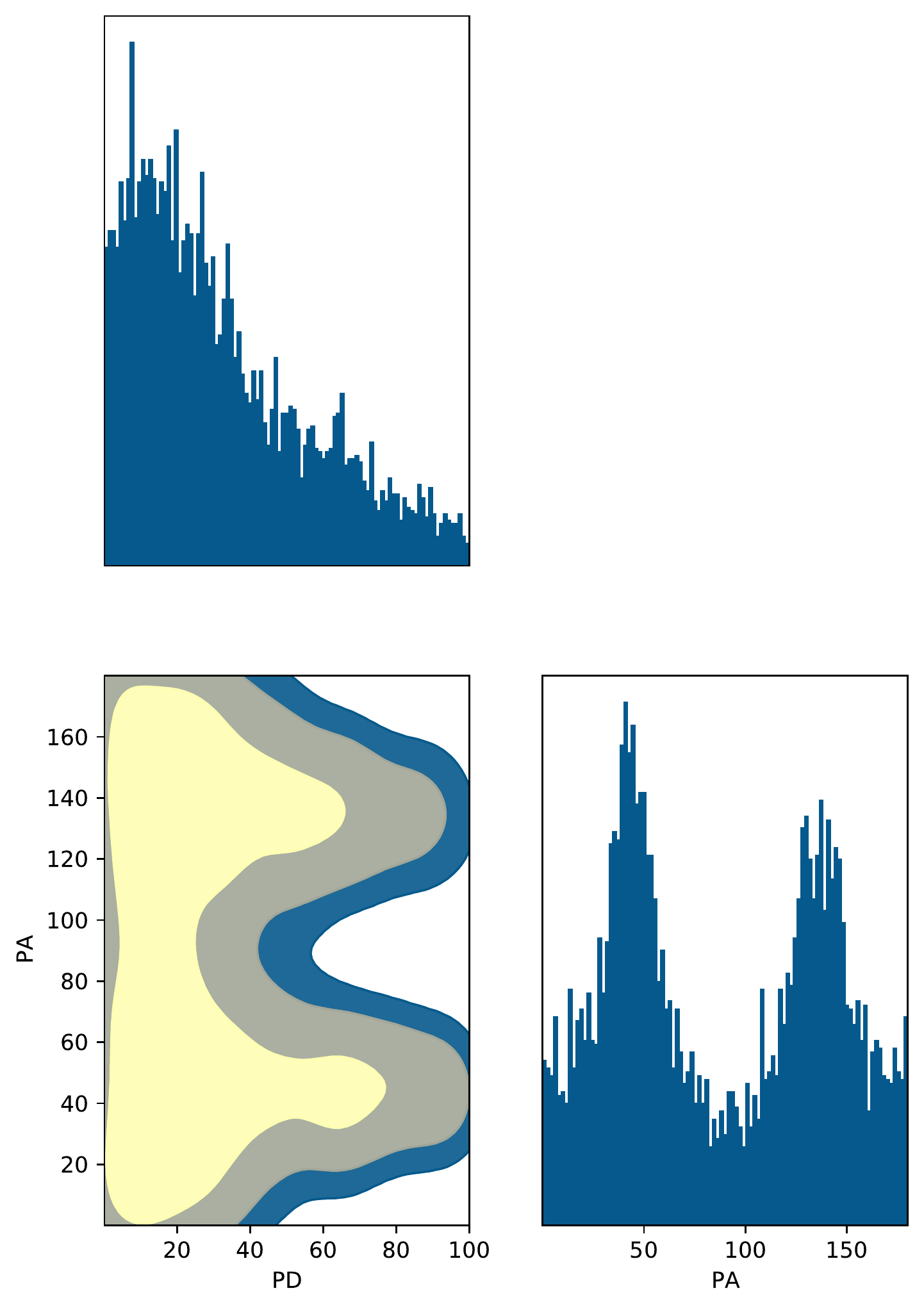}  
  \caption{The polarization posterior distributions for GRB 161203A with the 1 and 2 $\sigma$ credibility intervals as well as that corresponding to $99\%$ credibility.  The posterior distribution clearly reflects the lack of sensitivity for PAs of $45^\circ$ and $135^\circ$ in this measurement. The polarization angle shown here is in the POLAR coordinate system, a rotation in the positive direction of 98 degrees transforms this to the coordinate system as defined by the IAU.}
  \label{fig:161203A_PD}
\end{subfigure}
\newline
\begin{subfigure}{.5\textwidth}
  \centering
  % include second image
  \includegraphics[width=.85\linewidth]{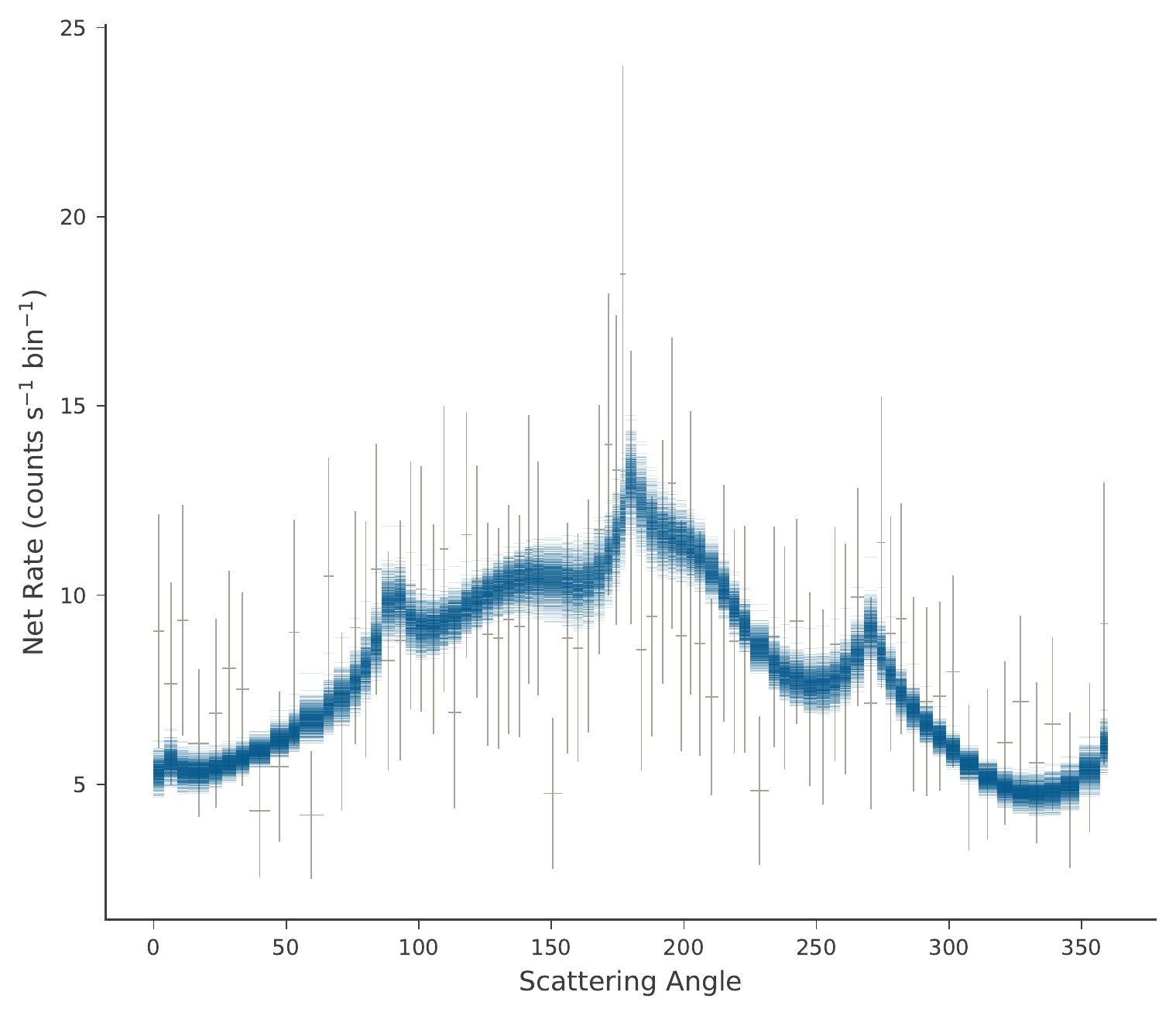}  
  \caption{The measured scattering angle distribution (gray data points with a 5 degree bin size) superimposed by the posterior model predictions (blue) is shown. The errors on the data points are the Poisson errors corrected for the background. }
  \label{fig:161203A_sd}
\end{subfigure}
\caption{The posterior distribution of the polarization parameters (a) together with the scattering angle distribution of GRB 161203A.}
\label{fig:161203A_PD_sd}
\end{figure}

The joint spectral and polarization fit for GRB 161203A was performed using POLAR data only. This makes GRB 161203A a GRB which makes use of POLAR data for the location, spectrum and polarization. As a result the systematic errors, which are all included in the final posterior distribution are relatively large. The posterior distributions of the spectral and polarization parameters are shown in figure \ref{fig:post_161203A}. Finally the posterior distribution of the polarization parameters is shown together with the measured scattering angle distribution superimposed by the posterior model predictions (blue) in figure \ref{fig:161203A_PD_sd}. The scattering angle distributions indicate clearly the large off-axis angle resulting in the $360^\circ$ modulation in the distribution. Additionally, the amplitude and phase of the $360^\circ$ modulation, which depend on the incoming direction of the GRB, can be seen to match between the model and the data, indicating that the localization method works properly. A PD of $16\substack{+29 \\ -15}\%$ is found. A $99\%$ credibility upper limit for PD of $98\%$ is found. Although these results exclude a very high polarization no proper constraining measurement was possible partly due to the lack of measurements by other instruments and partly due to the lack of sensitivity for a range of polarization angles due to the large incoming off-axis angle of the GRB. It should however be noted that while the polarization is found to be consistent with a PD of $0\%$ a high PD can be excluded with the exception of polarization angles for which POLAR was not sensitive due to the incoming angle. The signal to background count ratio in the POLAR data from this GRB do not allow for any time-resolved analysis.

 \clearpage

\section{161217C}

\begin{figure}[ht]
\begin{subfigure}{.5\textwidth}
  \centering
  % include first image
  \includegraphics[width=.95\linewidth]{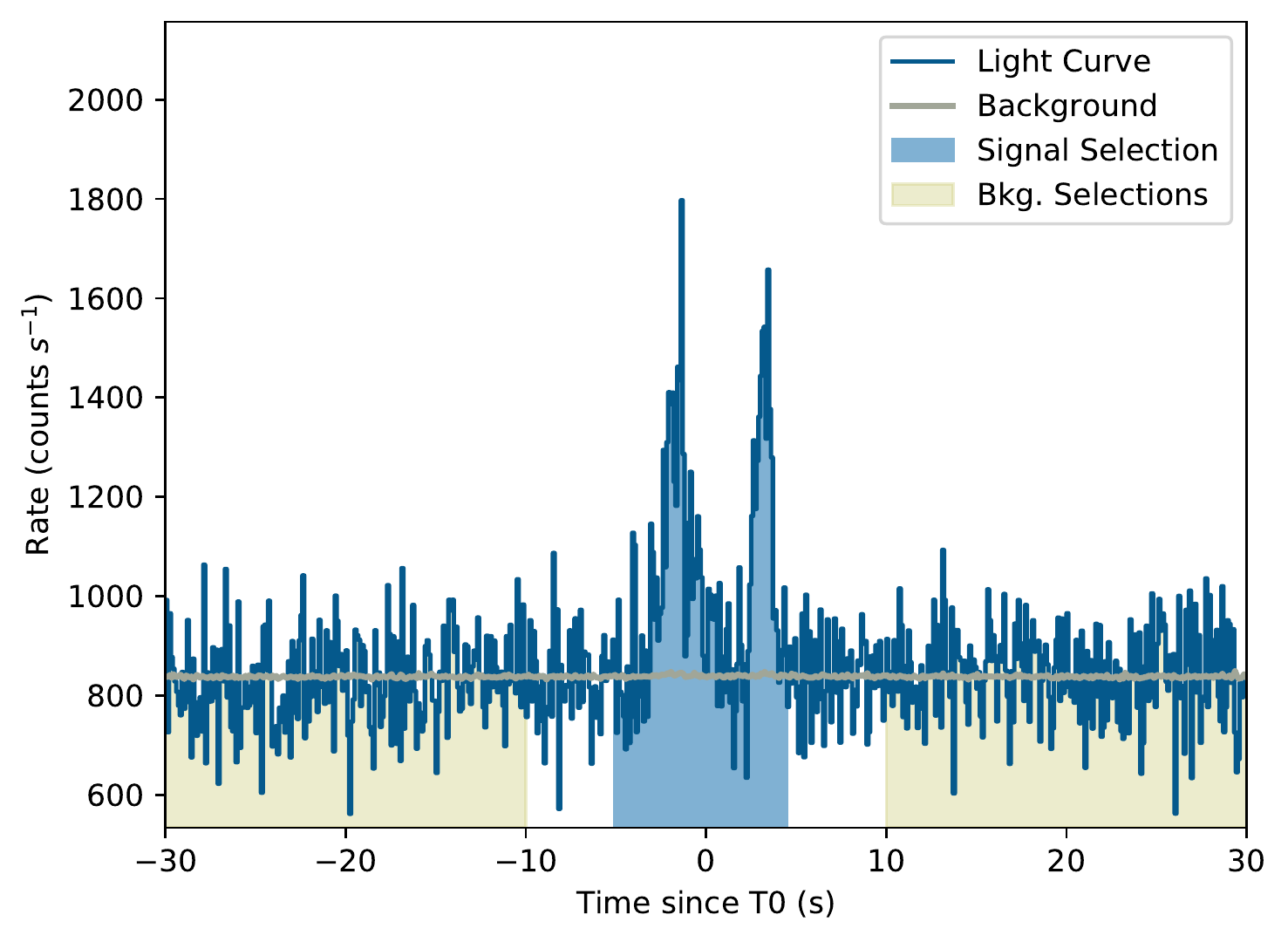}  
  \caption{The light curve of GRB 161217C as measured by POLAR, where $T=0\,\mathrm{s}$ is defined as the detection time by POLAR \citep{POLAR_GCN_161217C}.}
  \label{fig:161217C_lc}
\end{subfigure}
\newline
\begin{subfigure}{.5\textwidth}
  \centering
  % include second image
  \includegraphics[width=.95\linewidth]{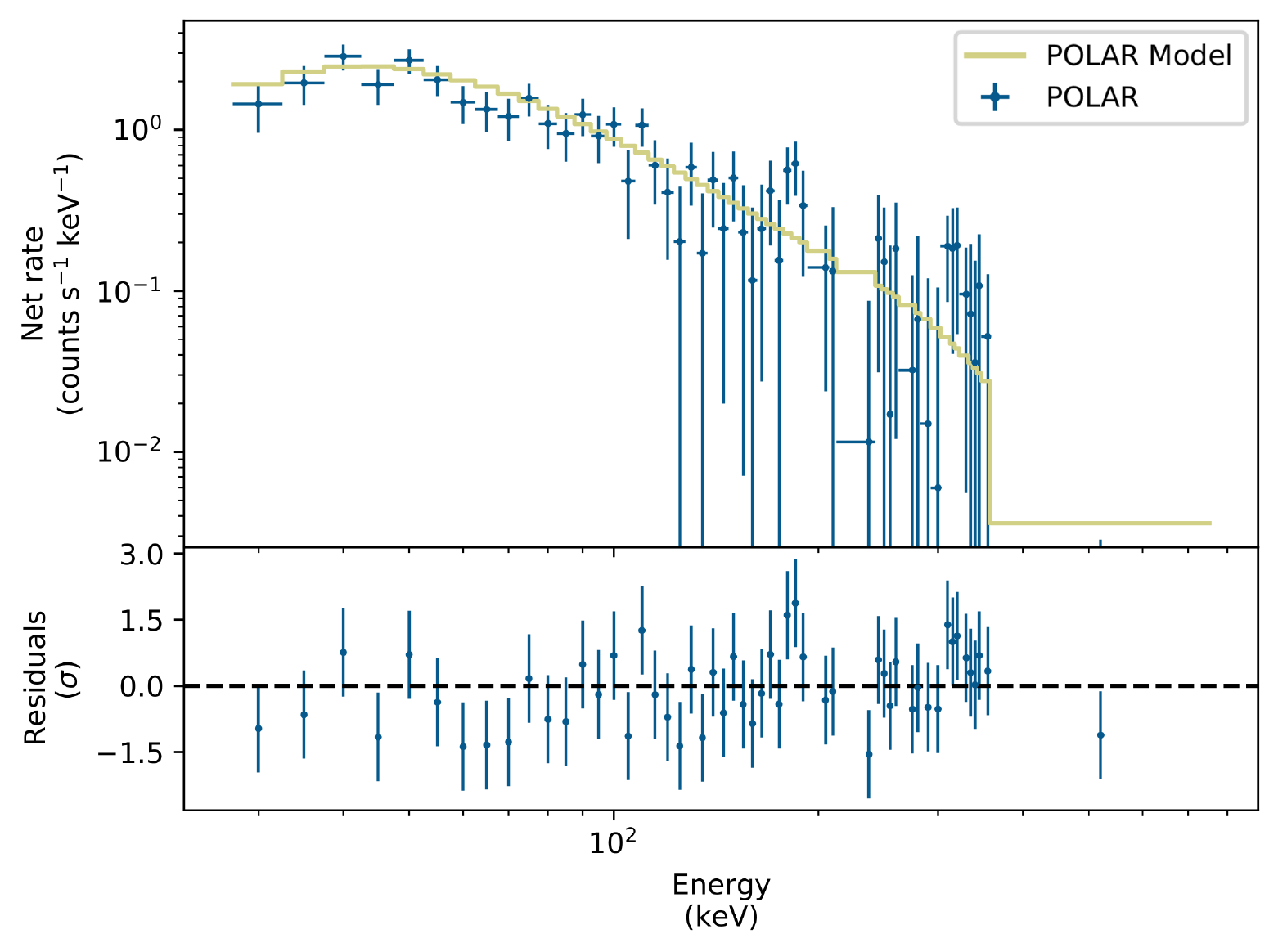}  
  \caption{The joint spectral fit result for 161217C. The number of counts as detected by POLAR (blue) is shown along with the best fitting spectrum folded through the instrument response in yellow. The residuals are shown in the bottom of the figure.}
  \label{fig:161217C_cs}
\end{subfigure}
\caption{The light curve as measured by POLAR for GRB 161217C (a) along with the joint spectral fit result of POLAR for the signal region indicated in yellow in figure (a).}
\label{fig:161217C_lc_cs}
\end{figure}

\begin{figure}[!ht]
   \centering
     \resizebox{\hsize}{!}{\includegraphics[width=11 cm]{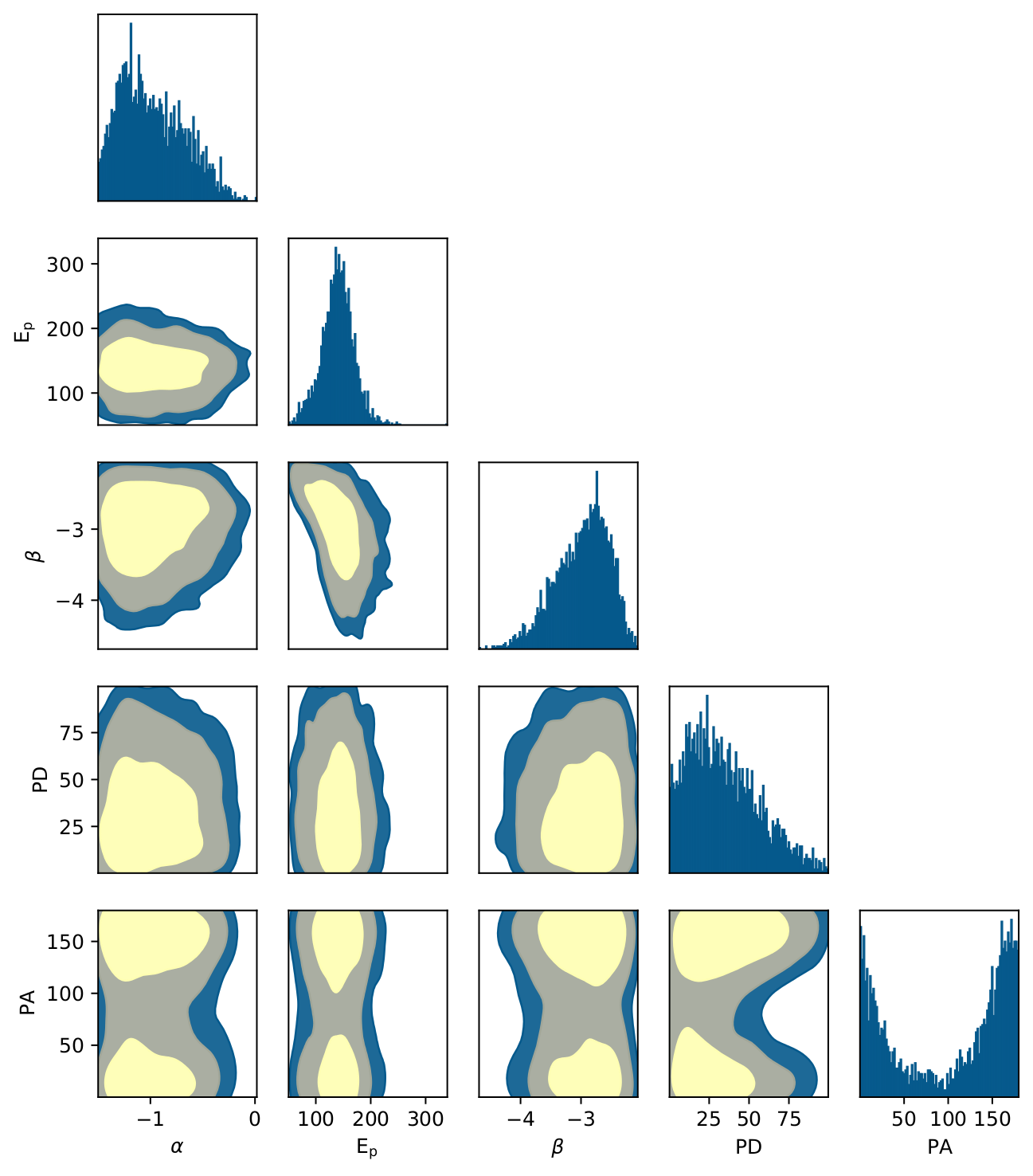}}
   \caption[Spectral and polarization posterior distributions.]
 {The spectral and polarization posterior distributions for GRB 161217C. The 1 and 2 $\sigma$ credibility intervals as well as that corresponding to $99\%$ are indicated. The polarization angle shown here is in the POLAR coordinate system,  a rotation in the positive direction of 34 degrees transforms this to the coordinate system as defined by the IAU.}
 \label{fig:post_161217C}
 \end{figure}

\begin{figure}[ht]
\begin{subfigure}{.5\textwidth}
  \centering
  % include first image
  \includegraphics[width=.85\linewidth]{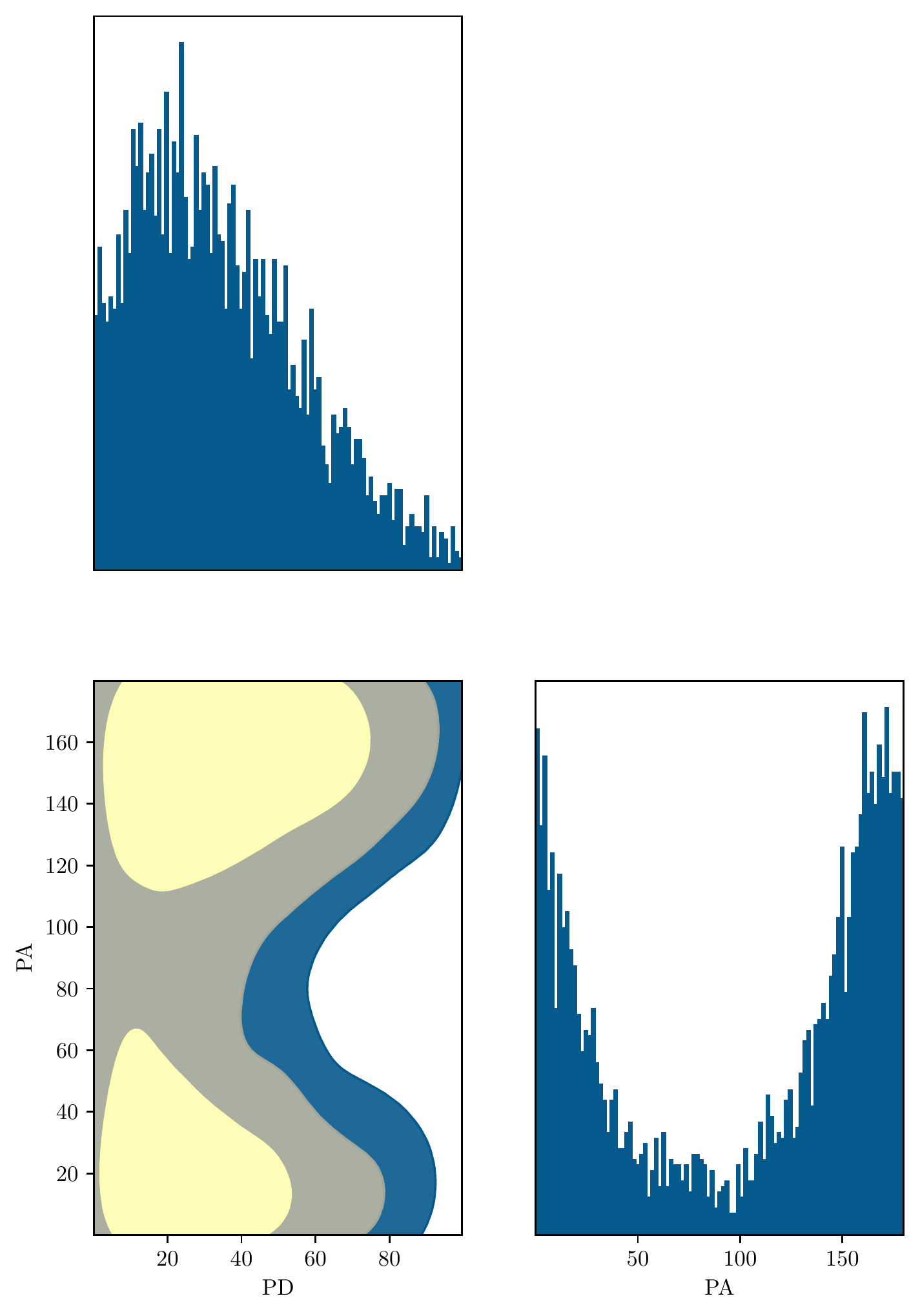}  
  \caption{The polarization posterior distributions for GRB 161217C with the 1 and 2 $\sigma$ credibility intervals as well as that corresponding to $99\%$ credibility. The PA shown here is in the POLAR coordinate system, a rotation in the positive direction of 34 degrees transforms this to the coordinate system as defined by the IAU.}
  \label{fig:161217C_PD}
\end{subfigure}
\newline
\begin{subfigure}{.5\textwidth}
  \centering
  % include second image
  \includegraphics[width=.85\linewidth]{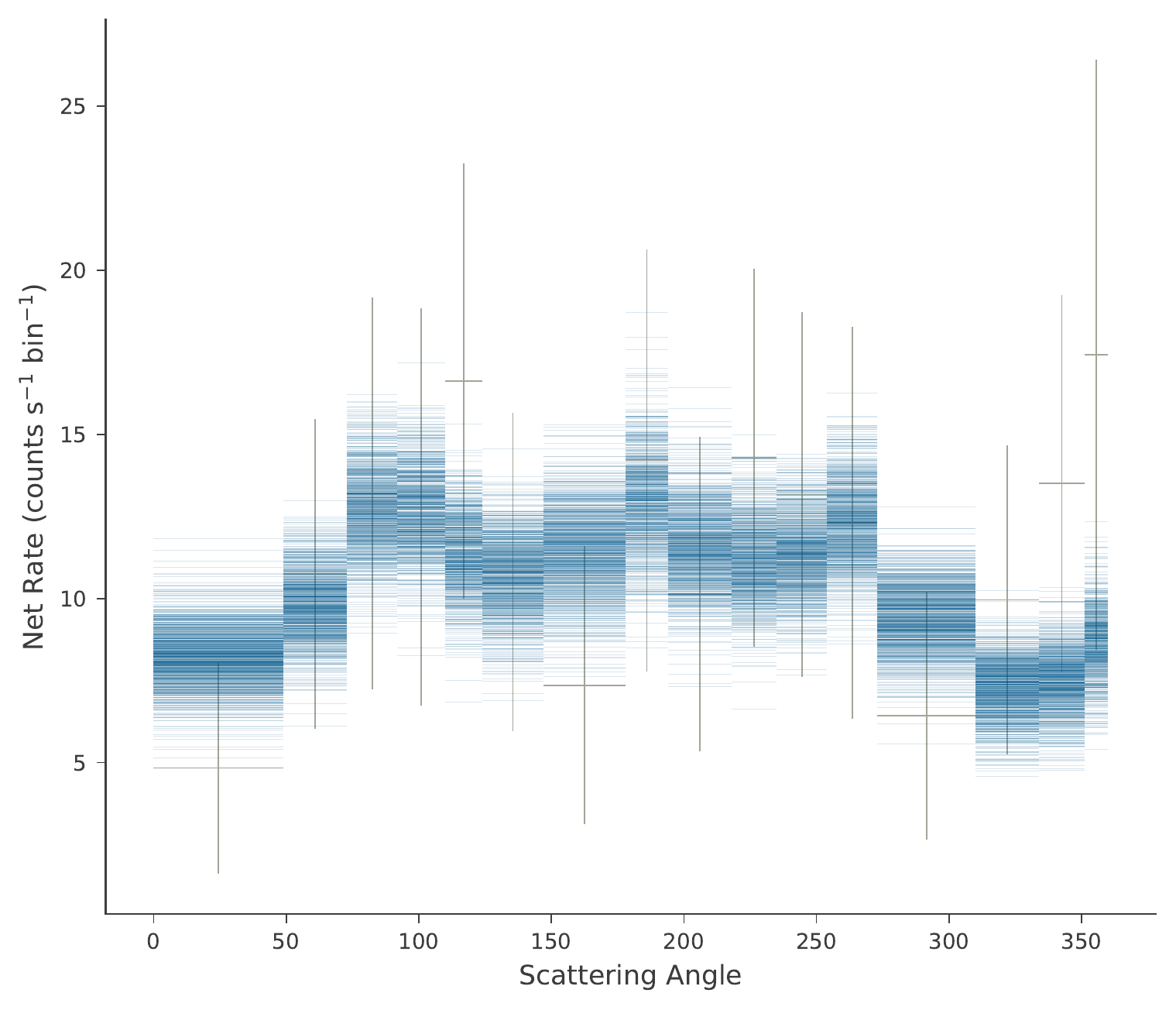}  
  \caption{The measured scattering angle distribution (gray data points with a $22.5$ degree bin size) superimposed by samples from the posterior model predictions (blue) is shown. The errors on the data points are the Poisson errors corrected for the background. }
  \label{fig:161217C_sd}
\end{subfigure}
\caption{The posterior distribution of the polarization parameters (a) together with the scattering angle distribution of GRB 161217C.}
\label{fig:161217C_PD_sd}
\end{figure}

GRB 161217C was detected by POLAR on 2016-12-17 at 03:53:15 (UT) \cite{POLAR_GCN_161217C} which will be used as $T0$ in the analysis presented here. A $T_{90}$ of $(6.3\pm0.25)\,\mathrm{s}$ was measured using POLAR data. The light curve, including the signal region (blue) and part of the background region (yellow) can be seen in figure \ref{fig:161217C_lc}. As it was not detected by \textit{Fermi}-GBM or \textit{Swift}-BAT, no spectral data from these instruments was available. Additionally, although the GRB was detected by Konus-Wind \citep{KONUS_cat_161217C}, no spectral parameters were reported and therefore no priors could be used. Furthermore, no location was reported for this GRB by any instrument. For this purpose the POLAR localization method reported in \cite{Yuanhao} was used to calculate a relative position for POLAR corresponding to $\theta=35^\circ$ and $\phi=172^\circ$ where $\theta$ is the angle with respect to the zenith of POLAR and $\phi$ is the angle with respect to the x-axis of POLAR. These coordinates correspond to RA (J2000)=$34^\circ$ and Dec (J2000)= $-4^\circ$ with an uncertainty of $7^\circ$. The polarization response of POLAR was produced using this location and includes the systematic error induced by the relatively large uncertainty. The joint spectral and polarization fit was performed using POLAR data only. This makes GRB 161217C, along with GRB 161203A, the only GRBs which only makes use of POLAR data for both the location, spectrum and polarization. As a result the systematic errors, which are all included in the final posterior distribution, are relatively large. The spectral results of the joint fit can be seen in figure \ref{fig:161217C_cs}. The posterior distributions of the spectral and polarization parameters are shown in figure \ref{fig:post_161217C}. Finally the posterior distribution of the polarization parameters is shown together with measured scattering angle distribution superimposed by the posterior model predictions (blue) in figure \ref{fig:161217C_PD_sd}. A PD of $21\substack{+30 \\ -16}\%$ is found. A $99\%$ credibility upper limit for PD of $94\%$ is found. Although the results exclude a very high polarization, no properly constraining measurement was possible, partly due to the lack of measurements by other instruments. The signal to background count ratio of POLAR for GRB do not allow for any time-resolved analysis.

% 
% 
% \begin{figure}[!ht]
%    \centering
%      \resizebox{\hsize}{!}{\includegraphics{161217C_lc.png}}
%    \caption[Light curve of GRB 161217C as measured by POLAR.]
%  {The light curve of GRB 161217C as measured by POLAR, where T0 is defined as the detection time by POLAR \cite{POLAR_GCN_161217C}. }
%  \label{fig:161217C_lc}
%  \end{figure}

\clearpage

\section{161218A}

\begin{figure}[ht]
\begin{subfigure}{.5\textwidth}
  \centering
  % include first image
  \includegraphics[width=.95\linewidth]{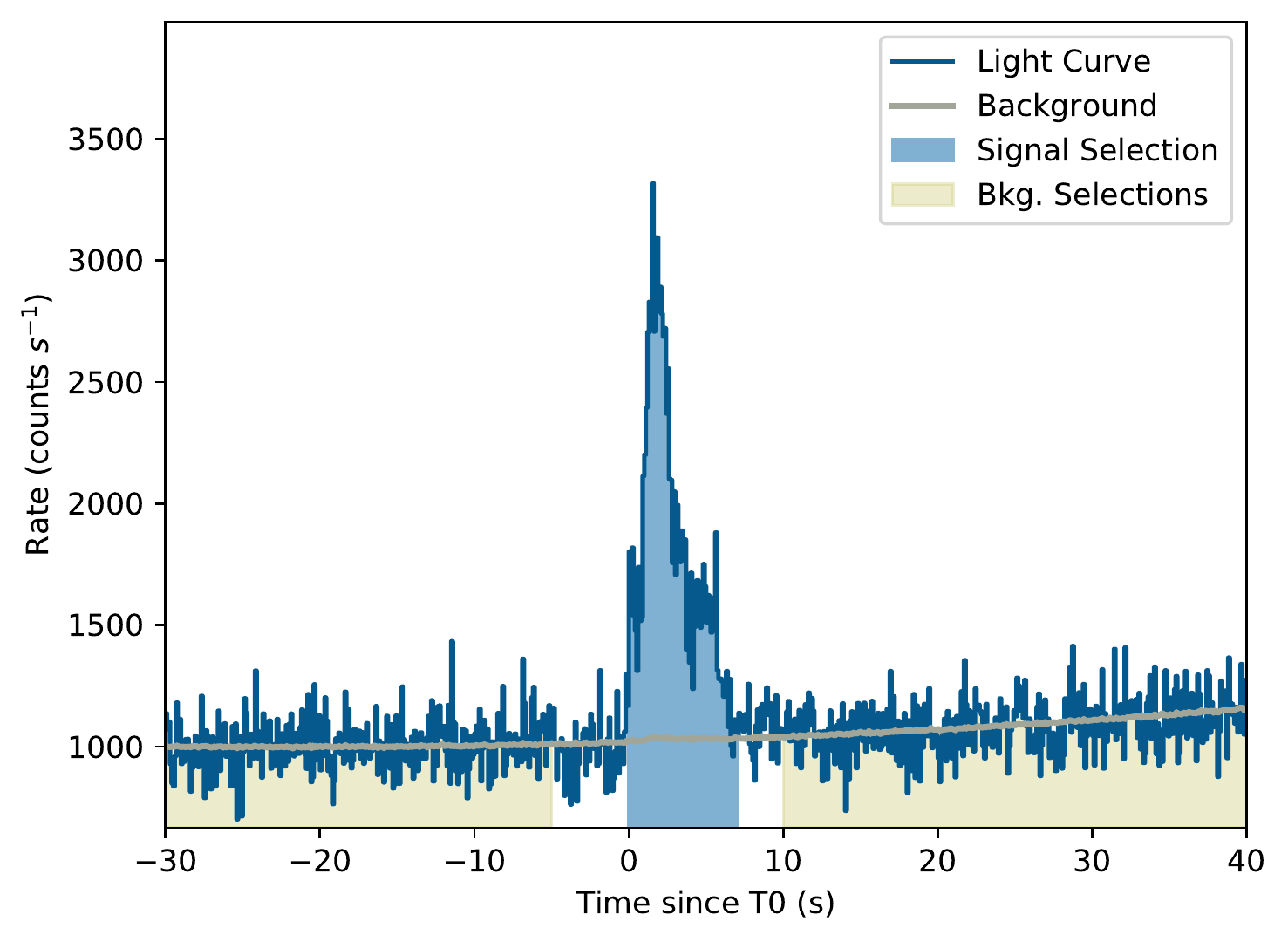}  
  \caption{The light curve of GRB 161218A as measured by POLAR, where $T=0\,\mathrm{s}$ is defined as the $T0$ employed by \textit{Swift}-BAT for this GRB in their data products.}
  \label{fig:161218A_lc}
\end{subfigure}
\newline
\begin{subfigure}{.5\textwidth}
  \centering
  % include second image
  \includegraphics[width=.95\linewidth]{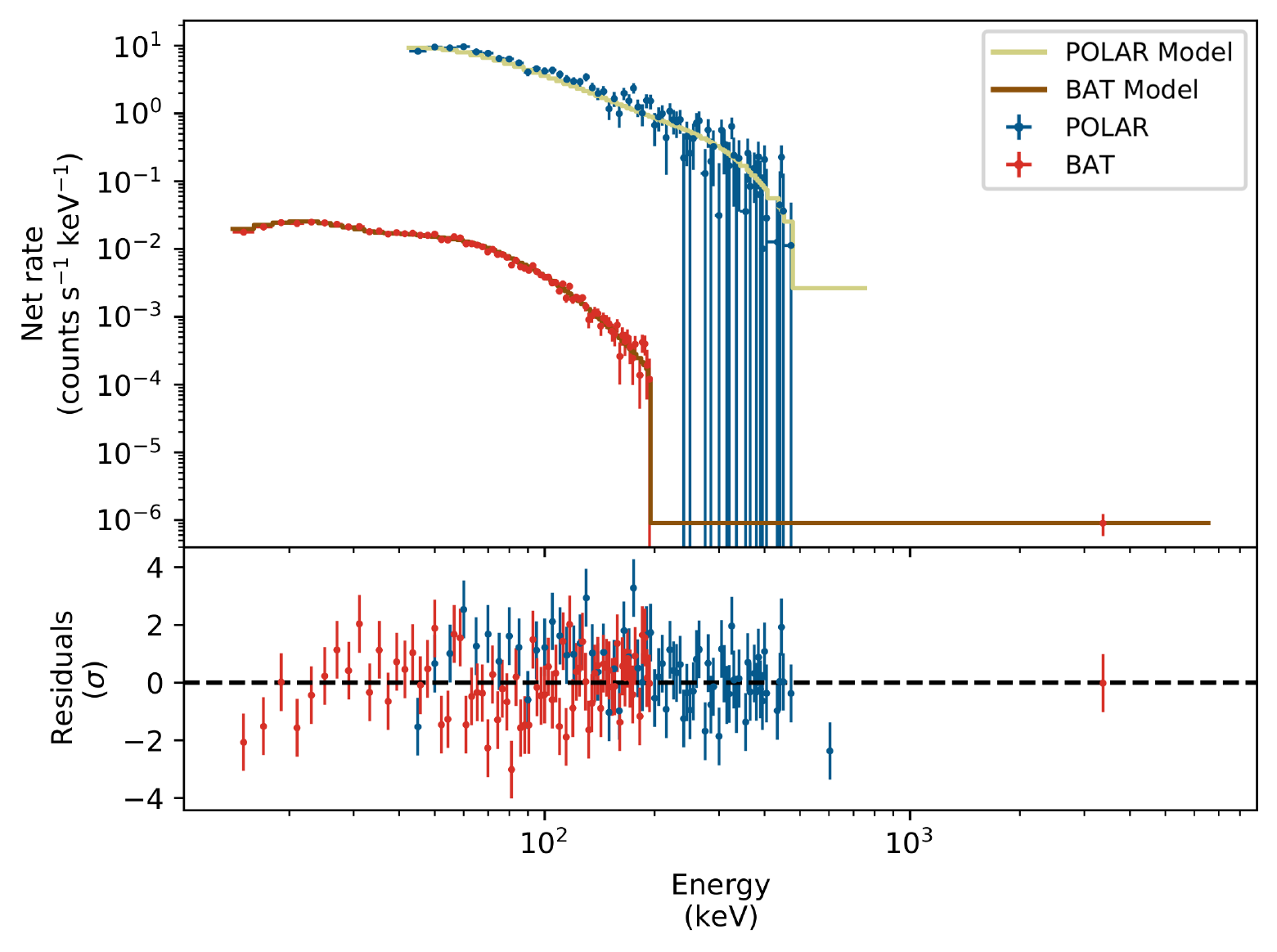}  
  \caption{The joint spectral fit result for 161218A. The number of counts as detected by both POLAR (blue) and \textit{Swift}-BAT (red) are shown along with the best fitting spectrum folded through the instrument responses in yellow for POLAR data and in brown for \textit{Swift}-BAT data. The residuals for both data sets are shown in the bottom of the figure.}
  \label{fig:161218A_cs}
\end{subfigure}
\caption{The light curve as measured by POLAR for GRB 161218A (a) along with the joint spectral fit results of POLAR and \textit{Swift}-BAT for the signal region indicated in yellow in  figure (a).}
\label{fig:161218A_lc_cs}
\end{figure}

\begin{figure}[!ht]
   \centering
     \resizebox{\hsize}{!}{\includegraphics{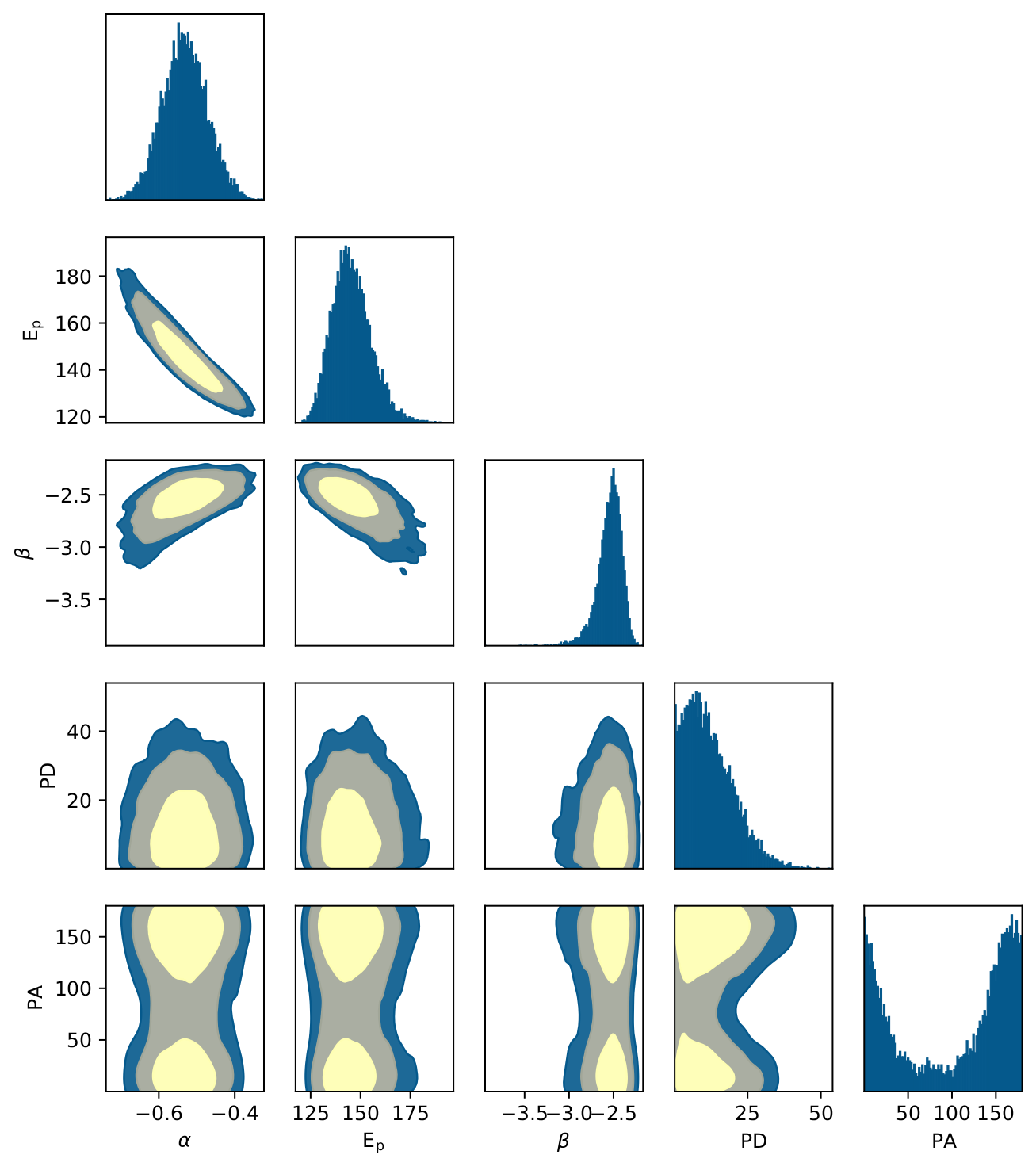}}
   \caption[Spectral and polarization posterior distributions.]
 {The spectral and polarization posterior distributions for GRB 161218A. The 1 and 2 $\sigma$ credibility intervals as well as that corresponding to $99\%$ are indicated. The polarization angle shown here is in the POLAR coordinate system, a rotation in the positive direction of 156 degrees transforms this to the coordinate system as defined by the IAU.}
 \label{fig:post_161218A}
 \end{figure}

\begin{figure}[ht]
\begin{subfigure}{.5\textwidth}
  \centering
  % include first image
  \includegraphics[width=.85\linewidth]{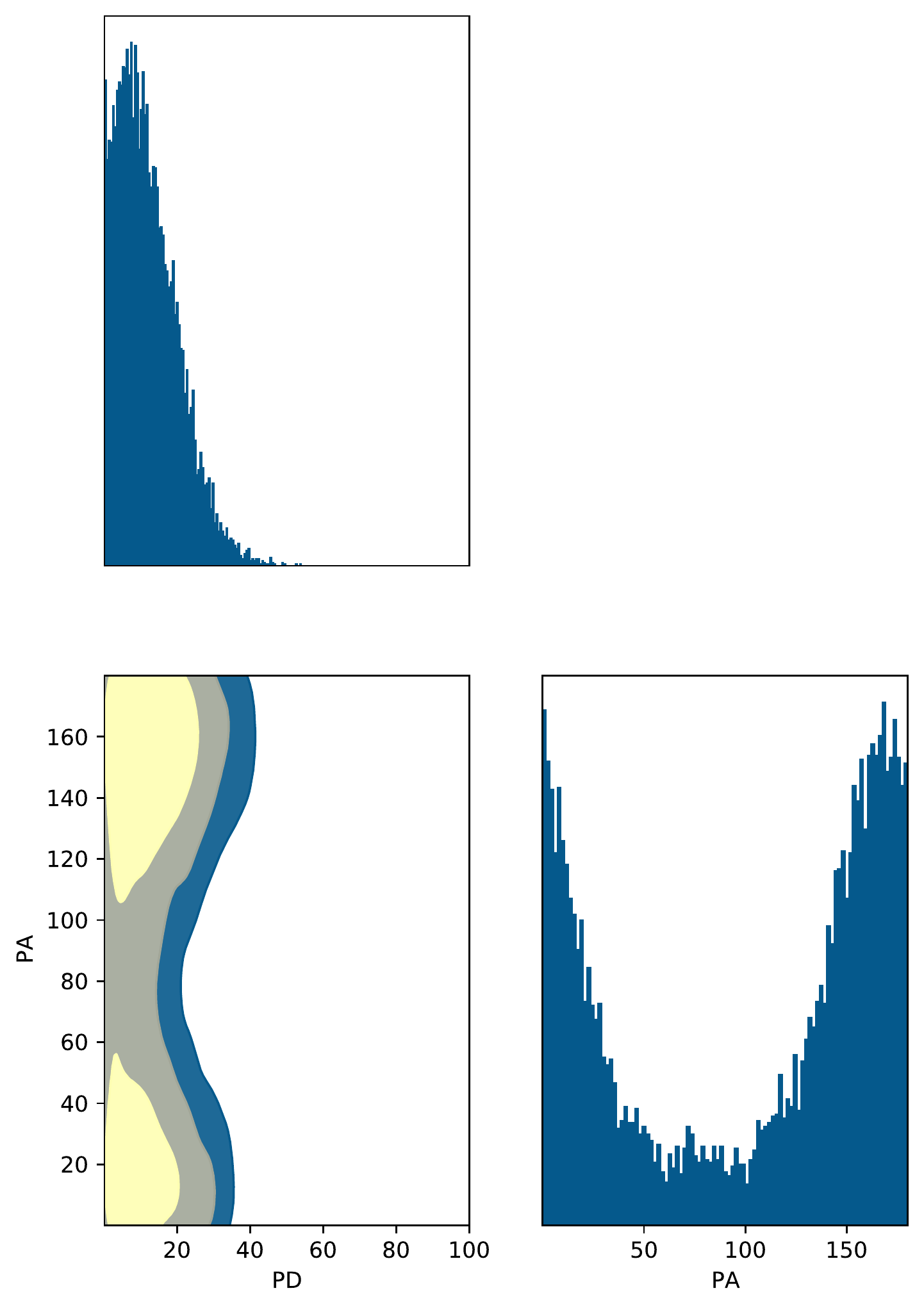}  
  \caption{The polarization posterior distributions for GRB 161218A with the 1 and 2 $\sigma$ credibility intervals as well as that corresponding to $99\%$. The polarization angle shown here is in the POLAR coordinate system, a rotation in the positive direction of 156 degrees transforms this to the coordinate system as defined by the IAU.}
  \label{fig:161218A_PD}
\end{subfigure}
\newline
\begin{subfigure}{.5\textwidth}
  \centering
  % include second image
  \includegraphics[width=.85\linewidth]{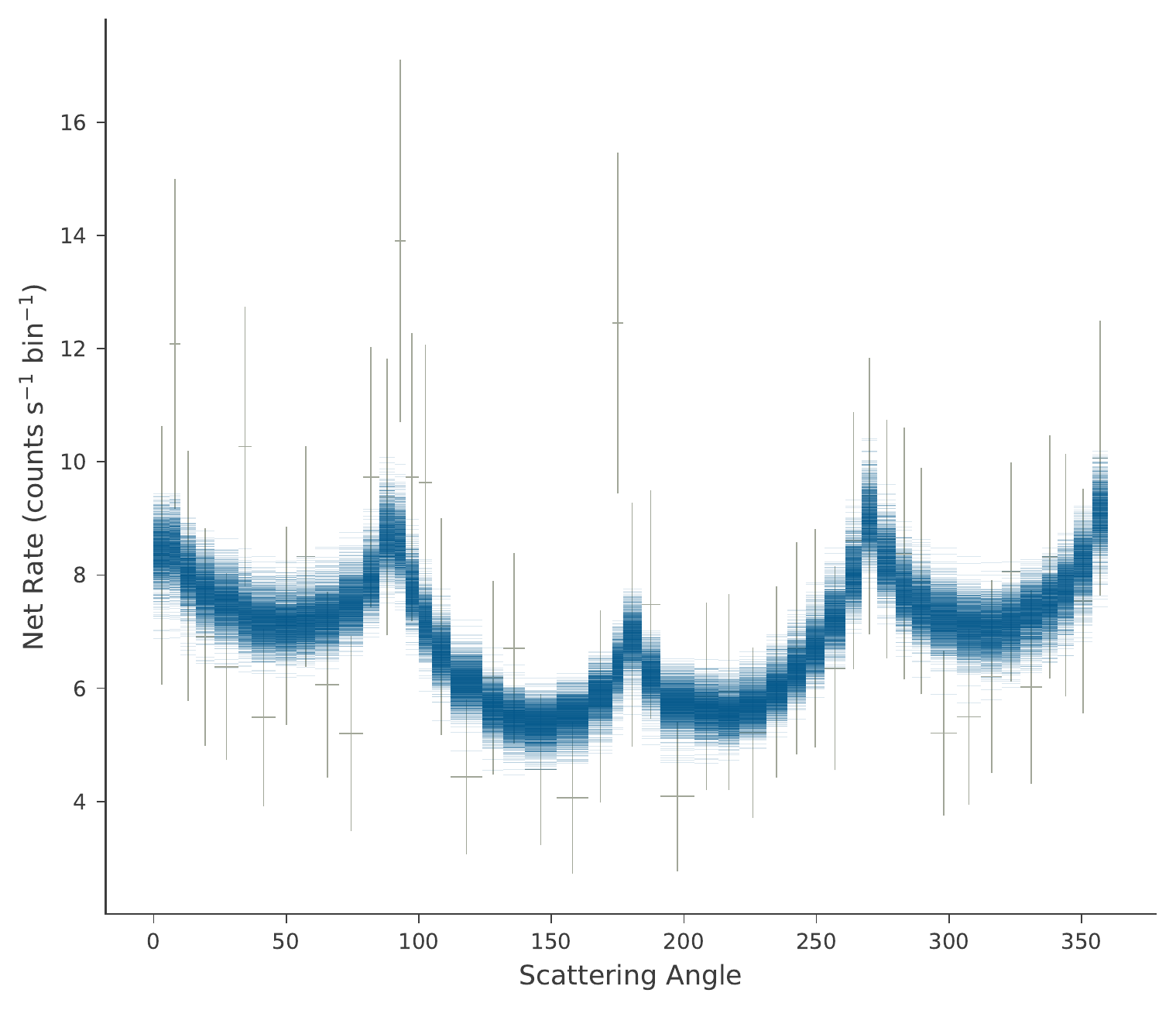}  
  \caption{The measured scattering angle distribution (gray data points with a 7.5 degree bin size) superimposed by samples from the posterior model predictions (blue) is shown. The errors on the data points are the Poisson errors corrected for the background.}
  \label{fig:161218A_sd}
\end{subfigure}
\caption{The posterior distribution of the polarization parameters (a) together with the scattering angle distribution of GRB 161218A.}
\label{fig:161218A_PD_sd}
\end{figure}

GRB 161218A was detected by POLAR and by \textit{Swift}-BAT \citep{GCN_161218A_SWIFT} which defined a $T0$ of 2016-12-18 at 03:47:34.75 (UT). For convenience this time will be used as $T0$ for the analysis presented here as well. A $T_{90}$ of $(11.48\pm0.14)\,\mathrm{s}$ was measured using POLAR data. The light curve from POLAR, including the signal region (blue) and part of the background region (yellow) can be seen in figure \ref{fig:161218A_lc}. As it was detected by \textit{Swift}-BAT, spectral data from this instrument was used to perform the joint fit. The spectral results of the joint fit for this GRB can be seen in figure \ref{fig:161218A_cs}. The effective area correction (applied to the POLAR data) found in the analysis was $1.08\pm0.05$. The spectral response of both instruments can be seen to be in good agreement.  The polarization response of POLAR was produced using the refined location provided by \textit{Swift}-BAT: RA (J2000) = $245.250^\circ$, Dec (J2000) = $-4.113^\circ$ \citep{GCN_161218A_SWIFT}. The posterior distributions of the spectral and polarization parameters are shown in figure \ref{fig:post_161218A}. Finally the posterior distribution of the polarization parameters is shown together with the measured scattering angle distribution superimposed by the posterior model predictions (blue) in figure \ref{fig:161218A_PD_sd}. A PD of $7.0\substack{+10.7 \\ -7.0}\%$ is found which is compatible with that reported in \cite{Zhang+Kole}. A $99\%$ credibility upper limit for PD of $38.0\%$ was found.

% \begin{figure}[!ht]
%    \centering
%      \resizebox{\hsize}{!}{\includegraphics{161218A_lc.png}}
%    \caption[Light curve of GRB 161218A as measured by POLAR.]
%  {The light curve of GRB 161218A as measured by POLAR where T0 is defined as that employed by \textit{Swift-BAT} for this GRB in their data products.}
%  \label{fig:161218A_lc}
%  \end{figure}

 \clearpage
 
\section{161218B}

\begin{figure}[ht]
\begin{subfigure}{.5\textwidth}
  \centering
  % include first image
  \includegraphics[width=.95\linewidth]{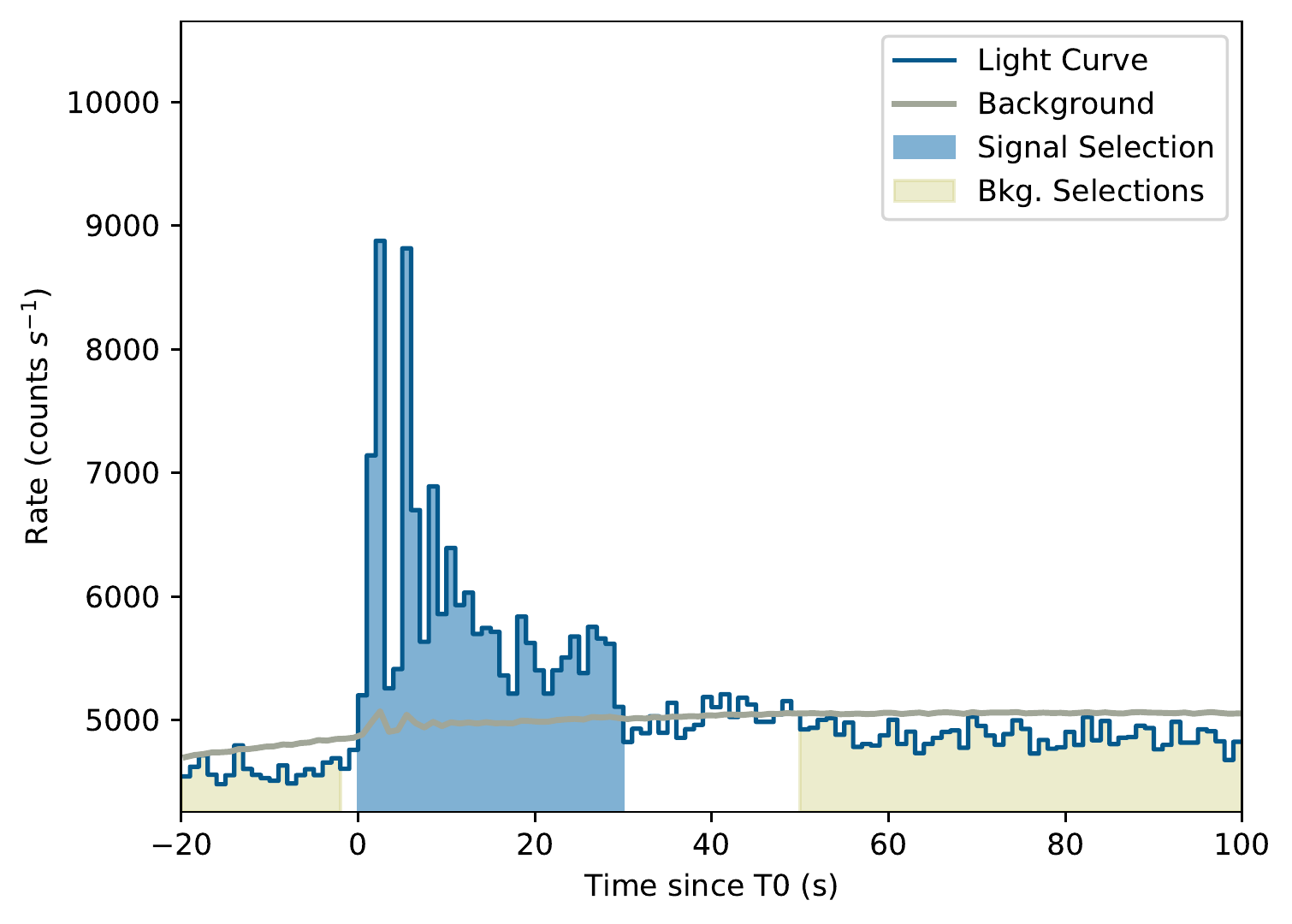}  
  \caption{The light curve of GRB 161218B as measured by POLAR, where $T=0\,\mathrm{s}$ is defined as the $T0$ used by \textit{Fermi}-GBM \citep{GCN_161218B_Fermi} for this GRB.}
  \label{fig:161218B_lc}
\end{subfigure}
\newline
\begin{subfigure}{.5\textwidth}
  \centering
  % include second image
  \includegraphics[width=.95\linewidth]{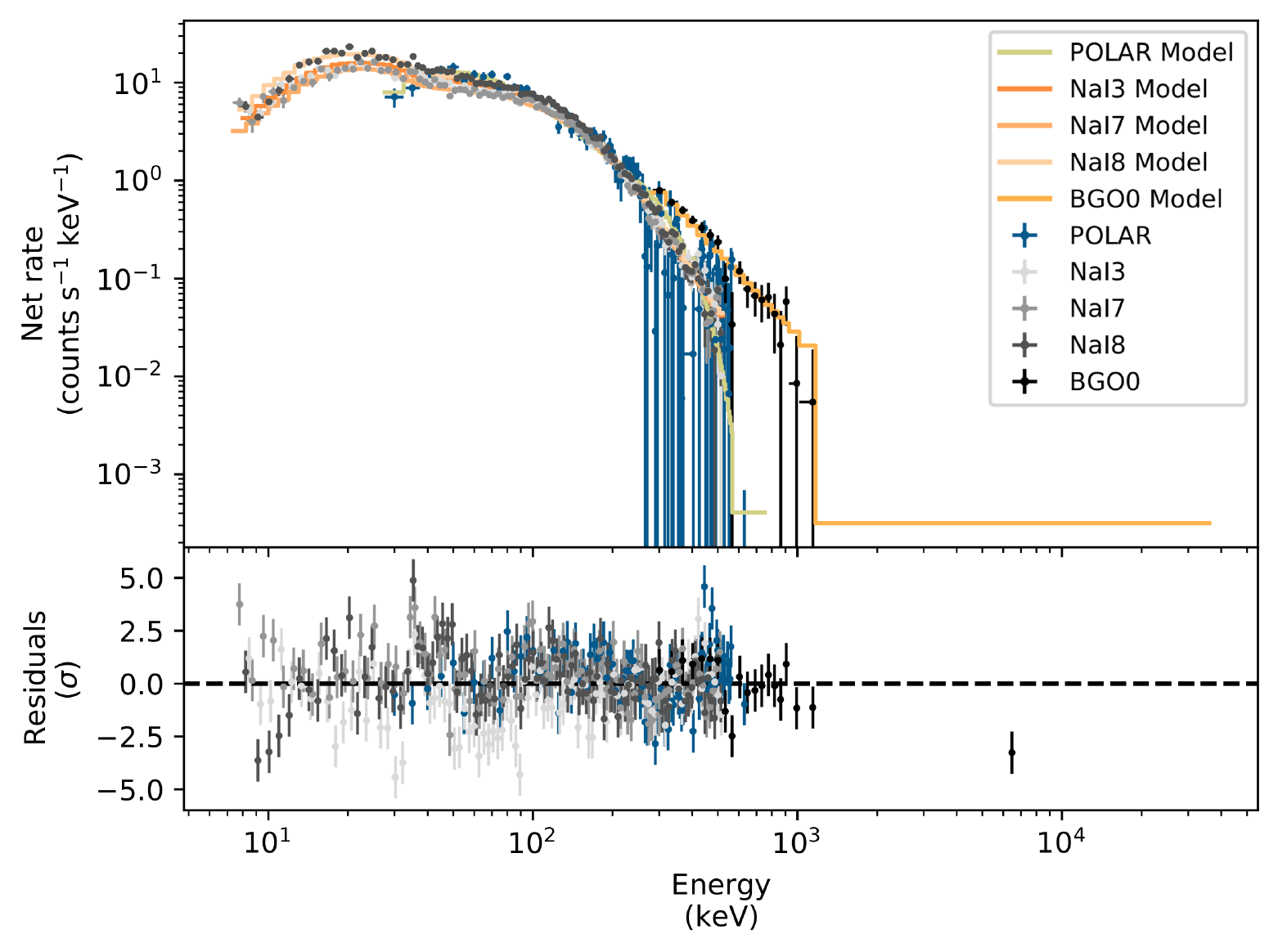}  
  \caption{The joint spectral fit result for 161218B. The number of counts as detected by both POLAR (blue) and the different NaI and BGO detectors of \textit{Fermi}-GBM (gray tints) are shown along with the best fitting spectrum folded through the instrument responses in yellow for POLAR data and in orange tints for the \textit{Fermi}-GBM data. The residuals for both data sets are shown in the bottom of the figure.}
  \label{fig:161218B_cs}
\end{subfigure}
\caption{The light curve as measured by POLAR for GRB 161218B (a) along with the joint spectral fit results of POLAR and \textit{Fermi}-GBM for the signal region indicated in yellow in  figure (a).}
\label{fig:161218B_lc_cs}
\end{figure}

\begin{figure}[!ht]
   \centering
     \resizebox{\hsize}{!}{\includegraphics{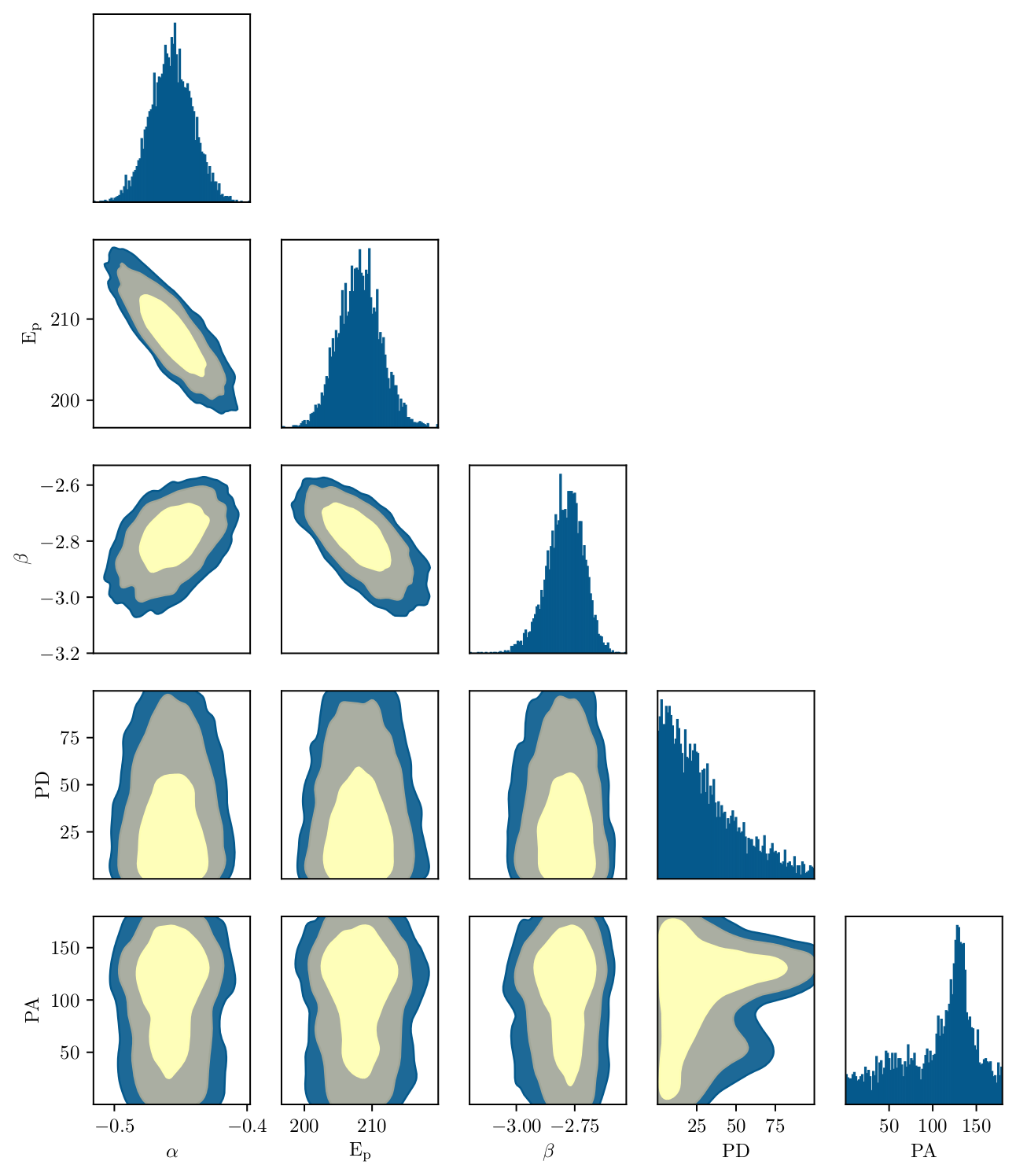}}
   \caption[Spectral and polarization posterior distributions.]
 {The spectral and polarization posterior distributions for GRB 161218B. The 1 and 2 $\sigma$ credibility intervals as well as that corresponding to $99\%$ are indicated. The polarization angle shown here is in the POLAR coordinate system, a rotation in the positive direction of 120 degrees transforms this to the coordinate system as defined by the IAU.}
 \label{fig:post_161218B}
 \end{figure}

\begin{figure}[ht]
\begin{subfigure}{.5\textwidth}
  \centering
  % include first image
  \includegraphics[width=.85\linewidth]{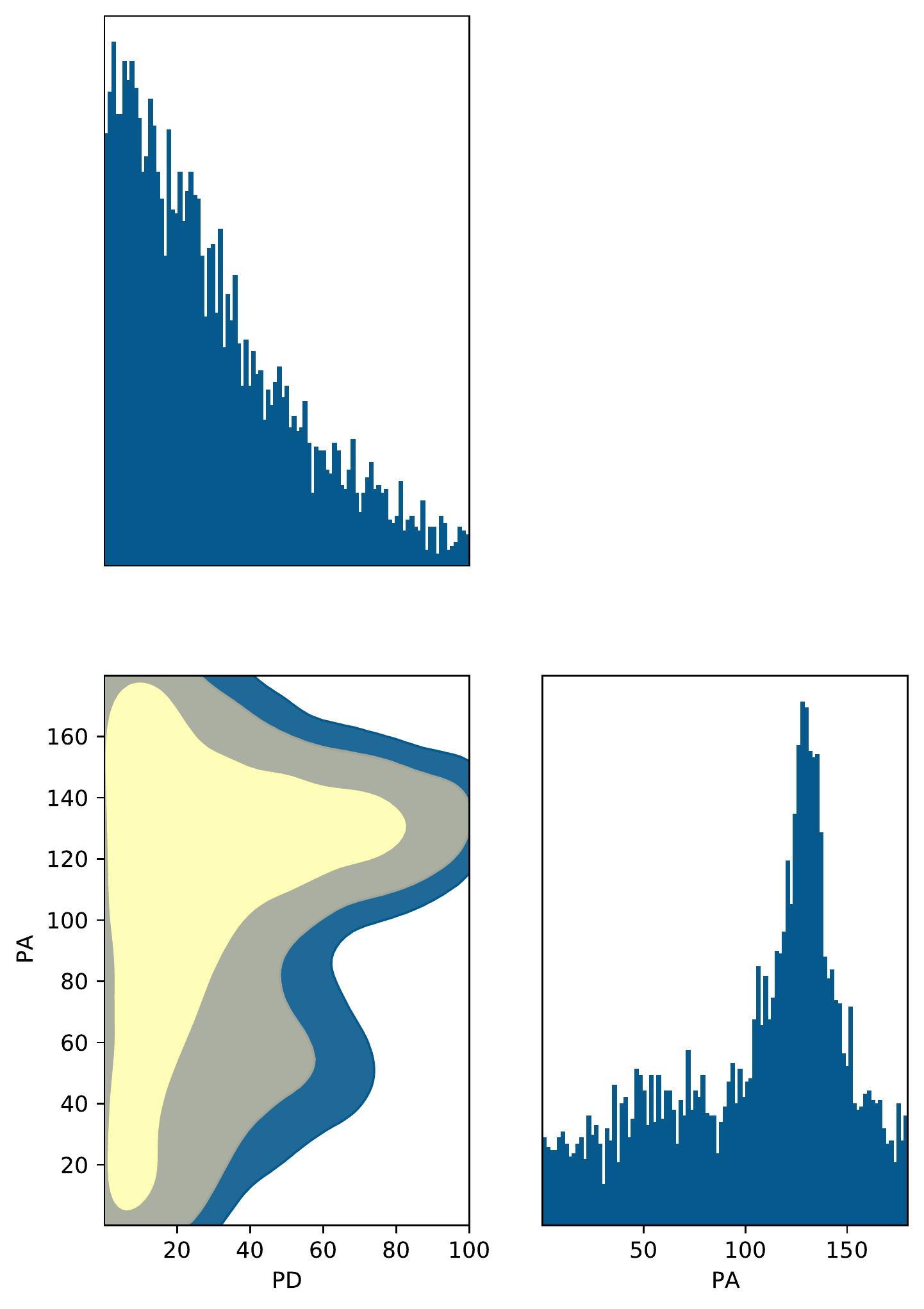}  
  \caption{The polarization posterior distributions for GRB 161218B with the 1 and 2 $\sigma$ credibility intervals as well as that corresponding to $99\%$ credibility. The posterior distribution clearly reflects the lack of sensitivity for polarization angles of $135^\circ$ in this measurement. The polarization angle shown here is in the POLAR coordinate system, a rotation in the positive direction of 120 degrees transforms this to the coordinate system as defined by the IAU.}
  \label{fig:161218B_PD}
\end{subfigure}
\newline
\begin{subfigure}{.5\textwidth}
  \centering
  % include second image
  \includegraphics[width=.85\linewidth]{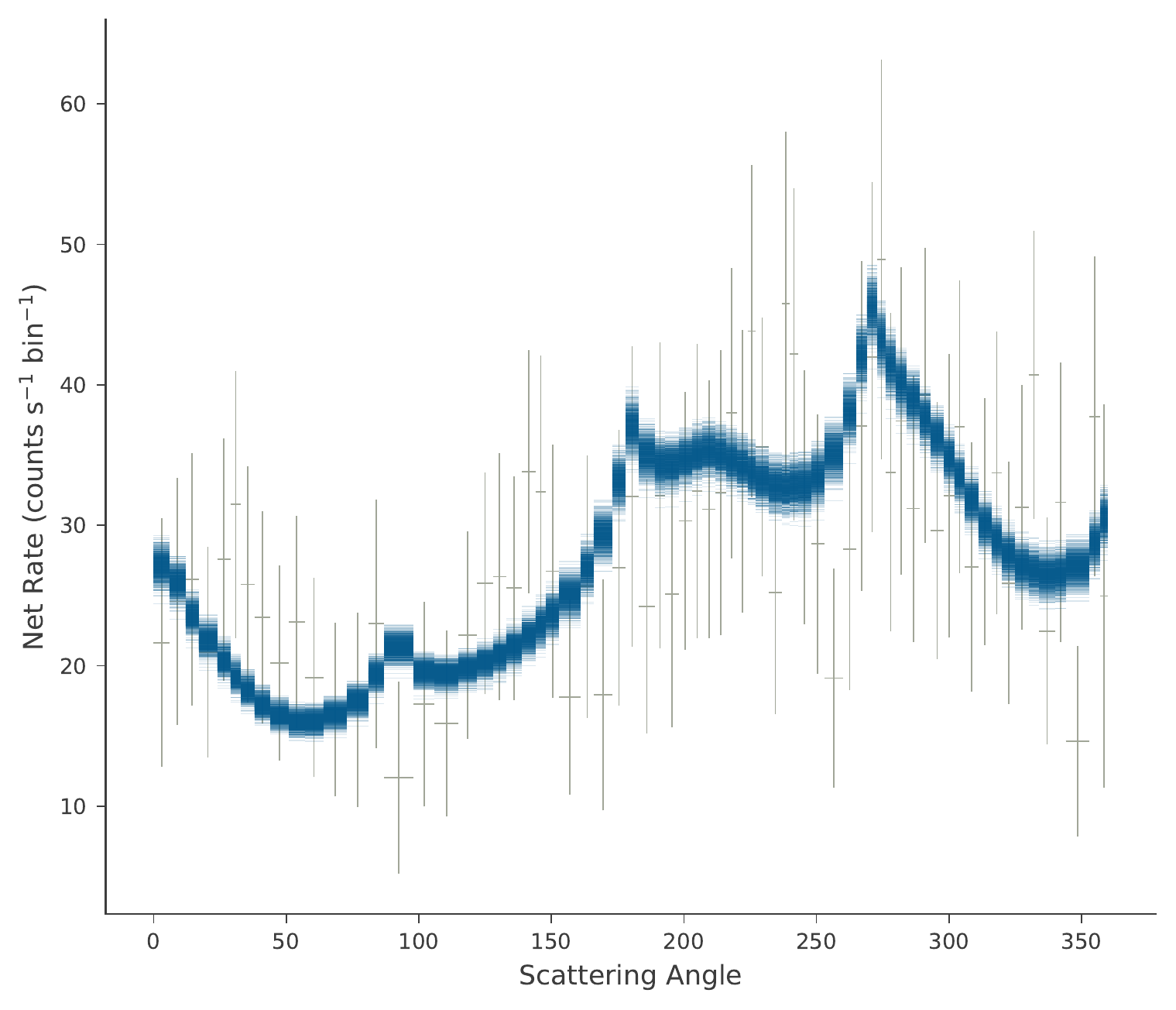}  
  \caption{The measured scattering angle distribution (gray data points with a 5 degree bin size) superimposed by samples from the posterior model predictions (blue) is shown.  The errors on the data points are the Poisson errors corrected for the background. }
  \label{fig:161218B_sd}
\end{subfigure}
\caption{The posterior distribution of the polarization parameters (a) together with the scattering angle distribution of GRB 161218B.}
\label{fig:161218B_PD_sd}
\end{figure}

GRB 161218B was detected by POLAR and by \textit{Fermi}-GBM. The latter reported a $T0$ of 2016-12-18 at 08:32:40.65 (UT) \citep{GCN_161218B_Fermi}, which, for convenience, is taken to be the $T0$ for the analysis presented here. A $T_{90}$ of $(25.1\pm0.2)\,\mathrm{s}$ was measured using POLAR data. The light curve, including the signal region (blue) and part of the the background region (yellow) can be seen in figure \ref{fig:161218B_lc}. It should be noted here that the GRB was detected by POLAR while the instrument was at a relatively high magnetic latitude. This resulted in an above average background rate which, additionally, showed some significant variations with time as can be seen in the light curve. Details on the background conditions of POLAR can be found in \cite{Suarez2010}. The fluctuations in the background could, however, be properly fitted using the \texttt{3ML} software. \textit{Fermi}-GBM detected this GRB \citep{GCN_161218B_Fermi}, therefore spectral data from \textit{Fermi}-GBM was used in the analysis. The joint spectral fit for this GRB can be seen in figure \ref{fig:161218B_cs}. The effective area correction (applied to the POLAR data) found in the analysis was $1.09\pm0.02$. The polarization response of POLAR was produced using the IPN location provided for this GRB: RA (J2000) = $0.888^\circ$, Dec (J2000) = $-14.700^\circ$ \citep{GCN_161218B_IPN}, a localization error of $4^\circ$ was assumed in the response. It should be noted that the error presented in \cite{GCN_161218B_IPN} is very asymmetric, while due to the methods used in the analysis here this cannot be taken properly into account as described earlier in this paper, and an average of the error is assumed. The GRB occurred far off-axis for POLAR at a $\theta$ angle of $80.5^\circ$ degrees. As a result the sensitivity of POLAR is reduced for specific polarization angles (at $45^\circ$ and $135^\circ$ as explained in the section regarding GRB 170207A) as is clear in the posterior distributions. The posterior distributions of the spectral and polarization parameters are shown in figure \ref{fig:post_161218B}. The posterior distribution of the polarization parameters is shown in \ref{fig:161218B_PD} and the measured scattering angle distribution superimposed by the posterior model predictions (blue) in figure \ref{fig:161218B_sd}. The scattering angle distribution indicates clearly the large off-axis angle resulting in the $360^\circ$ modulation in the distribution. A PD of $13.0\substack{+28.0 \\ -13.0}\%$ is found which is compatible with an unpolarized flux, especially when considering the lack of sensitivity in the PA range where the fit converges. No constraining $99\%$ upper limit is found. Although it appears that the posterior distribution is dominated by the lack of sensitivity, the results do clearly exclude high polarization degrees at any polarization angles away from $135^\circ$.
 
%  \begin{figure}[!ht]
%    \centering
%      \resizebox{\hsize}{!}{\includegraphics{161218B_lc2.png}}
%    \caption[Light curve of GRB 161218B as measured by POLAR.]
%  {Top: The light curve of GRB 161218B as measured by POLAR where T0 is defined as that used by \textit{Fermi-GBM} \citep{GCN_161218B_Fermi}. Note that the background fit is dead time corrected, which, due to the higher background during this GRB, makes it appear to be above the count rate. Bottom: the joint spectral fit for GRB 161218B in count space together with the residuals when using the data from POLAR, and the NaI detectors 2, 5 and 8 from \textit{Fermi-GBM} as well as BGO detectors 0 and 1.}
%  \label{fig:161218B_lc}
%  \end{figure}

 \clearpage

\section{161229A}

\begin{figure}[ht]
\begin{subfigure}{.5\textwidth}
  \centering
  % include first image
  \includegraphics[width=.95\linewidth]{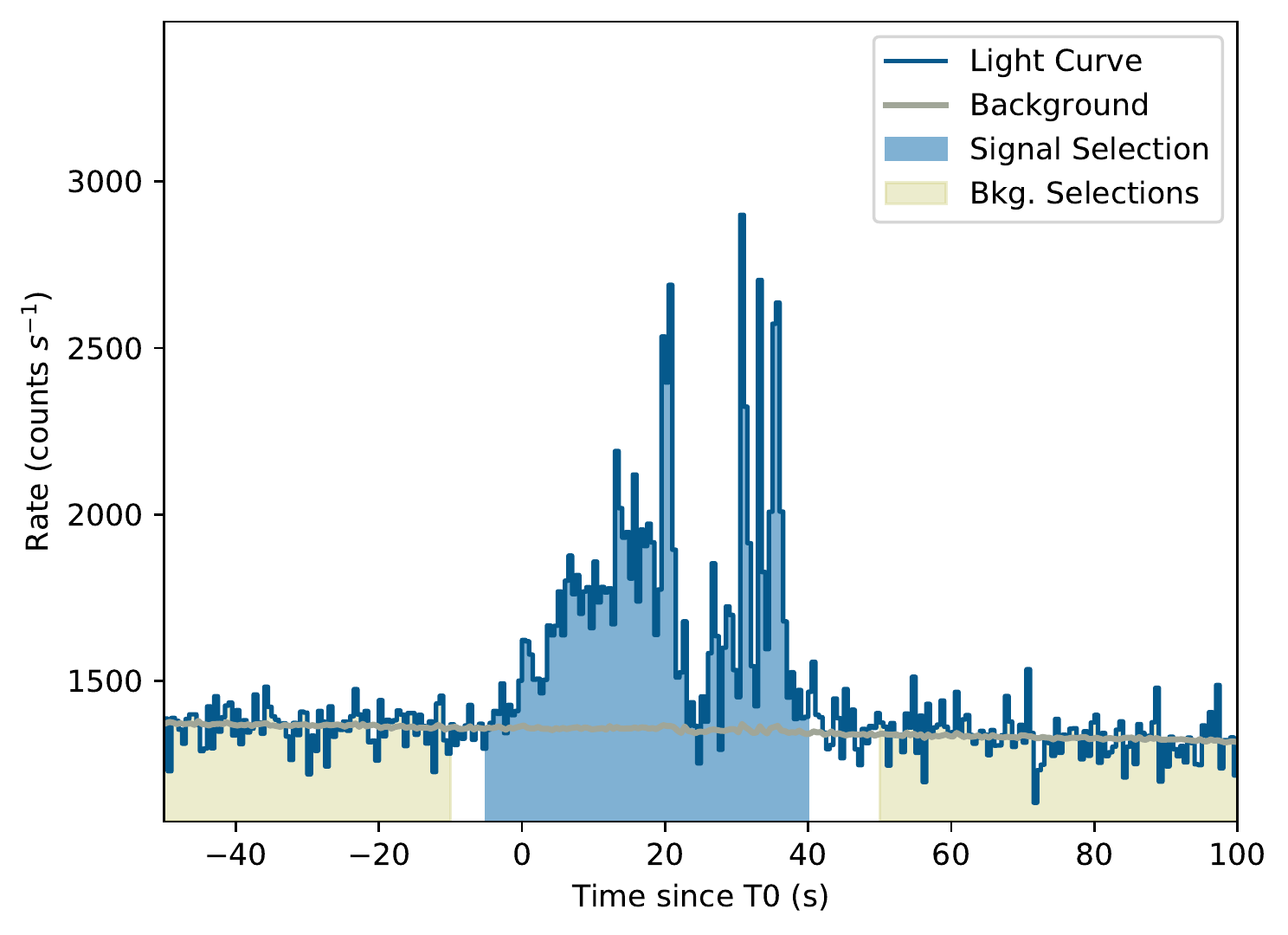}  
  \caption{The light curve of GRB 161229A as measured by POLAR, where $T=0\,\mathrm{s}$ is defined as the $T0$ employed by \textit{Fermi}-GBM in their data products for this GRB \citep{GCN_161229A_GCN}.}
  \label{fig:161229A_lc}
\end{subfigure}
\newline
\begin{subfigure}{.5\textwidth}
  \centering
  % include second image
  \includegraphics[width=.95\linewidth]{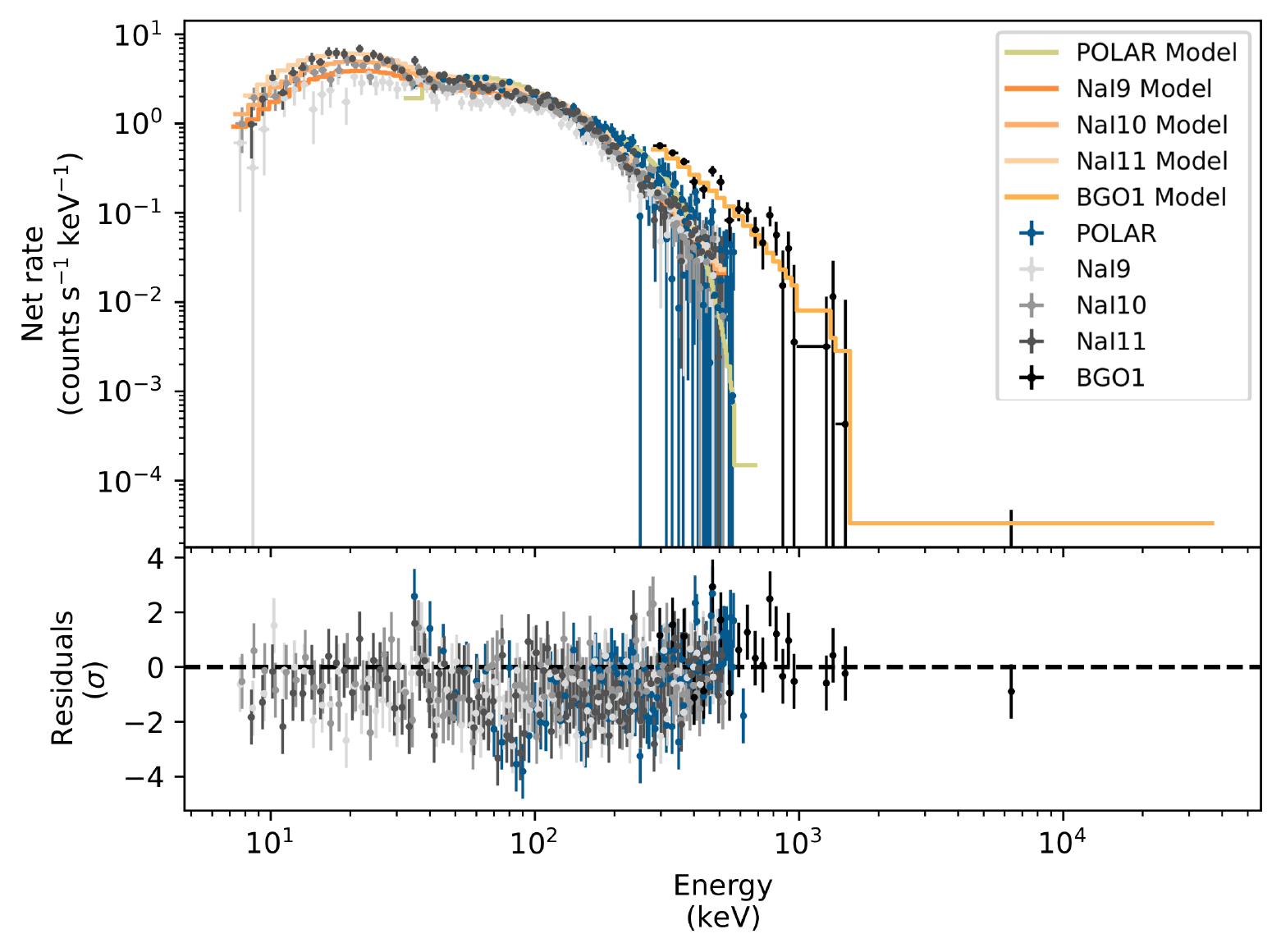}  
  \caption{The joint spectral fit result for 161229A. The number of counts as detected by both POLAR (blue) and the different NaI and BGO detectors of \textit{Fermi}-GBM (gray tints) are shown along with the best fitting spectrum folded through the instrument responses in yellow for POLAR data and in orange tints for the \textit{Fermi}-GBM data. The residuals for both data sets are shown in the bottom of the figure.}
  \label{fig:161229A_cs}
\end{subfigure}
\caption{The light curve as measured by POLAR for GRB 161229A (a) along with the joint spectral fit results of POLAR and \textit{Fermi}-GBM for the signal region indicated in yellow in  figure (a).}
\label{fig:161229A_lc_cs}
\end{figure}

\begin{figure}[!ht]
   \centering
     \resizebox{\hsize}{!}{\includegraphics{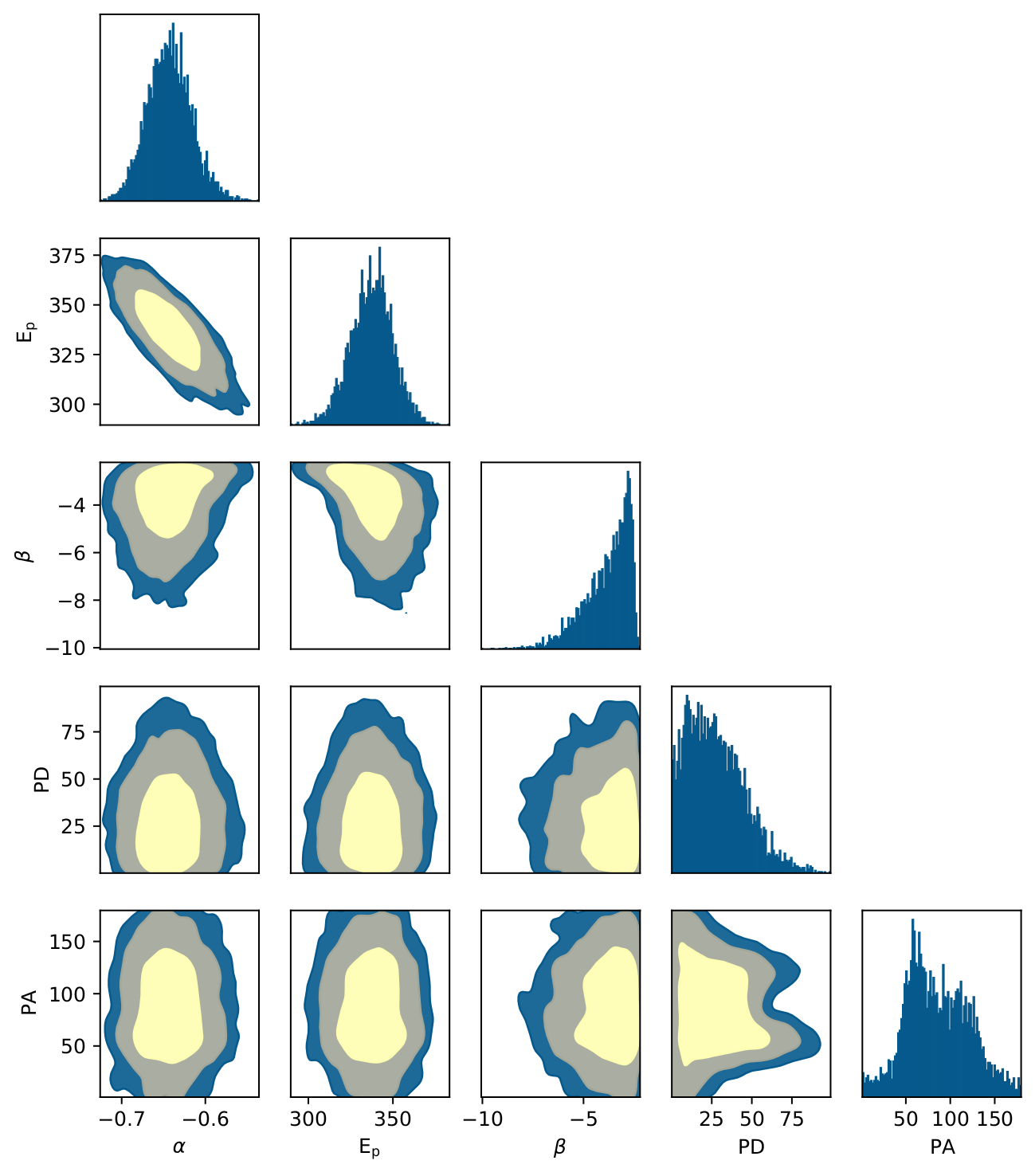}}
   \caption[Spectral and polarization posterior distributions.]
 {The spectral and polarization posterior distributions for GRB 161229A. The 1 and 2 $\sigma$ credibility intervals as well as that corresponding to $99\%$ are indicated. The polarization angle shown here is in the POLAR coordinate system, a rotation in the positive direction of 40 degrees transforms this to the coordinate system as defined by the IAU.}
 \label{fig:post_161229A}
 \end{figure}

\begin{figure}[ht]
\begin{subfigure}{.5\textwidth}
  \centering
  % include first image
  \includegraphics[width=.85\linewidth]{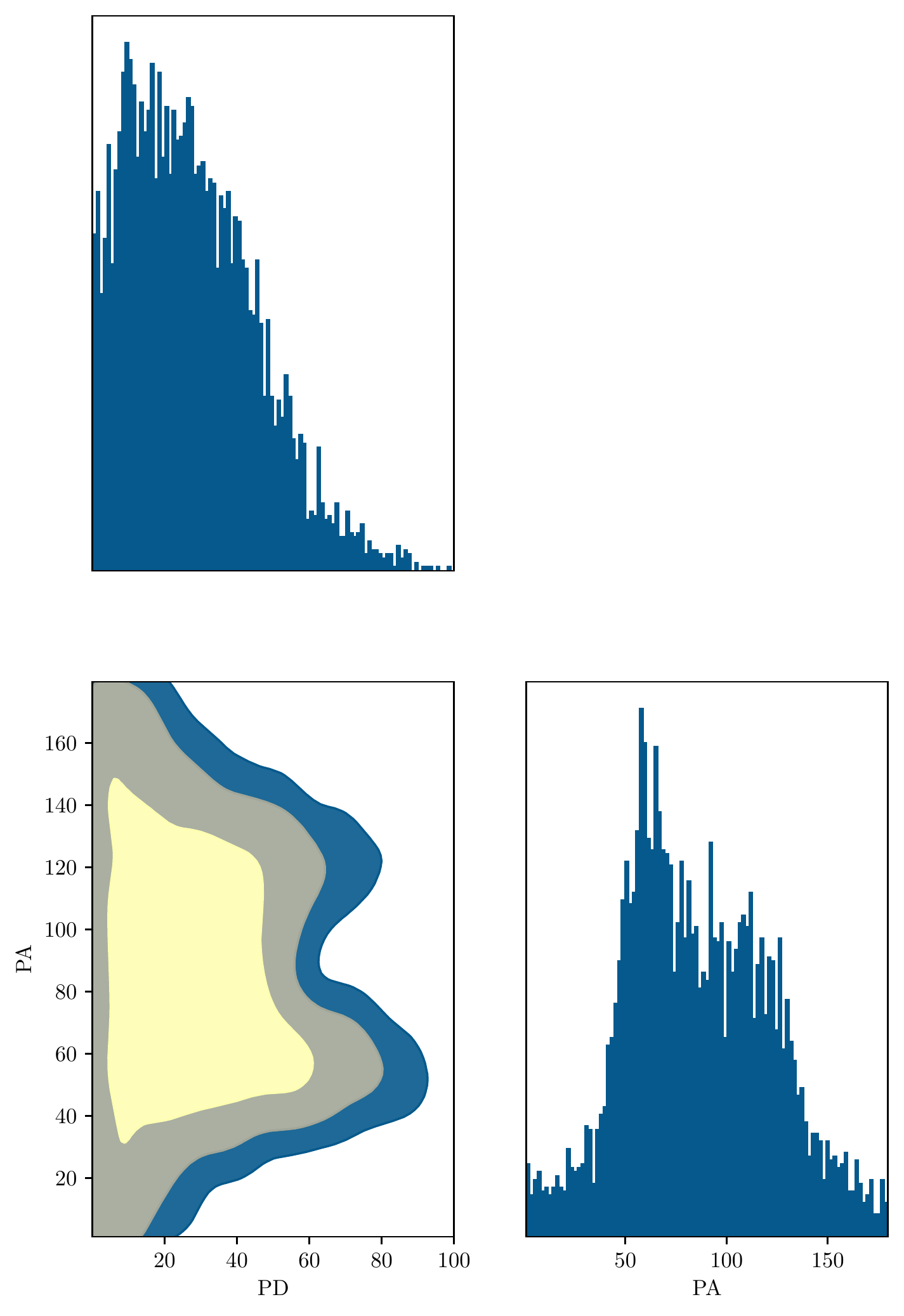}  
  \caption{The polarization posterior distributions for GRB 161229A with the 1 and 2 $\sigma$ credibility intervals as well as that corresponding to $99\%$. The posterior distribution clearly reflects the lack of sensitivity for polarization angles of $135^\circ$ in this measurement. The polarization angle shown here is in the POLAR coordinate system, a rotation in the positive direction of 40 degrees transforms this to the coordinate system as defined by the IAU.}
  \label{fig:161229A_PD}
\end{subfigure}
\newline
\begin{subfigure}{.5\textwidth}
  \centering
  % include second image
  \includegraphics[width=.85\linewidth]{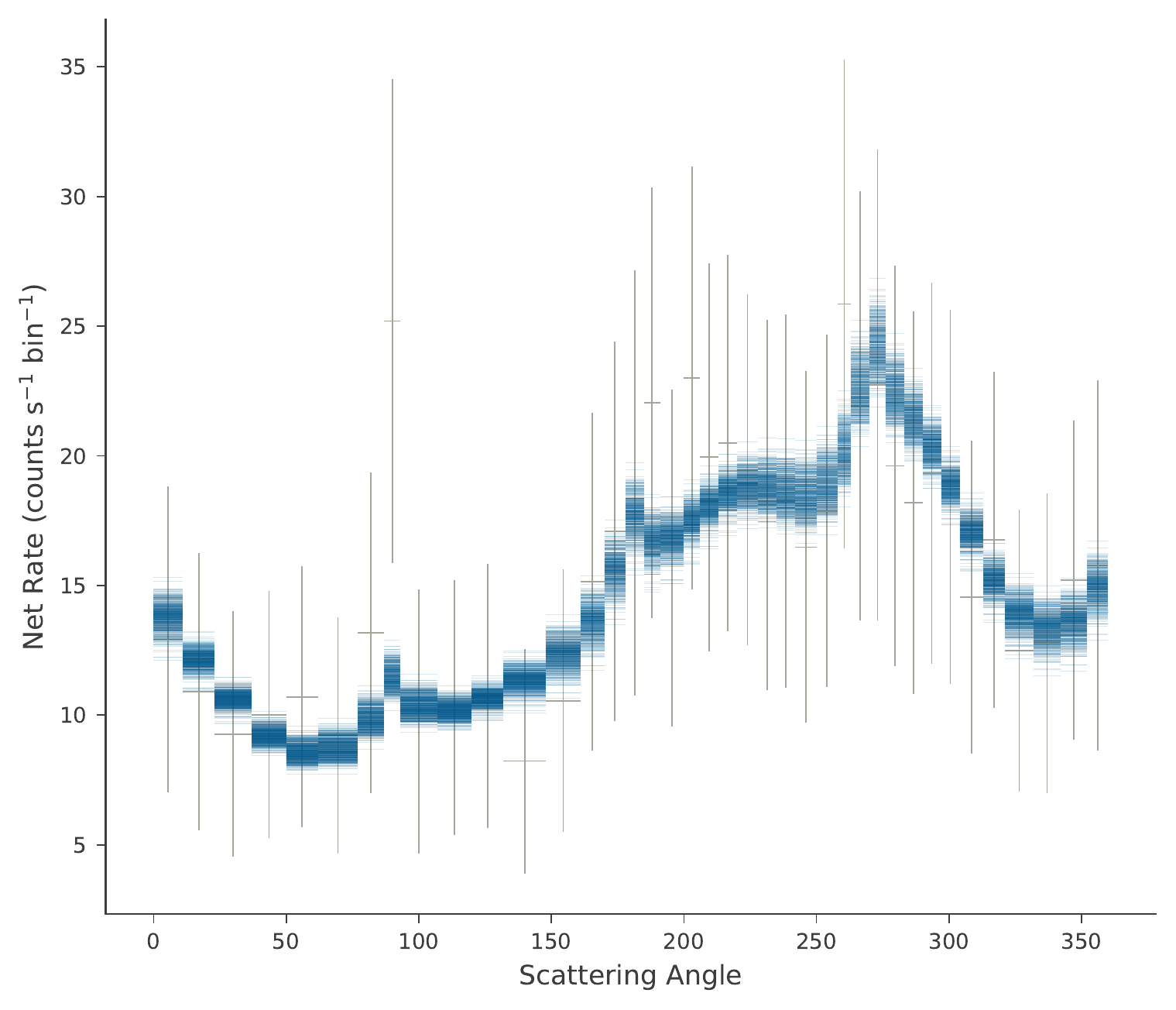}  
  \caption{The measured scattering angle distribution (gray data points with a 9 degree bin size) superimposed by samples from the posterior model predictions (blue) is shown.  The errors on the data points are the Poisson errors corrected for the background. }
  \label{fig:161229A_sd}
\end{subfigure}
\caption{The posterior distribution of the polarization parameters (a) together with the scattering angle distribution of GRB 161229A.}
\label{fig:161229A_PD_sd}
\end{figure}

GRB 161229A was detected by POLAR and by \textit{Fermi}-GBM. The latter define $T0$ as 2016-12-29 at 21:03:48.82 (UT) \citep{GCN_161229A_GCN}, which, for convenience is taken as $T0$ for the analysis presented here. A $T_{90}$ of $(31.26\pm0.44)\,\mathrm{s}$ was measured using POLAR data. The light curve, including the signal region (blue) and part of the the background region (yellow) can be seen on the left in figure \ref{fig:161229A_lc}. The GRB was detected by \textit{Fermi}-GBM \citep{GCN_161229A_GCN} as trigger number 504738232. Data from \textit{Fermi}-GBM was therefore used in the analysis. The spectral results from the joint fit for this GRB can be seen in figure \ref{fig:161229A_cs}. The effective area correction (applied to the POLAR data) found in the analysis was $0.94\pm0.02$. The polarization response of POLAR was produced using the location calculated using the BALROG method, detailed in \cite{BALROG} using \textit{Fermi}-GBM data for this GRB: RA (J2000) = $78.9^\circ$, Dec (J2000) = $6.2^\circ$, a localization error of $2^\circ$ was assumed in the response. Using this location it is found that this GRB occurred far off-axis for POLAR at a $\theta$ angle of $87.6^\circ$ degrees. As a result, POLAR is almost fully insensitive for specific polarization angles (as described in the section regarding GRB 170207A ) while it remains unaffected for other polarization angles as is clear in the posterior distributions. The posterior distributions of the spectral and polarization parameters are shown in figure \ref{fig:post_161229A}. Finally, the posterior distribution of the polarization parameters is shown together with measured scattering angle distribution superimposed by the posterior model predictions (blue) in figure \ref{fig:161229A_PD_sd}. A PD of $17\substack{+24 \\ -13}\%$ is found which is compatible with an unpolarized flux especially when considering the lack of sensitivity in the polarization space for the polarization angle range where the fit converges. A $99\%$ credibility upper limit for PD of $81.0\%$ was found.
 
\clearpage

\section{170101B}

\begin{figure}[ht]
\begin{subfigure}{.5\textwidth}
  \centering
  % include first image
  \includegraphics[width=.95\linewidth]{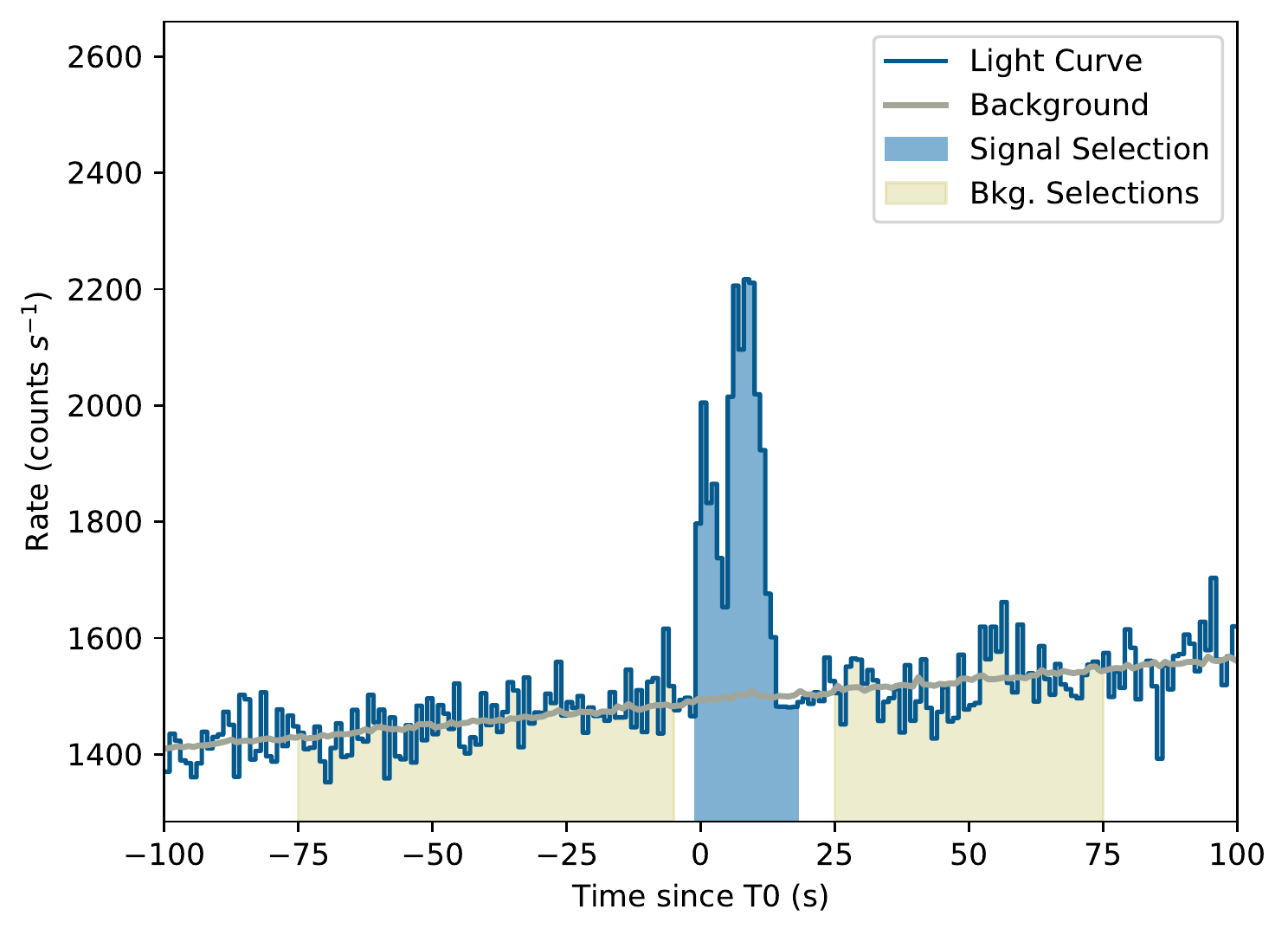}  
  \caption{The light curve of GRB 170101B as measured by POLAR, where $T=0\,\mathrm{s}$ is defined as the $T0$ employed by \textit{Fermi}-GBM in their data products for this GRB \citep{GCN_170101B_Fermi}.}
  \label{fig:170101B_lc}
\end{subfigure}
\newline
\begin{subfigure}{.5\textwidth}
  \centering
  % include second image
  \includegraphics[width=.95\linewidth]{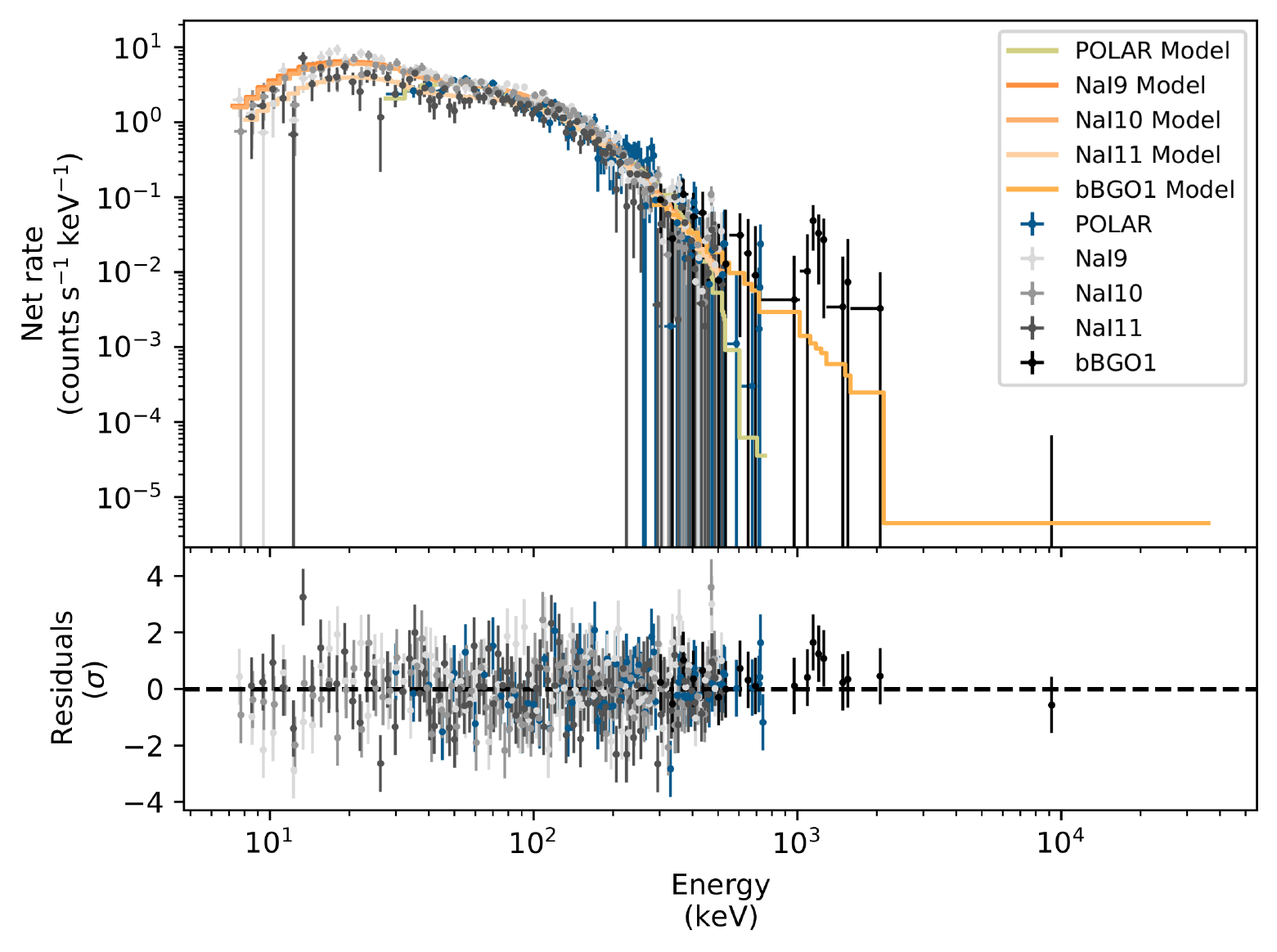}  
  \caption{The joint spectral fit result for 170101B. The number of counts as detected by both POLAR (blue) and the different NaI and BGO detectors of \textit{Fermi}-GBM (gray tints) are shown along with the best fitting spectrum folded through the instrument responses in yellow for POLAR data and in orange tints for the \textit{Fermi}-GBM data. The residuals for both data sets are shown in the bottom of the figure.}
  \label{fig:170101B_cs}
\end{subfigure}
\caption{The light curve as measured by POLAR for GRB 170101B (a) along with the joint spectral fit results of POLAR and \textit{Fermi}-GBM for the signal region indicated in yellow in  figure (a).}
\label{fig:170101B_lc_cs}
\end{figure}

\begin{figure}[!ht]
   \centering
     \resizebox{\hsize}{!}{\includegraphics{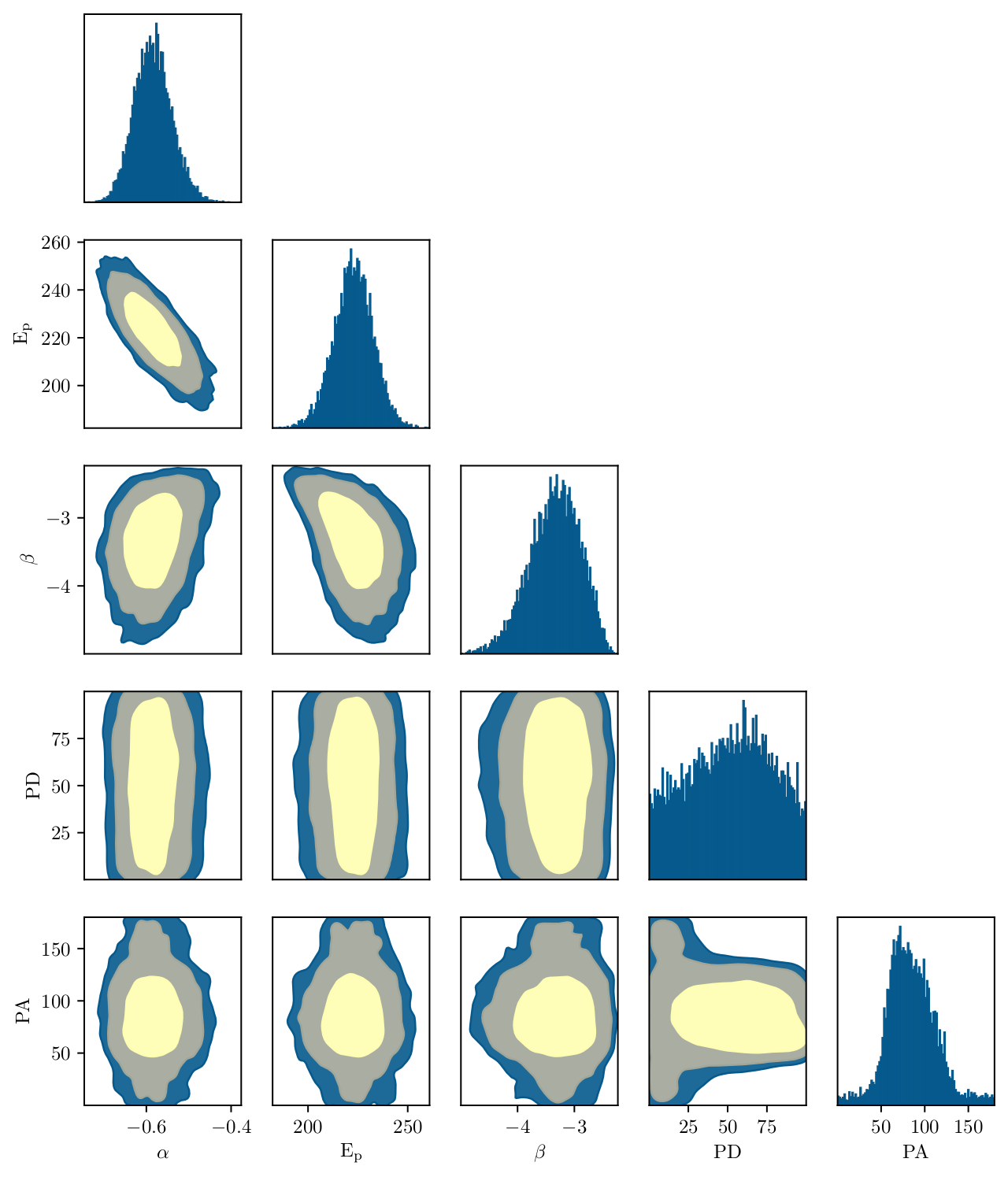}}
   \caption[Spectral and polarization posterior distributions.]
 {The spectral and polarization posterior distributions for GRB 170101B. The 1 and 2 $\sigma$ credibility intervals as well as that corresponding to $99\%$ are indicated. The polarization angle shown here is in the POLAR coordinate system, a rotation in the positive direction of 33 degrees transforms this to the coordinate system as defined by the IAU.}
 \label{fig:post_170101B}
 \end{figure}

\begin{figure}[ht]
\begin{subfigure}{.5\textwidth}
  \centering
  % include first image
  \includegraphics[width=.85\linewidth]{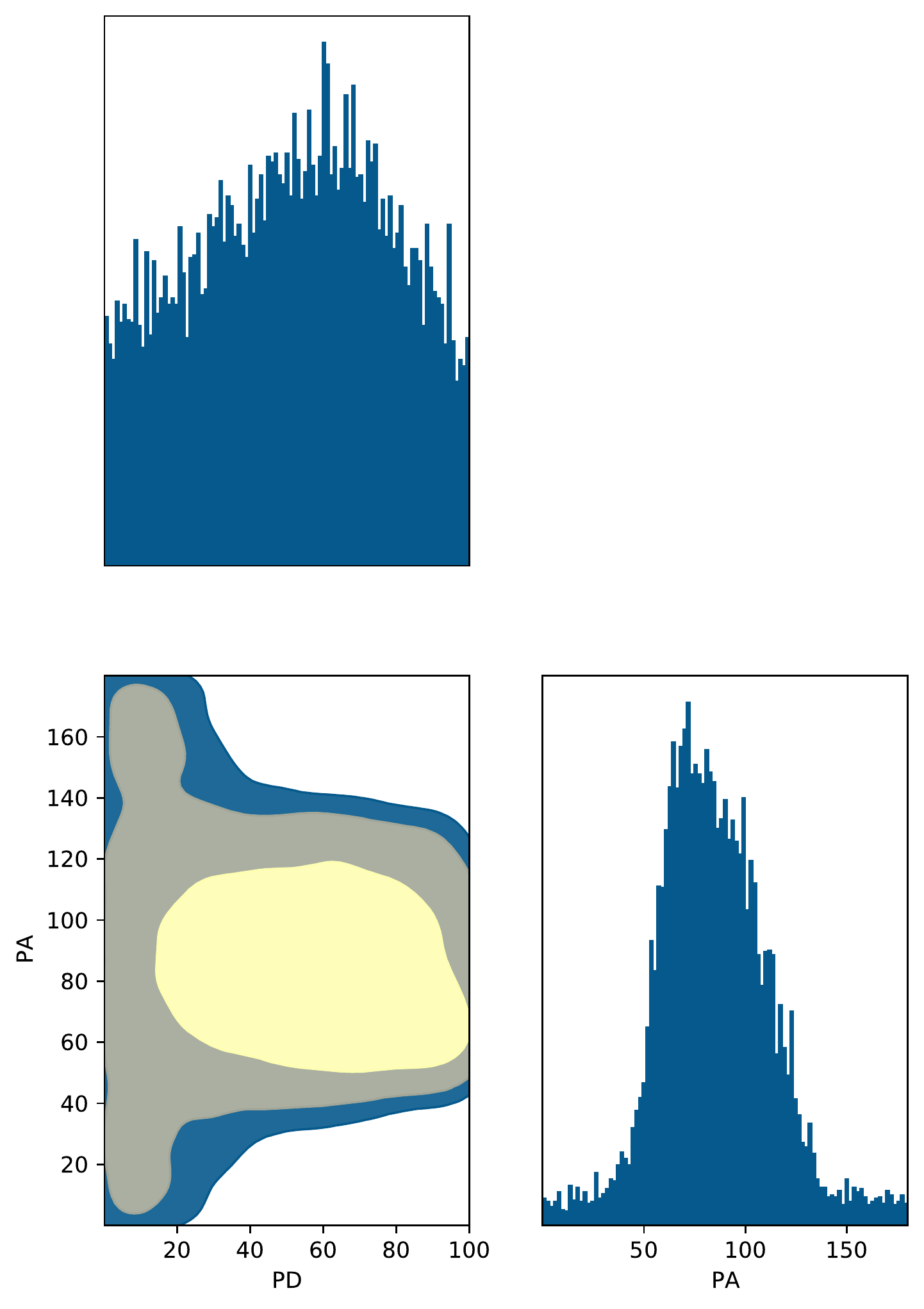}  
  \caption{The polarization posterior distributions for GRB 170101B with the 1 and 2 $\sigma$ credibility intervals as well as that corresponding to $99\%$ credibility. The polarization angle shown here is in the POLAR coordinate system, a rotation in the positive direction of 33 degrees transforms transforms this to the coordinate system as defined by the IAU.}
  \label{fig:170101B_PD}
\end{subfigure}
\newline
\begin{subfigure}{.5\textwidth}
  \centering
  % include second image
  \includegraphics[width=.85\linewidth]{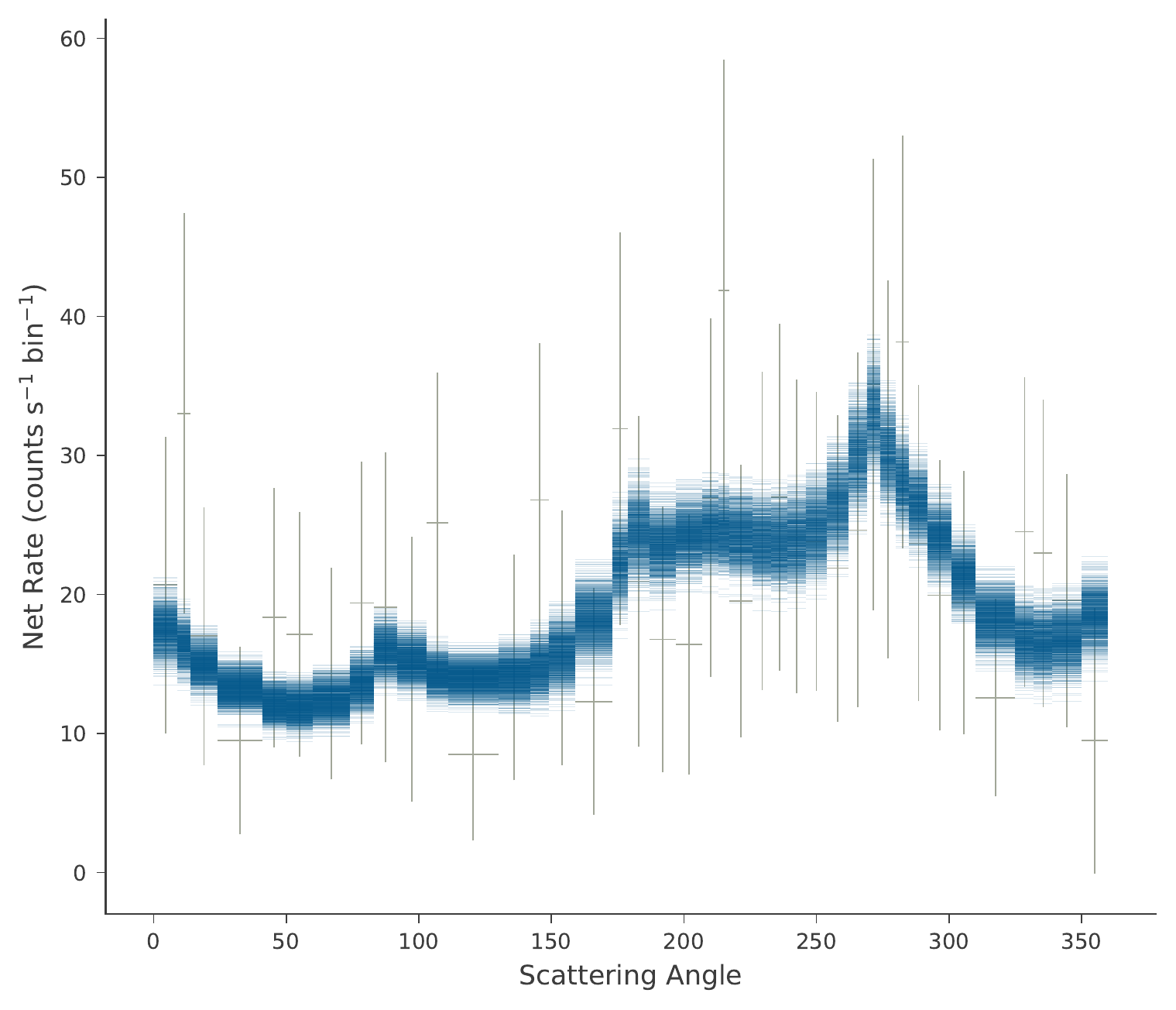}  
  \caption{The measured scattering angle distribution (gray data points with a 9 degree bin size) superimposed by samples from the posterior model predictions (blue) is shown.  The errors on the data points are the Poisson errors corrected for the background. }
  \label{fig:170101B_sd}
\end{subfigure}
\caption{The posterior distribution of the polarization parameters (a) together with the scattering angle distribution of GRB 170101B.}
\label{fig:170101B_PD_sd}
\end{figure}

GRB 170101B, a long GRB with several overlapping pulses, was detected by POLAR and by \textit{Fermi}-GBM. The latter reported a $T0$ of 2017-01-01 at 02:47:17.87 (UT) \citep{GCN_170101B_Fermi}, which, for convenience, will be used as the $T0$ for the analysis presented here as well. A $T_{90}$ of $(11.14\pm0.38)\,\mathrm{s}$ was measured using POLAR data. The light curve, including the signal region (blue) and part of the background region (yellow) can be seen on the left in figure \ref{fig:170101B_lc}. Data from \textit{Fermi}-GBM was used in the analysis. The spectral results of the joint fit for this GRB using both POLAR and \textit{Fermi}-GBM data can be seen in figure \ref{fig:170101B_cs}. The effective area correction (applied to the POLAR data) found in the analysis was $1.19\pm0.04$. The spectral fits for both instruments can be seen to agree well. The polarization response of POLAR was produced using the location calculated using the BALROG method described in \cite{BALROG} with \textit{Fermi}-GBM data. The best fitting location was: RA (J2000) = $69.6^\circ$, Dec (J2000) = $-1.0^\circ$, a localization error of $2^\circ$ was assumed. Using this location it is found that this GRB occurred relatively far off-axis for POLAR at a $\theta$ angle of $75.0^\circ$ degrees. As a result POLAR is not very sensitive for specific polarization angles as described in the section describing the analysis of GRB 170207A. The posterior distributions of the spectral and polarization parameters are shown in figure \ref{fig:post_170101B}. Finally the posterior distribution of the polarization parameters is shown together with the measured scattering angle distribution superimposed by the posterior model predictions (blue) in figure \ref{fig:170101B_PD_sd}. The scattering angle distributions indicate clearly the large off-axis angle resulting in the $360^\circ$ modulation in the distribution. A PD of $60\substack{+24 \\ -36}\%$ is found which, although not being fully incompatible with an unpolarized flux, gives a hint of polarization. It should additionally be noted that the best fitting PA of $76\substack{+34 \\ -21}^\circ$ (as measured in the POLAR coordinate system) is largely unaffected by the before mentioned loss of sensitivity due to the off-axis angle. The best fitting PD is therefore likely not a result of this effect. 
 
\clearpage

\section{170114A}

GRB 170114A was detected by POLAR and \textit{Fermi}-GBM. The latter reported a $T0$ of 2017-01-14 at 22:01:09.50 (UT) \citep{GCN_170114A_GBM} which, for convenience, is also taken as $T0$ for the analysis presented here. A $T_{90}$ of $(10.48\pm0.16)\,\mathrm{s}$ was measured using POLAR data. The polarization properties of this GRB has been extensively studied in a previous publication \citep{170114A_BALROG} using the same methodology used here. There, a time integrated analysis found a polarization consistent with an unpolarized flux, while time-resolved analysis results in a PD of around $30\%$ with a changing polarization angle after an initial unpolarized flux during the start of the GRB. Additionally, it was found that the spectrum of this GRB is consistent with that from synchrotron radiation. In this work, however, we perform the analysis using a Band function instead of a physical synchrotron model. We include only a summary of our results from this analysis here for completeness, and to show that consistent results are found when using the Band function instead of the more physical synchrotron model. For details on this GRB and its polarization characteristics we refer to reader to \cite{170114A_BALROG}.

\begin{figure}[ht]
\begin{subfigure}{.5\textwidth}
  \centering
  % include first image
  \includegraphics[width=.95\linewidth]{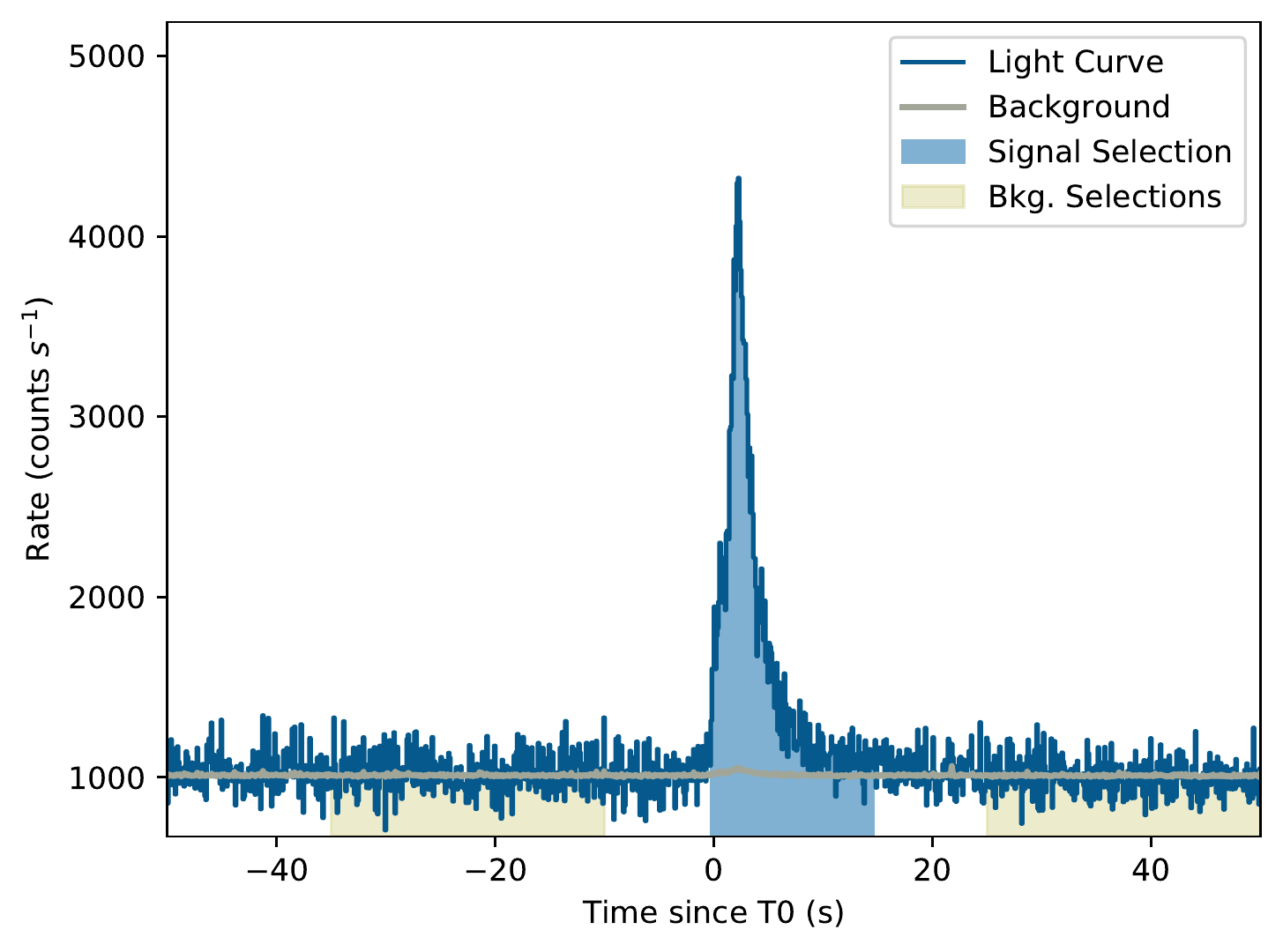}  
  \caption{The light curve of GRB 170114A as measured by POLAR where $T=0\,\mathrm{s}$ is defined as the $T0$ employed by \textit{Fermi}-GBM \citep{GCN_170114A_GBM} for this GRB.}
  \label{fig:170114A_lc}
\end{subfigure}
\newline
\begin{subfigure}{.5\textwidth}
  \centering
  % include second image
  \includegraphics[width=.95\linewidth]{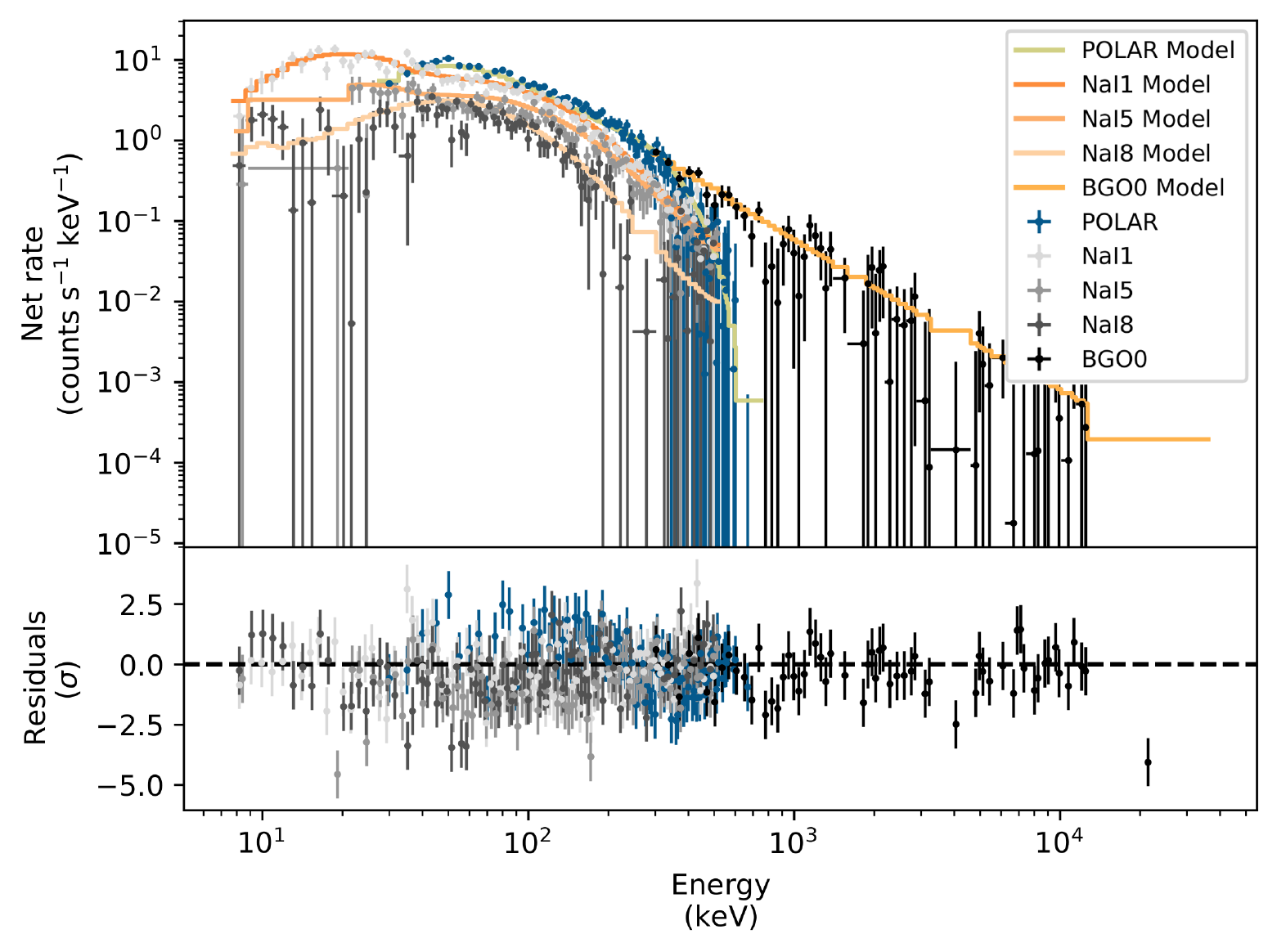}  
  \caption{The joint spectral fit result for 170114A. The number of counts as detected by both POLAR (blue) and the different NaI and BGO detectors of \textit{Fermi}-GBM (gray tints) are shown along with the best fitting spectrum folded through the instrument responses in yellow for POLAR data and in orange tints for the \textit{Fermi}-GBM data. The residuals for both data sets are shown in the bottom of the figure.}
  \label{fig:170114A_cs}
\end{subfigure}
\caption{The light curve as measured by POLAR for GRB 170114A (a) along with the joint spectral fit results of POLAR and \textit{Fermi}-GBM for the signal region indicated in yellow in  figure (a).}
\label{fig:170114A_lc_cs}
\end{figure}

\begin{figure}[!ht]
   \centering
     \resizebox{\hsize}{!}{\includegraphics{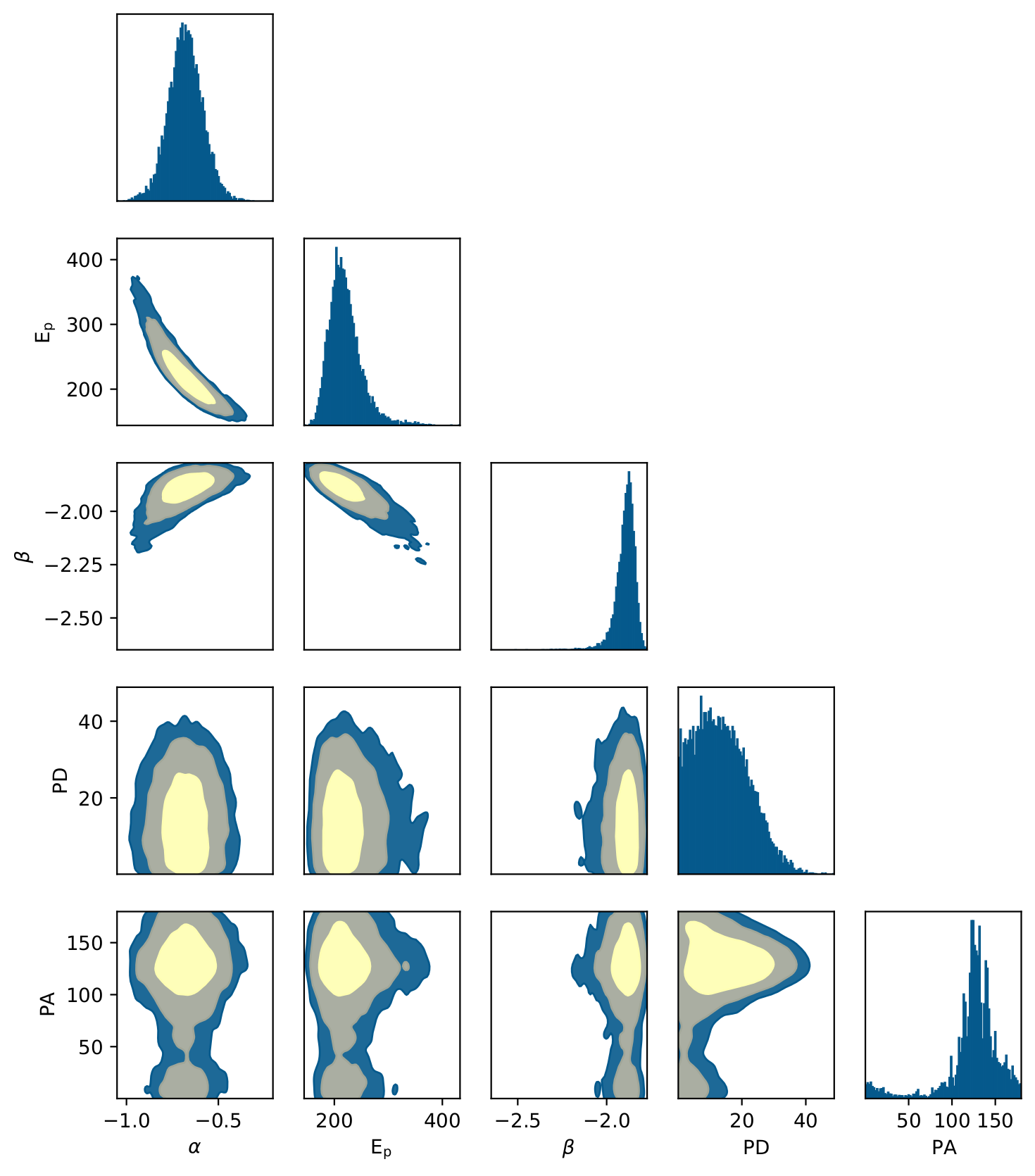}}
   \caption[Spectral and polarization posterior distributions.]
 {The spectral and polarization posterior distributions for GRB 170114A. The 1 and 2 $\sigma$ credibility intervals as well as that corresponding to $99\%$ are indicated. The polarization angle shown here is in the POLAR coordinate system, a rotation in the positive direction of 38 degrees transforms this to the coordinate system as defined by the IAU.}
 \label{fig:post_170114A}
 \end{figure}

\begin{figure}[ht]
\begin{subfigure}{.5\textwidth}
  \centering
  % include first image
  \includegraphics[width=.85\linewidth]{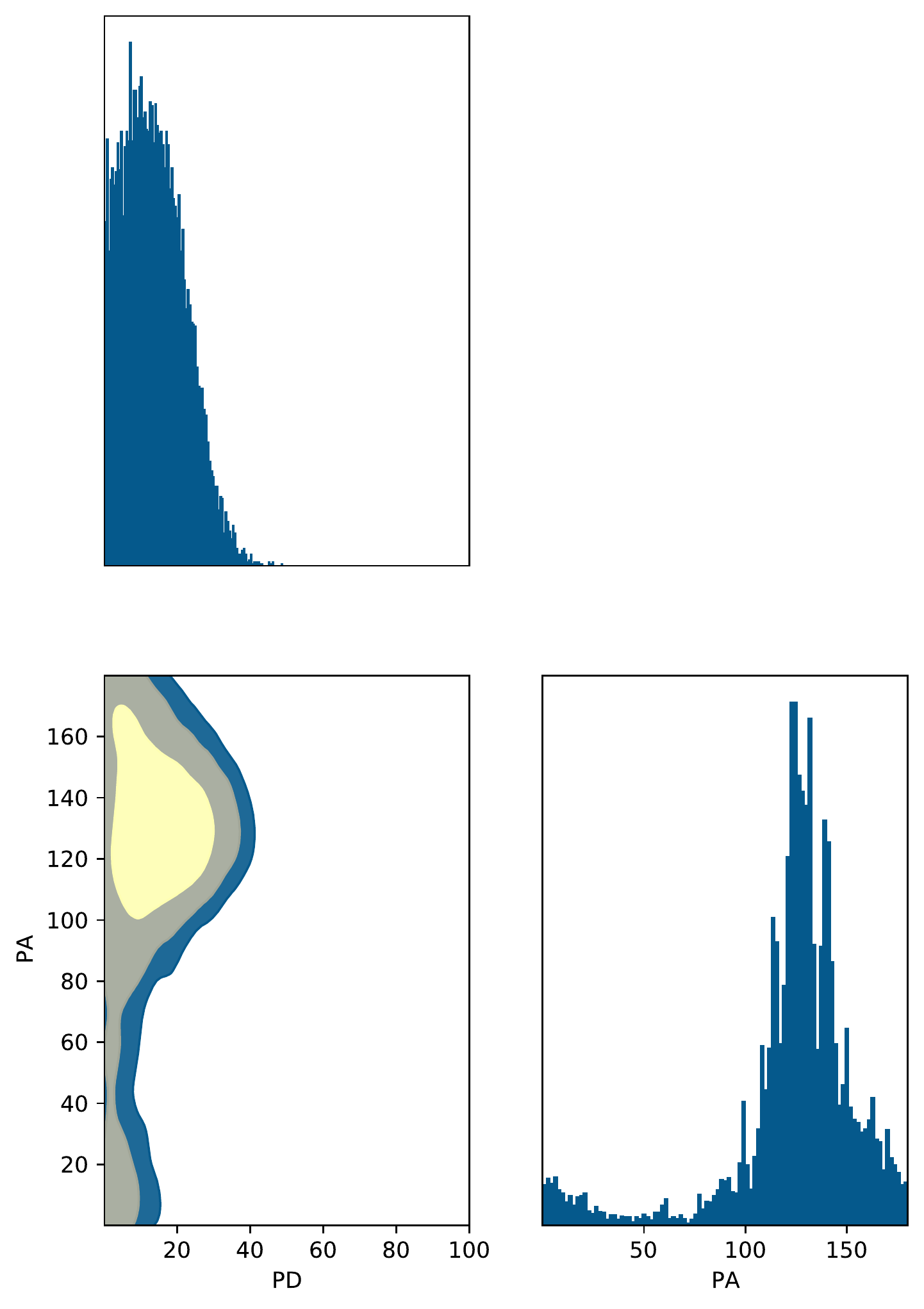}  
  \caption{The polarization posterior distributions for GRB 170114A with the 1 and 2 $\sigma$ credibility intervals as well as that corresponding to $99\%$. The polarization angle shown here is in the POLAR coordinate system, a rotation in the positive direction of 38 degrees transforms this to the coordinate system as defined by the IAU. }
  \label{fig:170114A_PD}
\end{subfigure}
\newline
\begin{subfigure}{.5\textwidth}
  \centering
  % include second image
  \includegraphics[width=.85\linewidth]{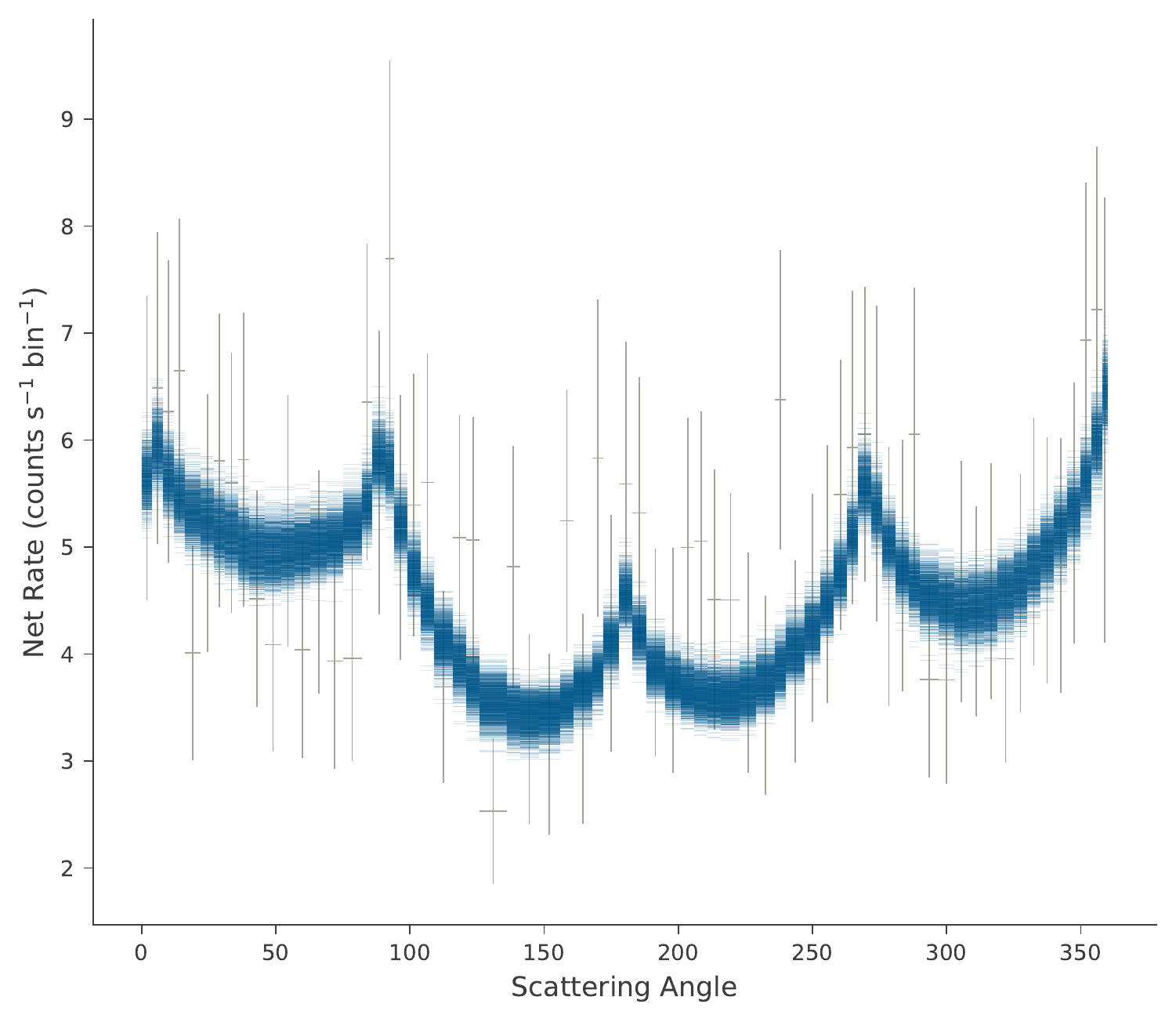}  
  \caption{The measured scattering angle distribution (gray data points with a 9 degree bin size) superimposed by the posterior model predictions (blue) is shown. The errors on the data points are the Poisson errors corrected for the background. }
  \label{fig:170114A_sd}
\end{subfigure}
\caption{The posterior distribution of the polarization parameters (a) together with the scattering angle distribution of GRB 170114A.}
\label{fig:170114A_PD_sd}
\end{figure}

The light curve, including the signal region (blue) and the background region (yellow) can be seen in figure \ref{fig:170114A_lc}. The GRB was detected by \textit{Fermi}-GBM \citep{GCN_170114A_GBM} allowing spectral data from this instrument to be used in the analysis. The spectral results of the joint fit can be seen in figure \ref{fig:170114A_cs}. The fit results of both instruments can be seen to be in good agreement for the full energy range. The effective area correction (applied to the POLAR data) found in the analysis was $1.24\pm0.04$. The polarization response of POLAR was produced using the location calculated using the BALROG method, detailed in \cite{BALROG}, using \textit{Fermi}-GBM data for this GRB. The location of the GRB found using this method corresponds to: RA (J2000) = $13.10^\circ$, Dec (J2000) = $-13.0^\circ$, a localization error of $2^\circ$ was assumed in the response. The posterior distributions of the spectral and polarization parameters are shown in figure \ref{fig:post_170114A}. Finally the posterior distribution of the polarization parameters is shown together with measured scattering angle distribution superimposed by the posterior model predictions (blue) in figure \ref{fig:170114A_PD_sd}. A PD of $10.1\substack{+10.5 \\ -7.4}\%$ is found which is compatible with that reported in \cite{Zhang+Kole}, it should be noted that the PA used in \cite{Zhang+Kole} is presented following the IAU convention while here we measure it in the POLAR coordinate system, an rotation of $142$ degrees in the negative direction is required to move to the coordinate system as defined by the IAU. A $99\%$ credibility upper limit of $35.9\%$ is found for this GRB.

\begin{figure}[!ht]
   \centering
      \resizebox{\hsize}{!}{\includegraphics{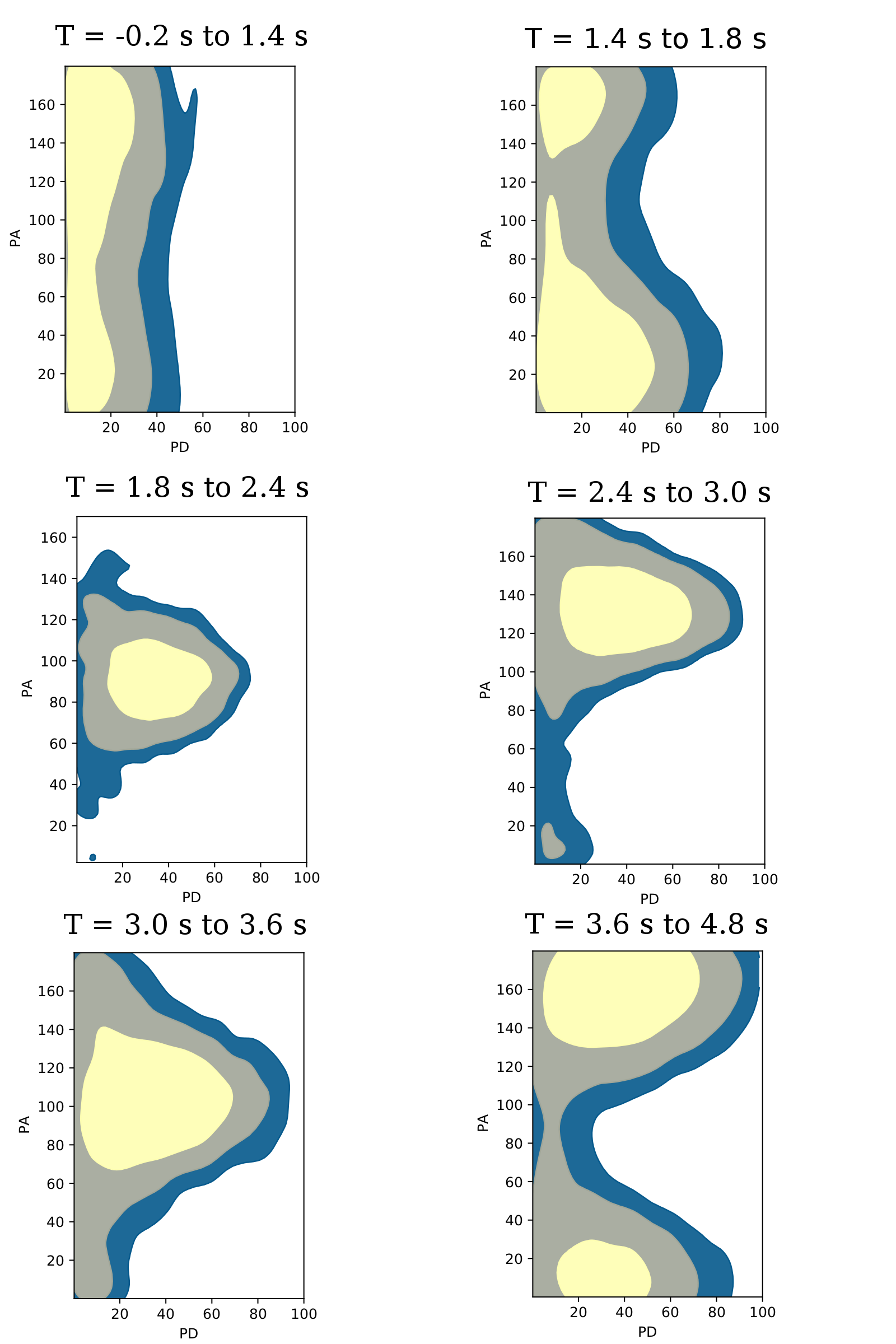}}
   \caption[Light curve of GRB 170114A as measured by POLAR.]
 {The polarization posterior distributions for the  time bins of GRB 170114A are shown with the 1 and 2 $\sigma$ credibility intervals as well as that corresponding to $99\%$ credibility. The polarization angle shown here is in the POLAR coordinate system, a rotation (in the negative direction) of 142 degrees transforms this to the coordinate system as defined by the IAU.}
 \label{time_res_170114A}
 \end{figure}

In order to verify if the current analysis yields results consistent with that found in \cite{170114A_BALROG} for a time-resolved analysis, the same data from the time bins as those calculated using the method in \cite{170114A_BALROG} were studied here. In figure \ref{time_res_170114A} we present the posteriors for the polarization for the first 6 time bins, those corresponding to $T=-0.2-1.4\,\mathrm{s}$, $ T=1.4-1.8\,\mathrm{s}$,  $T=1.8-2.4\,\mathrm{s}$, $T=2.4-3.0\,\mathrm{s}$, $T=3.0-3.6\,\mathrm{s}$ and $T=3.6-4.8\,\mathrm{s}$. It was found that the polarization parameters found here are consistent with those found in \cite{170114A_BALROG}. The results are point to an initially unpolarized flux followed by 4 time bins with a PD around $30\%$ where the PA varies from time bin to time bin. The polarization angles shown here are in the POLAR coordinate system again. It can be concluded that the same polarization angles and changes are found between both analyses.

% 
% \begin{figure}[!ht]
%    \centering
%      \resizebox{\hsize}{!}{\includegraphics[width=14 cm]{170114A_lc2.png}}
%    \caption[Light curve of GRB 170114A as measured by POLAR.]
%  {On top we see the light curve of GRB 170114A as measured by POLAR where T=0 is defined as the detection time by \textit{Fermi-GBM} \citep{GCN_170114A_GBM}. Below we see the joint spectral fit for the full GRB is shown using data from POLAR (purple) and data from the NaI1, NaI5, NaI8 and BGO0 detectors from \textit{Fermi-GBM} in blue, magenta, orange and red respectively.}
%  \label{fig:170114A_lc}
%  \end{figure}

\clearpage

\section{170127C}

\begin{figure}[ht]
\begin{subfigure}{.5\textwidth}
  \centering
  % include first image
  \includegraphics[width=.95\linewidth]{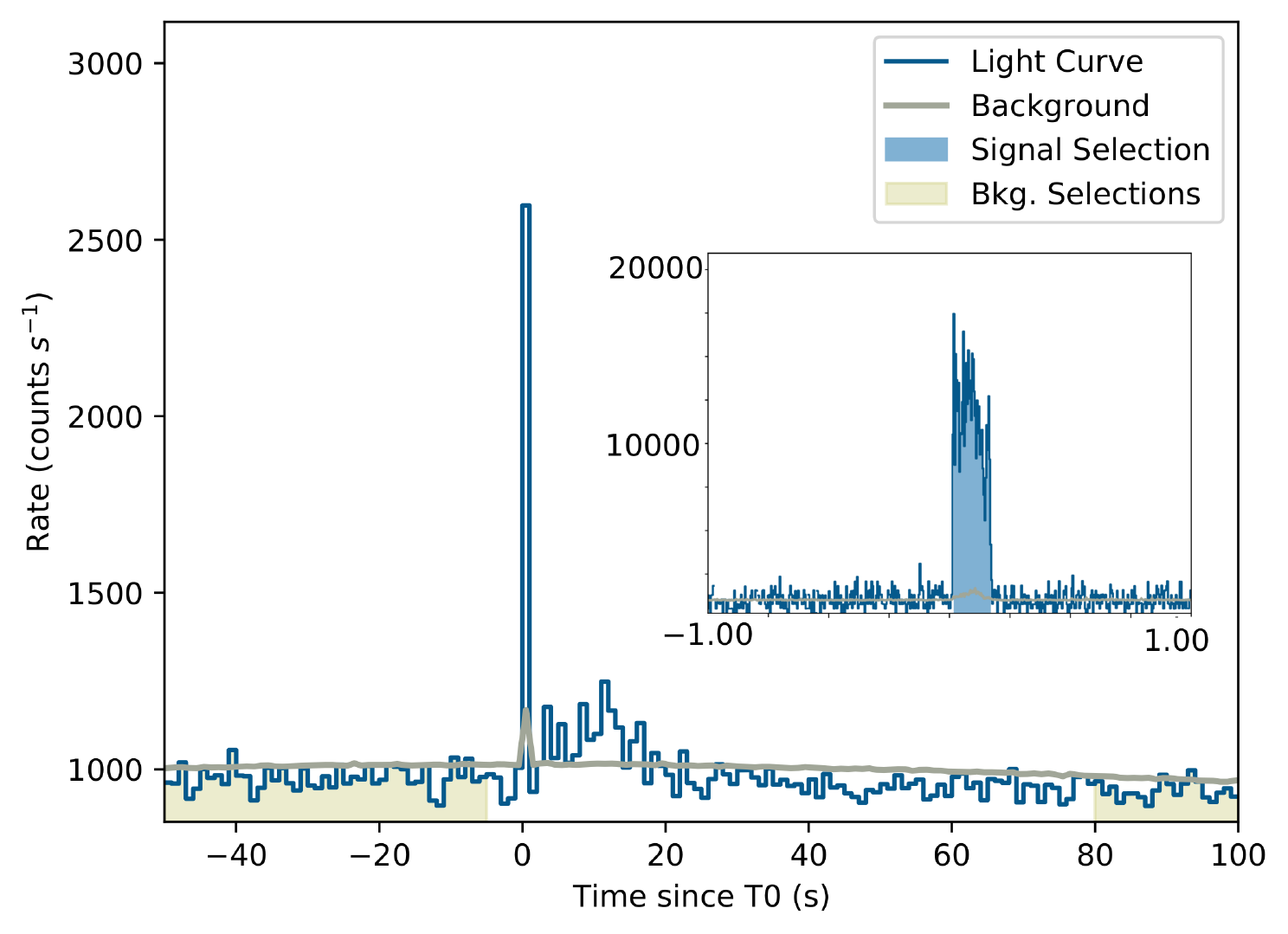}  
  \caption{The light curve of GRB 170127C as measured by POLAR with an inlay of a zoomed in version around the peak, where $T=0\,\mathrm{s}$ is defined as the $T0$  employed by \textit{Fermi}-GBM in their data products for this GRB \citep{GCN_170127C_Fermi}. The spectrum of the late emission, which can be seen until approximately $20\,\mathrm{s}$ was fitted separately. The results of this are reported in table \ref{tab:spectra}. }
  \label{fig:170127C_lc}
\end{subfigure}
\newline
\begin{subfigure}{.5\textwidth}
  \centering
  % include second image
  \includegraphics[width=.95\linewidth]{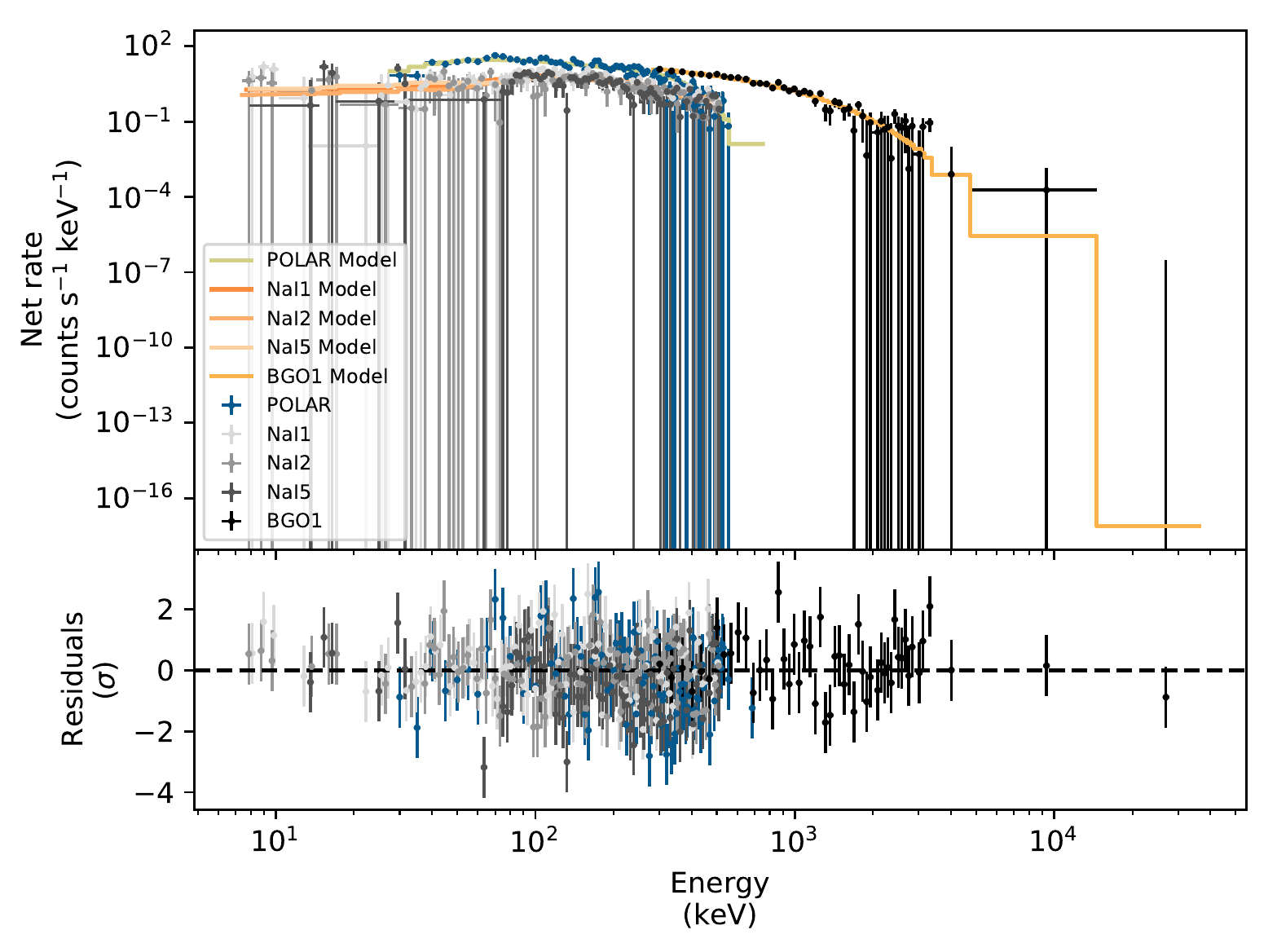}  
  \caption{The joint spectral fit result for GRB 170127C. The number of counts as detected by both POLAR (blue) and the different NaI and BGO detectors of \textit{Fermi}-GBM (gray tints) are shown along with the best fitting spectrum folded through the instrument responses in yellow for POLAR data and in orange tints for the \textit{Fermi}-GBM data. The residuals for both data sets are shown in the bottom of the figure.}
  \label{fig:170127C_cs}
\end{subfigure}
\caption{The light curve as measured by POLAR for GRB 170127C (a) along with the joint spectral fit results of POLAR and \textit{Fermi}-GBM for the signal region indicated in yellow in  figure (a).}
\label{fig:170127C_lc_cs}
\end{figure}

\begin{figure}[!ht]
   \centering
     \resizebox{\hsize}{!}{\includegraphics{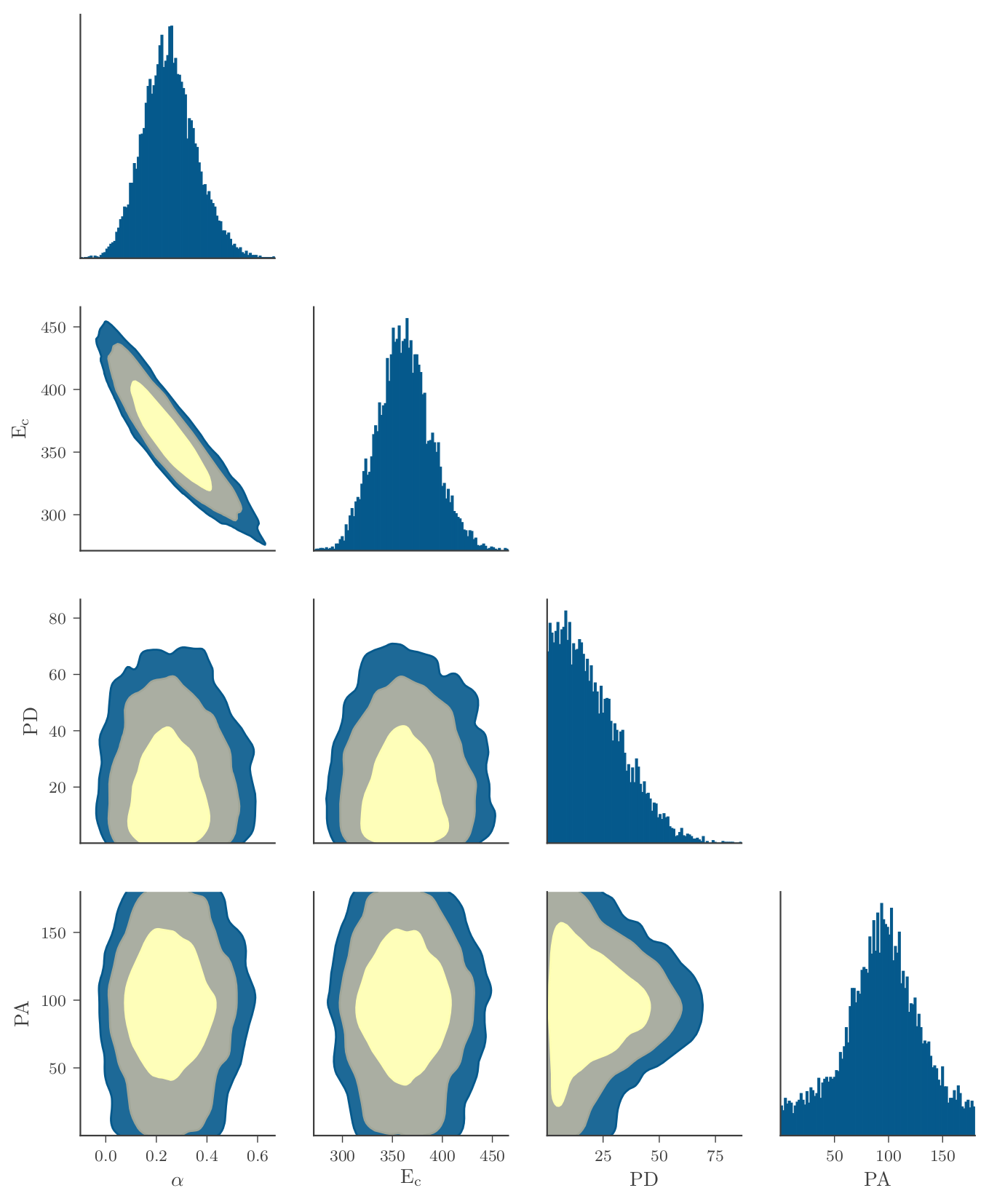}}
   \caption[Spectral and polarization posterior distributions.]
 {The spectral and polarization posterior distributions for GRB 170127C. The 1 and 2 $\sigma$ credibility intervals as well as that corresponding to $99\%$ are indicated. The polarization angle shown here is in the POLAR coordinate system, a rotation in the positive direction of 130 degrees transforms transforms this to the coordinate system as defined by the IAU.} 
 \label{fig:post_170127C}
 \end{figure}

\begin{figure}[ht]
\begin{subfigure}{.5\textwidth}
  \centering
  % include first image
  \includegraphics[width=.85\linewidth]{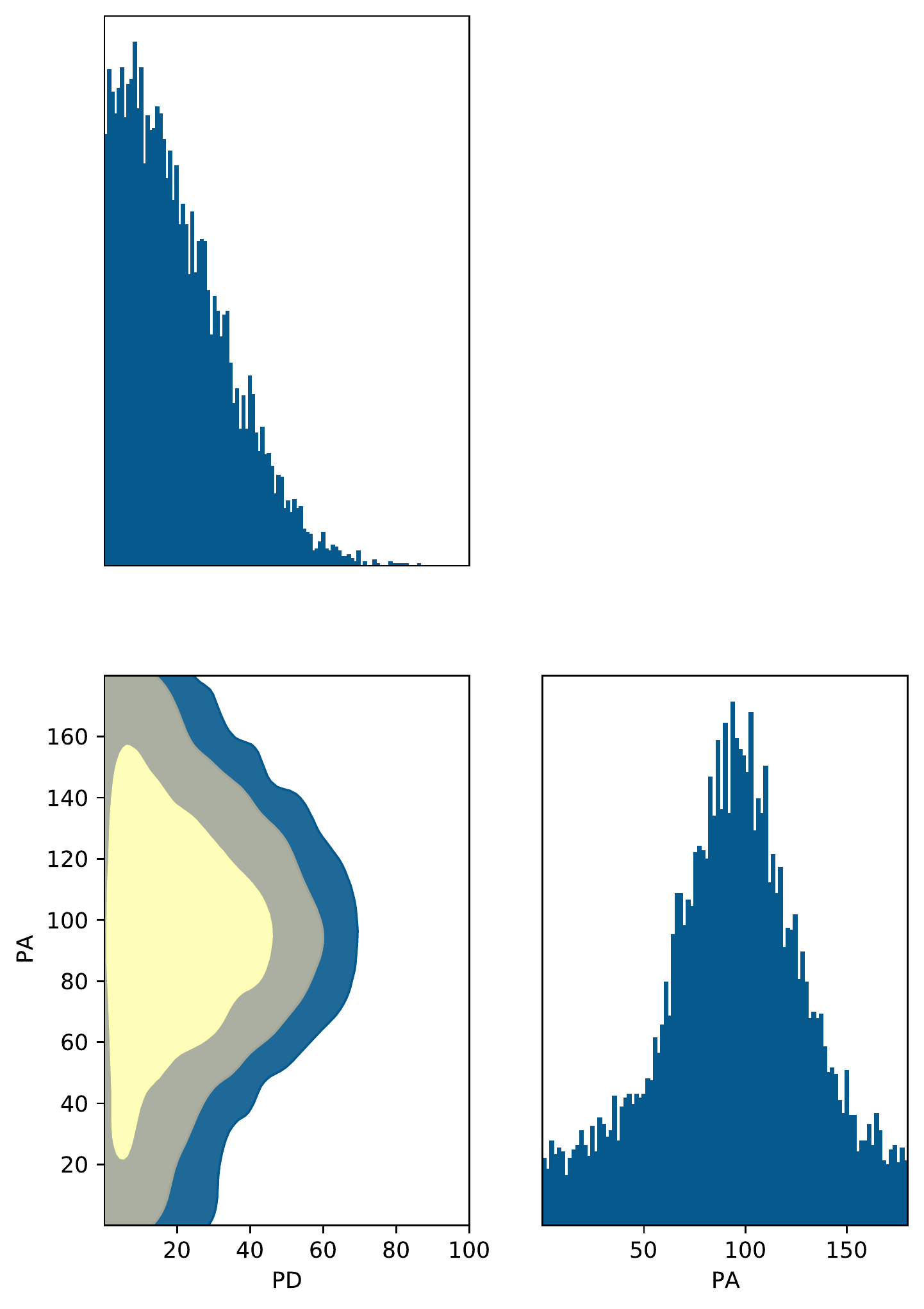}  
  \caption{The polarization posterior distributions for GRB 170127C with the 1 and 2 $\sigma$ credibility intervals as well as that corresponding to $99\%$ credibility. The polarization angle shown here is in the POLAR coordinate system, a rotation in the positive direction of 130 degrees transforms this to the coordinate system as defined by the IAU. }
  \label{fig:170127C_PD}
\end{subfigure}
\newline
\begin{subfigure}{.5\textwidth}
  \centering
  % include second image
  \includegraphics[width=.85\linewidth]{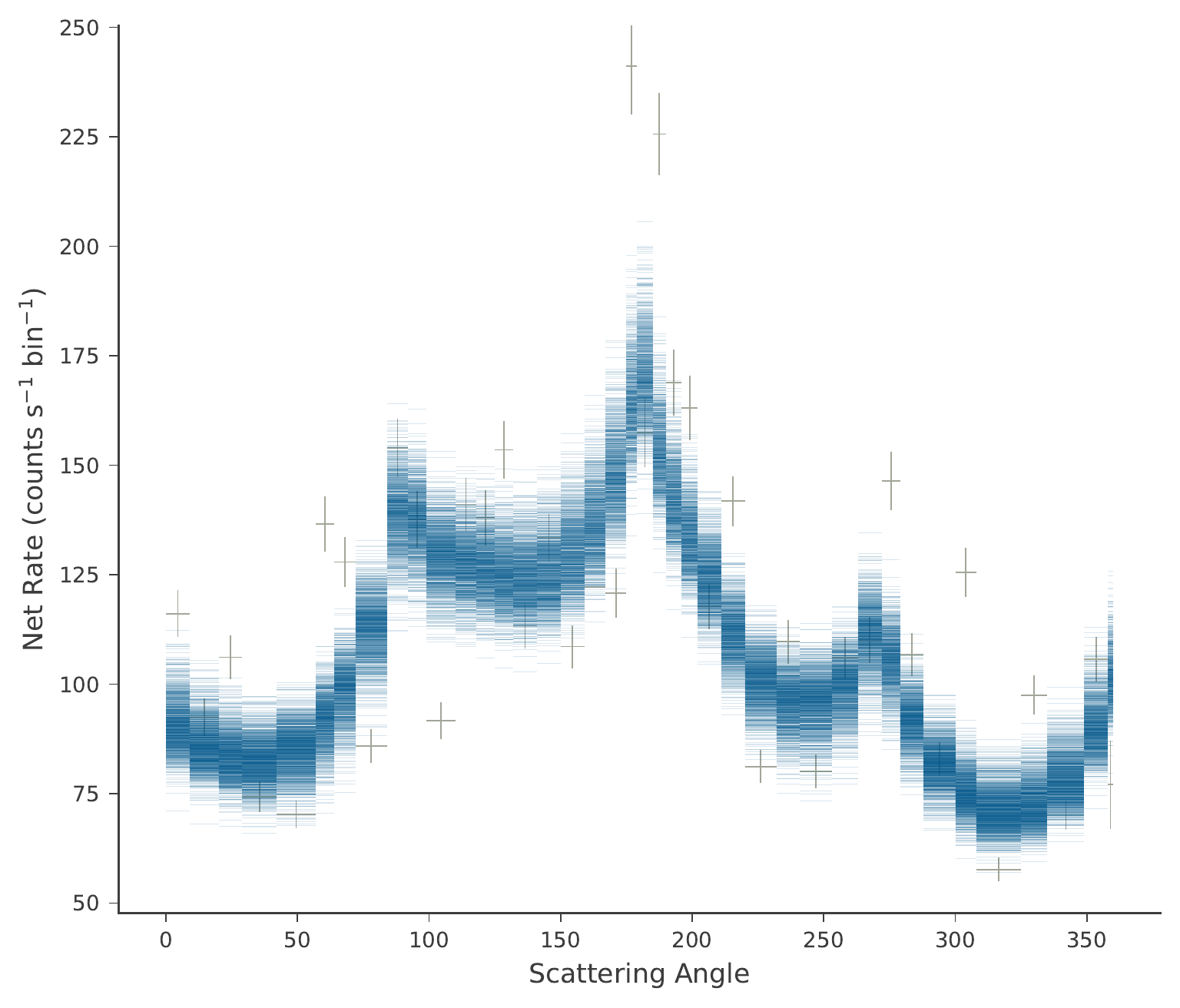}  
  \caption{The measured scattering angle distribution (gray data points with a 9 degree bin size) superimposed by samples from the posterior model predictions (blue) is shown. The errors on the data points are the Poisson errors corrected for the background. }
  \label{fig:170127C_sd}
\end{subfigure}
\caption{The posterior distribution of the polarization parameters (a) together with the scattering angle distribution of GRB 170127C.}
\label{fig:170127C_PD_sd}
\end{figure}

GRB 170127C, a bright short GRB with several overlapping pulses, was detected by POLAR and by \textit{Fermi}-GBM who report a $T0$ of 2017-01-27 at 01:35:47.79 (UT) \citep{GCN_170127C_Fermi}. For convenience this will be used as the $T0$ for the analysis presented here as well. A $T_{90}$ of $(0.14\pm0.01)\,\mathrm{s}$ was measured using POLAR data. The light curve, including the signal region (blue) and the background region (yellow) can be seen in figure \ref{fig:170127C_lc} along with a more detailed inset of the pulse region. It should be noted here that a large period of the time after the GRB was excluded as background as a longer lasting afterglow was reported by Konus-Wind \citep{GCN_170127C_KONUS} for this GRB. This late emission is additionally visible in the POLAR data. A separate spectral fit was performed on this afterglow using both POLAR and \textit{Fermi}-GBM data. The results indicate that the emission is significantly softer than the rather hard emission in the bright first peak. The spectral parameters for both periods are reported in table \ref{tab:spectra}. As, based on the spectral results, we assume this afterglow to have a different origin we exclude it from the polarization analysis here. Data from \textit{Fermi}-GBM was used for the joint analysis. The spectral results of the joint fit can be seen in figure \ref{fig:170127C_cs}. The effective area correction (applied to the POLAR data) found in the analysis was $0.60\pm0.23$. The response was produced using the location provided by \textit{Fermi}-LAT: RA (J2000) = $339.3^\circ$, Dec (J2000) = $-63.9^\circ$ \citep{GCN_170127C_LAT}. The posterior distributions of the spectral and polarization parameters are shown in figure \ref{fig:post_170127C}. Finally the posterior distribution of the polarization parameters is shown together with measured scattering angle distribution superimposed by the posterior model predictions (blue) in figure \ref{fig:170127C_PD_sd}. A PD of $9.9\substack{+19.3 \\ -9.9}\%$ is found which is compatible with that reported in \cite{Zhang+Kole}. It should be noted that the PA used in \cite{Zhang+Kole} is measured in the IAU coordinate system while here we measure it in the POLAR coordinate system. A rotation of $50^\circ$ in the negative direction transforms the POLAR coordinate system to that as defined by the IAU for this GRB. A $99\%$ credibility upper limit for PD of $61.2\%$ is found.

\clearpage
 \section{170206A}

\begin{figure}[ht]
\begin{subfigure}{.5\textwidth}
  \centering
  % include first image
  \includegraphics[width=.95\linewidth]{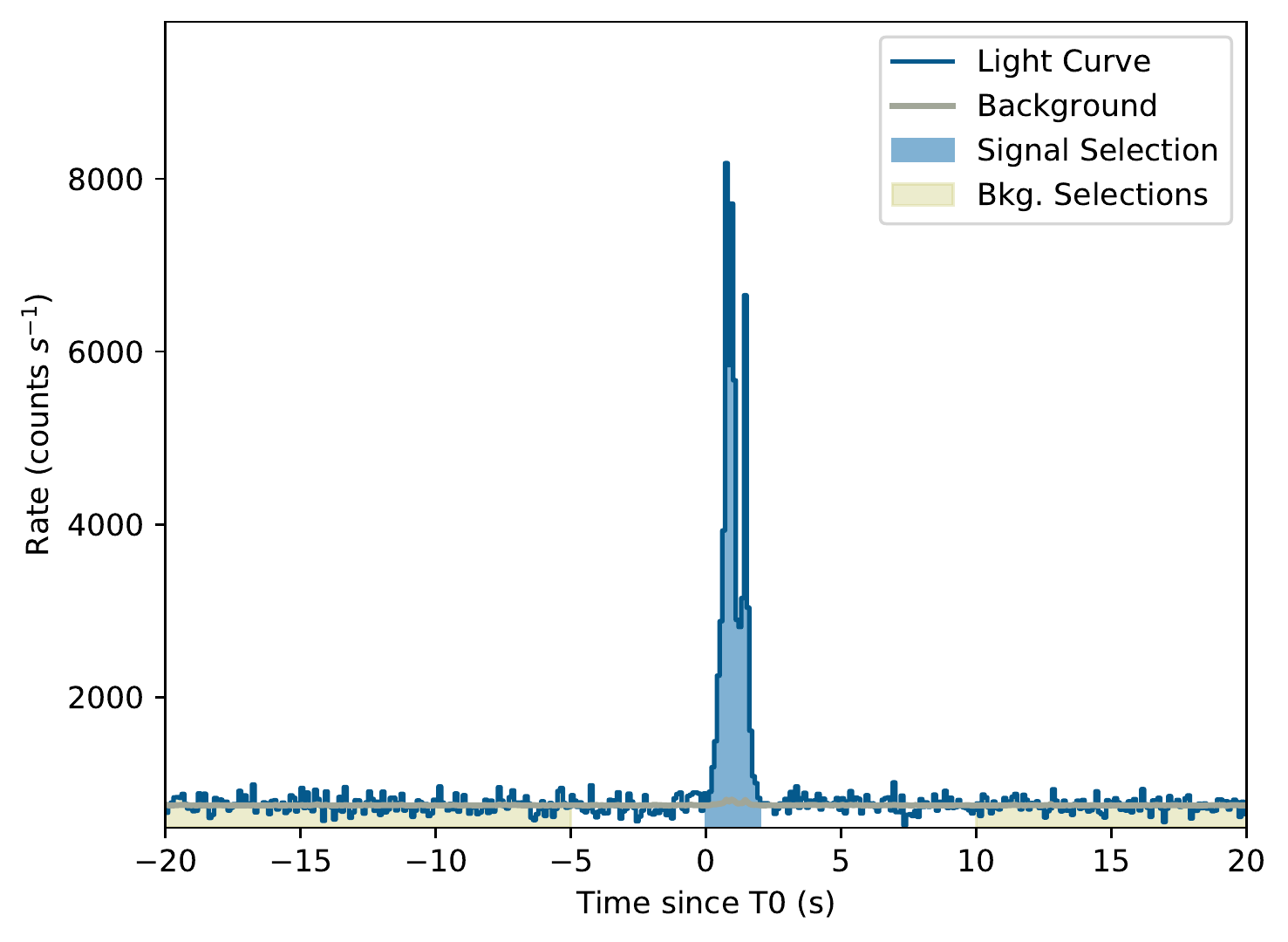}  
  \caption{The light curve of GRB 170206A as measured by POLAR, where $T=0\,\mathrm{s}$ is defined as the $T0$ employed by \textit{Fermi}-GBM in their data products \citep{GCN_170206A_Fermi}.}
  \label{fig:170206A_lc}
\end{subfigure}
\newline
\begin{subfigure}{.5\textwidth}
  \centering
  % include second image
  \includegraphics[width=.95\linewidth]{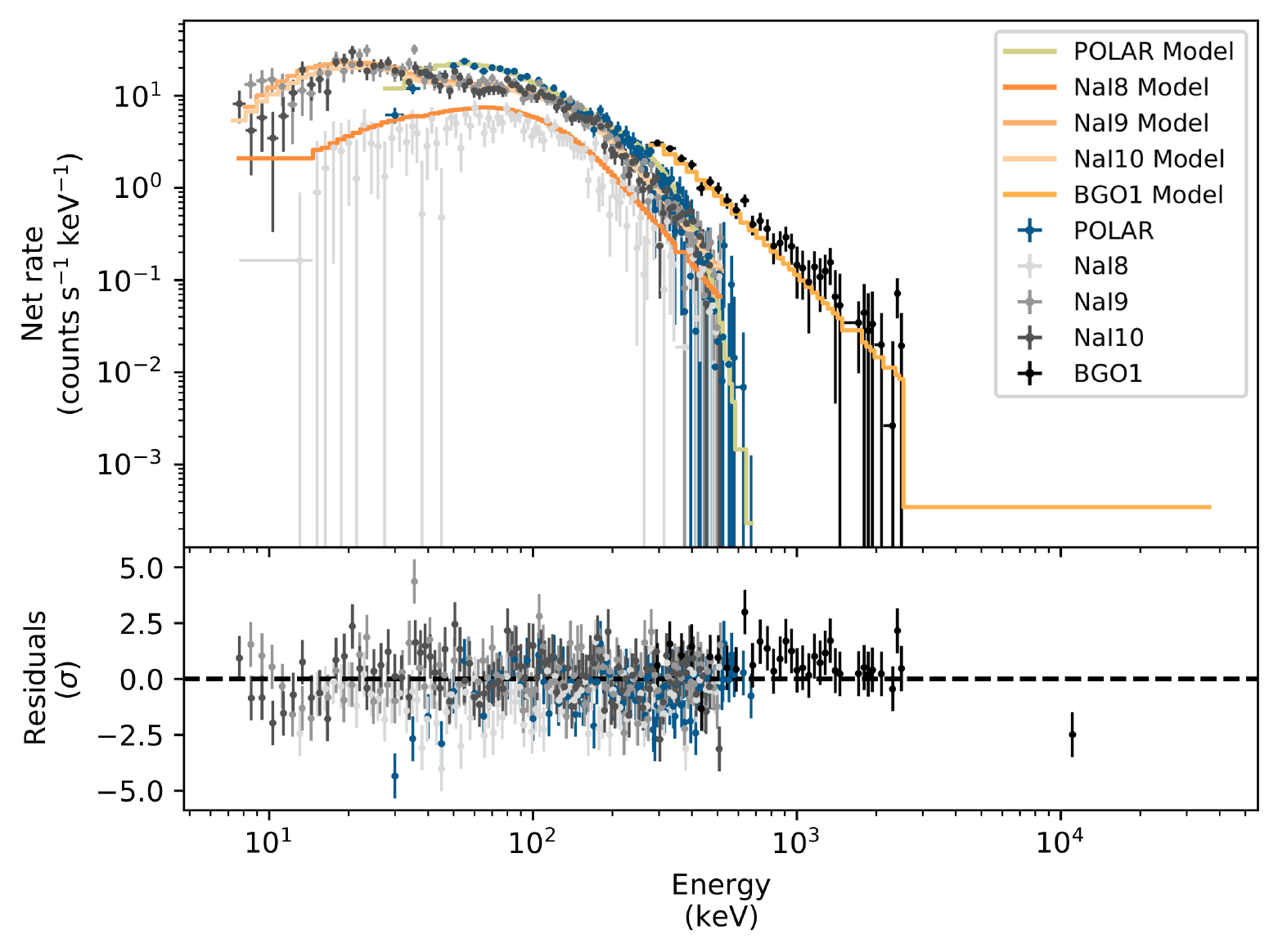}  
  \caption{The joint spectral fit result for 170206A. The number of counts as detected by both POLAR (blue) and the different NaI and BGO detectors of \textit{Fermi}-GBM (gray tints) are shown along with the best fitting spectrum folded through the instrument responses in yellow for POLAR data and in orange tints for the \textit{Fermi}-GBM data. The residuals for both data sets are shown in the bottom of the figure.}
  \label{fig:170206A_cs}
\end{subfigure}
\caption{The light curve as measured by POLAR for GRB 170206A (a) along with the joint spectral fit results of POLAR and \textit{Fermi}-GBM for the signal region indicated in yellow in  figure (a).}
\label{fig:170206A_lc_cs}
\end{figure}

\begin{figure}[!ht]
   \centering
     \resizebox{\hsize}{!}{\includegraphics{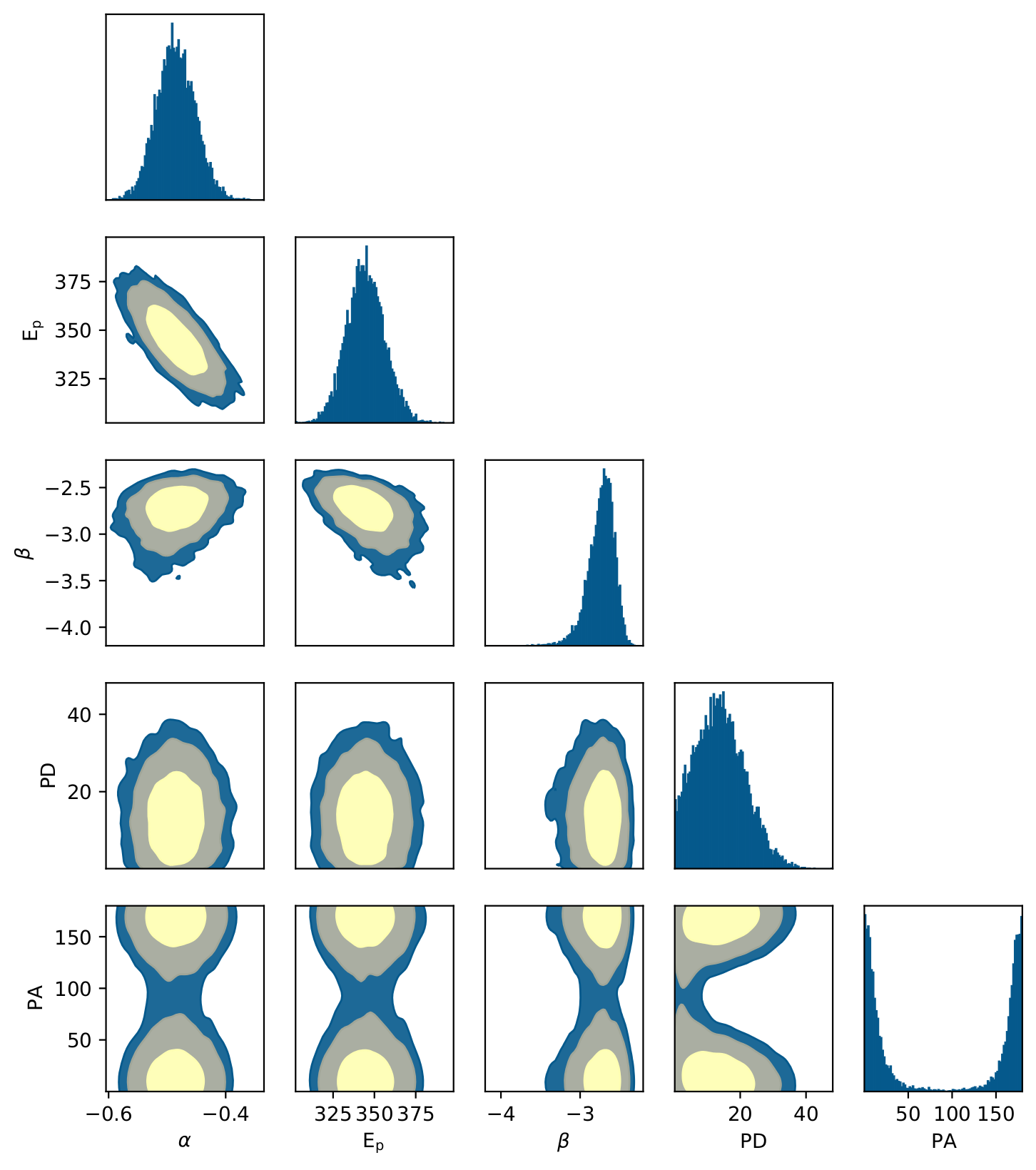}}
   \caption[Spectral and polarization posterior distributions.]
 {The spectral and polarization posterior distributions for GRB 170206A. The 1 and 2 $\sigma$ credibility intervals as well as that corresponding to $99\%$ are indicated. The polarization angle shown here is in the POLAR coordinate system, a rotation in the positive direction of 106 degrees transforms this to the coordinate system as defined by the IAU.}
 \label{fig:post_170206A}
 \end{figure}

\begin{figure}[ht]
\begin{subfigure}{.5\textwidth}
  \centering
  % include first image
  \includegraphics[width=.85\linewidth]{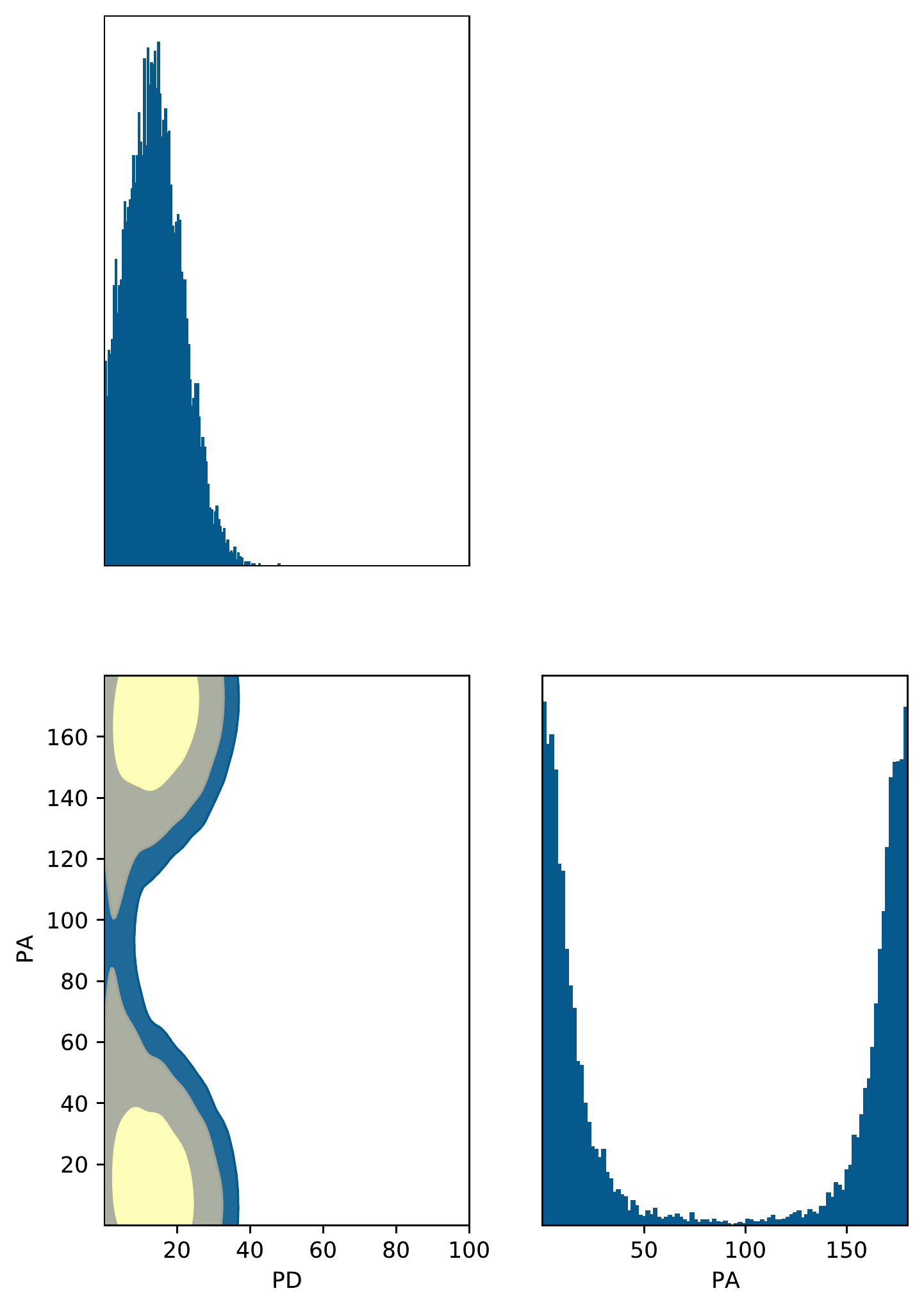}  
  \caption{The polarization posterior distributions for GRB 170206A with the 1 and 2 $\sigma$ credibility intervals as well as that corresponding to $99\%$ credibility. The polarization angle shown here is in the POLAR coordinate system, a rotation in the positive direction of 106 degrees transforms this to the coordinate system as defined by the IAU. }
  \label{fig:170206A_PD}
\end{subfigure}
\newline
\begin{subfigure}{.5\textwidth}
  \centering
  % include second image
  \includegraphics[width=.85\linewidth]{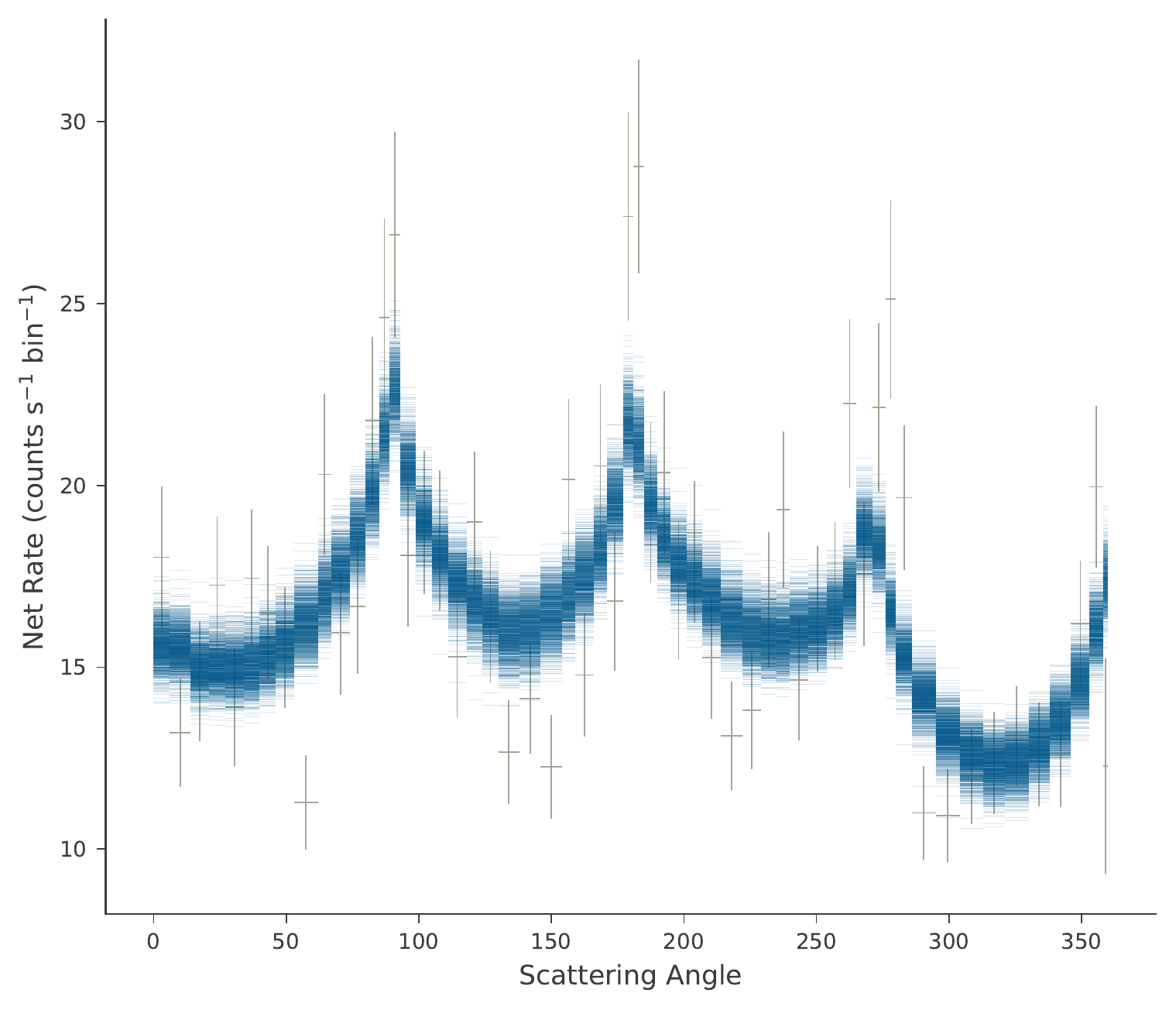}  
  \caption{The measured scattering angle distribution (gray data points with a 6 degree bin size) superimposed by samples from the posterior model predictions (blue) is shown.  The errors on the data points are the Poisson errors corrected for the background. }
  \label{fig:170206A_sd}
\end{subfigure}
\caption{The posterior distribution of the polarization parameters (a) together with the scattering angle distribution of GRB 170206A.}
\label{fig:170206A_PD_sd}
\end{figure}

GRB 170206A, a bright multi-pulse GRB was detected by POLAR and by \textit{Fermi}-GBM who report a $T0$ of 2017-02-06 at 10:51:57.70 (UT) \citep{GCN_170206A_Fermi}. This is for convenience taken as $T0$ for the analysis presented here as well. A $T_{90}$ of $(1.26\pm0.01)\,\mathrm{s}$ was measured using POLAR data. The light curve, including the signal region (blue) and part of the background region (yellow) can be seen in figure \ref{fig:170206A_lc}. As it was also detected by \textit{Fermi}-GBM \citep{GCN_170206A_Fermi} the spectral data from this instrument was used for the joint analysis. The spectral results of the joint fit results can be seen in figure \ref{fig:170206A_cs}. The effective area correction (applied to the POLAR data) found in the analysis was $0.93\pm0.02$. The POLAR response was produced using the location provided by \textit{Fermi}-LAT: RA (J2000) = $212.79^\circ$, Dec (J2000) = $14.48^\circ$ \citep{GCN_170206A_LAT}. The posterior distributions of the spectral and polarization parameters are shown in figure \ref{fig:post_170206A}. Finally, the posterior distribution of the polarization parameters is shown together with measured scattering angle distribution superimposed by the posterior model predictions (blue) in figure \ref{fig:170206A_PD_sd}. A PD of $13.5\substack{+7.4 \\ -8.6}\%$ is found which is compatible with that reported in \cite{Zhang+Kole}. It should be noted that the PA used in \cite{Zhang+Kole} is measured in the GRB coordinate system while here we measure it in the POLAR coordinate system. A $99\%$ credibility upper limit for PD of $34.0\%$ is found.

Additionally, time-resolved studies were performed for this GRB. The 3 pulses in the emission were studied independently. The first selected time bin is from $T=0.0\,\mathrm{s}$ to $T=0.8\,\mathrm{s}$, the second from $T=0.8\,\mathrm{s}$ to $T=1.2\,\mathrm{s}$ and the third from $T=1.2\,\mathrm{s}$ to $T=2.0\,\mathrm{s}$. The polarization of the three time intervals was found to be $8.9\substack{+15.0 \\ -7.3}$, $7.4\substack{+13.9 \\ -5.7}$ and $14\substack{+16 \\ -10}$ and therefore compatible with an unpolarized flux. The following three upper limits for PD were found for the 3 respective time intervals: $46.2\%$, $40.0\%$ and $56.0\%$. 

%   \begin{figure}[!ht]
%    \centering
%      \resizebox{\hsize}{!}{\includegraphics{170206A_time_res.png}}
%    \caption[Light curve of GRB 170206A as measured by POLAR.]
%  {The polarization posterior distributions for the 3 time intervals studied for GRB 170206A are shown with the 1 and 2 $\sigma$ credibility intervals as well as that corresponding to $99\%$ indicated. For the 1d distributions only the 68\% credibility region is shown. On the top left the posterior distribution for time interval T0+0.0 to T0+0.8 can be seen, top right shows this for T0+0.8 to T0+1.2 and the bottom shows this for T0+1.2 to T0+2.0. The polarization angle shown here is in the POLAR coordinate system, a rotation (in the positive direction) of xxx degrees transforms this to the coordinate system as defined by the International Astronomical Union. }
%  \label{170206A_time_res}
%  \end{figure} 
\clearpage

\section{170210A}

\begin{figure}[ht]
\begin{subfigure}{.5\textwidth}
  \centering
  % include first image
  \includegraphics[width=.95\linewidth]{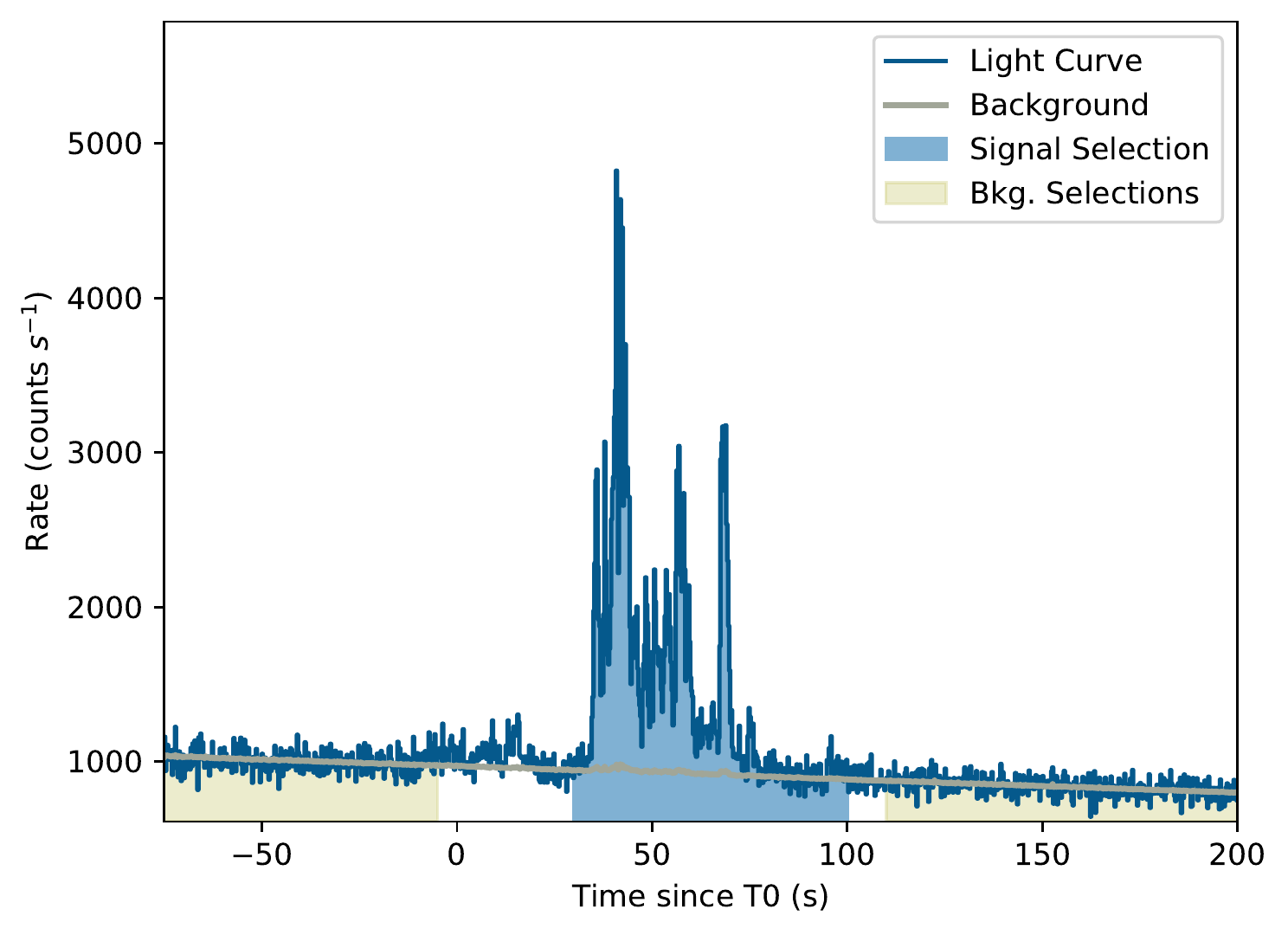}  
  \caption{The light curve of GRB 170210A as measured by POLAR, where $T=0\,\mathrm{s}$ is defined as the $T0$ employed by \textit{Fermi}-GBM in their data products for this GRB  \citep{GCN_170210A_Fermi}.}
  \label{fig:170210A_lc}
\end{subfigure}
\newline
\begin{subfigure}{.5\textwidth}
  \centering
  % include second image
  \includegraphics[width=.95\linewidth]{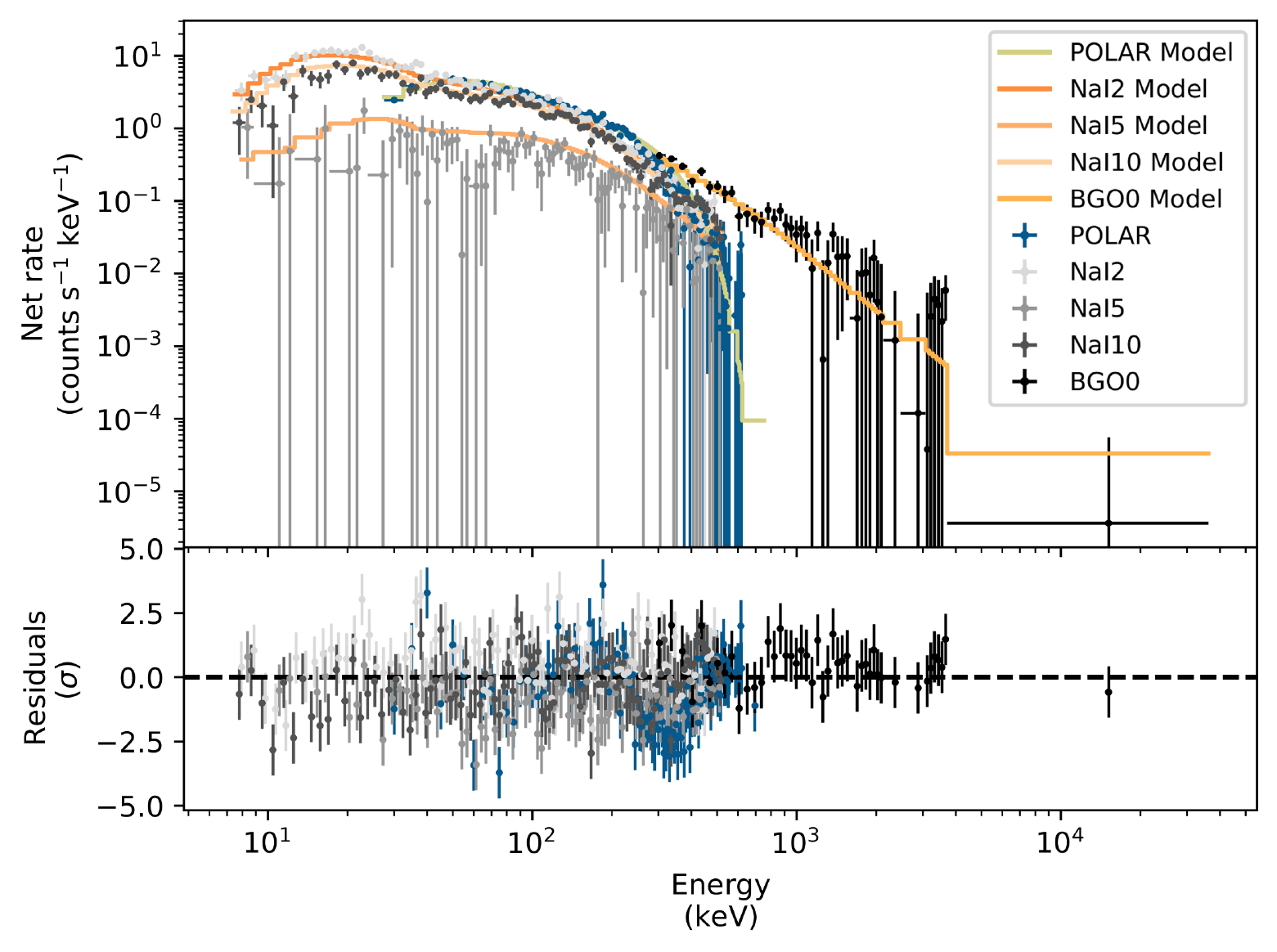}  
  \caption{The joint spectral fit result for 170210A. The number of counts as detected by both POLAR (blue) and the different NaI and BGO detectors of \textit{Fermi}-GBM (gray tints) are shown along with the best fitting spectrum folded through the instrument responses in yellow for POLAR data and in orange tints for the \textit{Fermi}-GBM data. The residuals for both data sets are shown in the bottom of the figure.}
  \label{fig:170210A_cs}
\end{subfigure}
\caption{The light curve as measured by POLAR for GRB 170210A (a) along with the joint spectral fit results of POLAR and \textit{Fermi}-GBM for the signal region indicated in yellow in  figure (a).}
\label{fig:170210A_lc_cs}
\end{figure}

\begin{figure}[!ht]
   \centering
     \resizebox{\hsize}{!}{\includegraphics{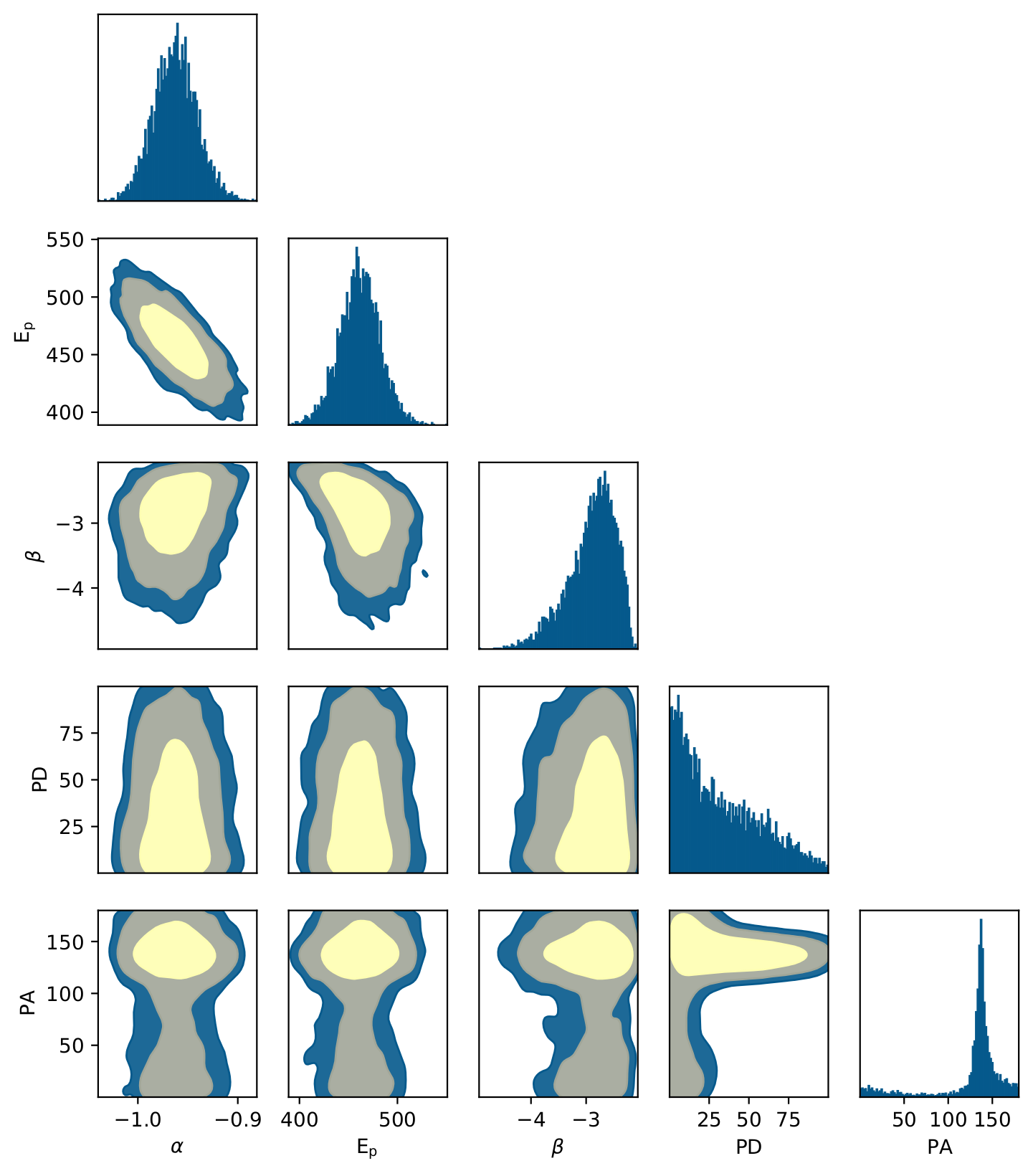}}
   \caption[Spectral and polarization posterior distributions.]
 {The spectral and polarization posterior distributions for GRB 170210A. The 1 and 2 $\sigma$ credibility intervals as well as that corresponding to $99\%$ are indicated. The polarization angle shown here is in the POLAR coordinate system, a rotation in the positive direction of 55 degrees transforms this to the coordinate system as defined by the IAU.}
 \label{fig:post_170210A}
 \end{figure}

\begin{figure}[ht]
\begin{subfigure}{.5\textwidth}
  \centering
  % include first image
  \includegraphics[width=.85\linewidth]{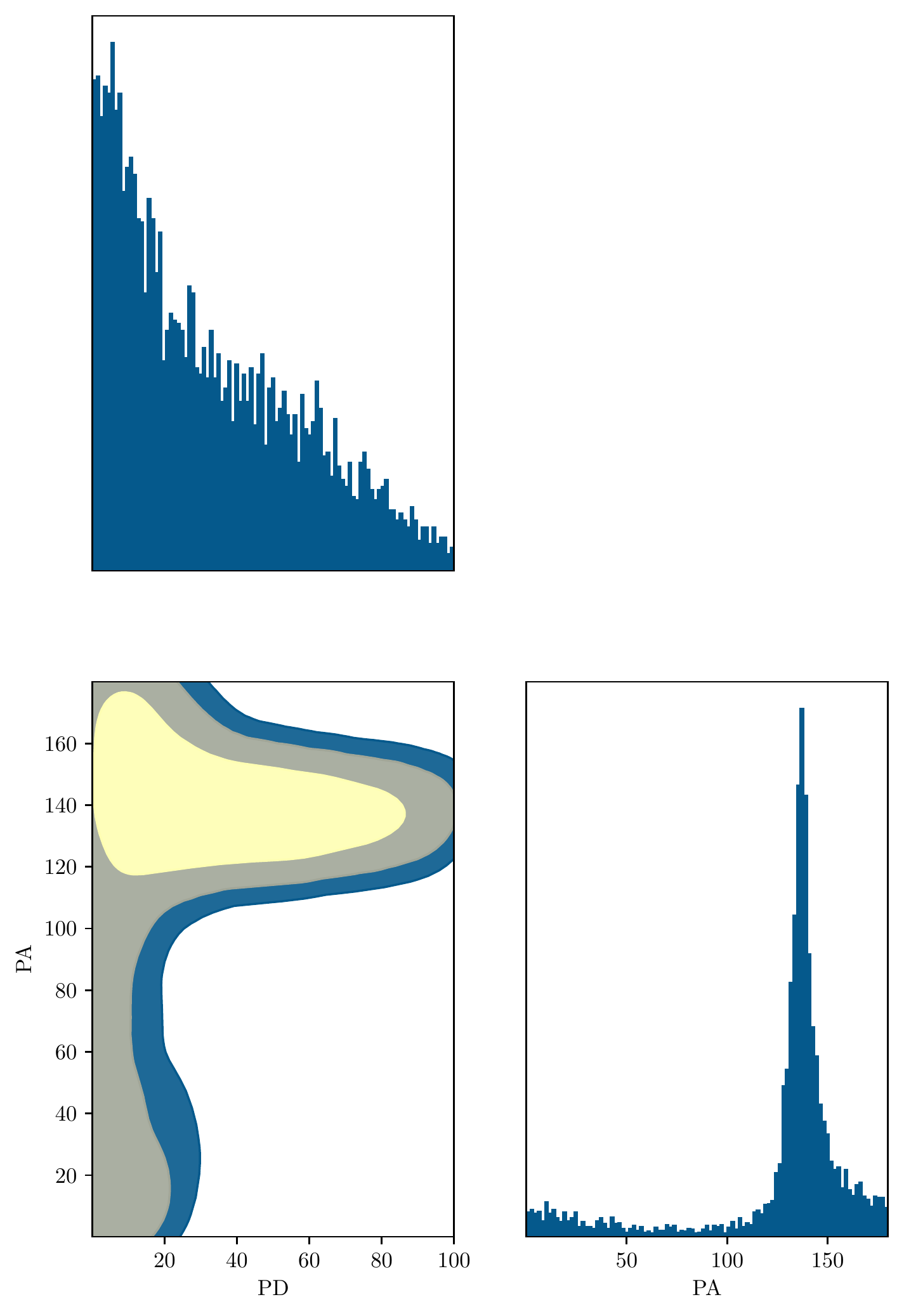}  
  \caption{The polarization posterior distributions for GRB 170210A with the 1 and 2 $\sigma$ credibility intervals as well as that corresponding to $99\%$ credibility. The posterior distribution clearly indicates the lack of sensitivity for this GRB for a polarization angle of $135^\circ$. The polarization angle shown here is in the POLAR coordinate system, a rotation in the positive direction of 55 degrees transforms this to the coordinate system as defined by the IAU.}
  \label{fig:170210A_PD}
\end{subfigure}
\newline
\begin{subfigure}{.5\textwidth}
  \centering
  % include second image
  \includegraphics[width=.85\linewidth]{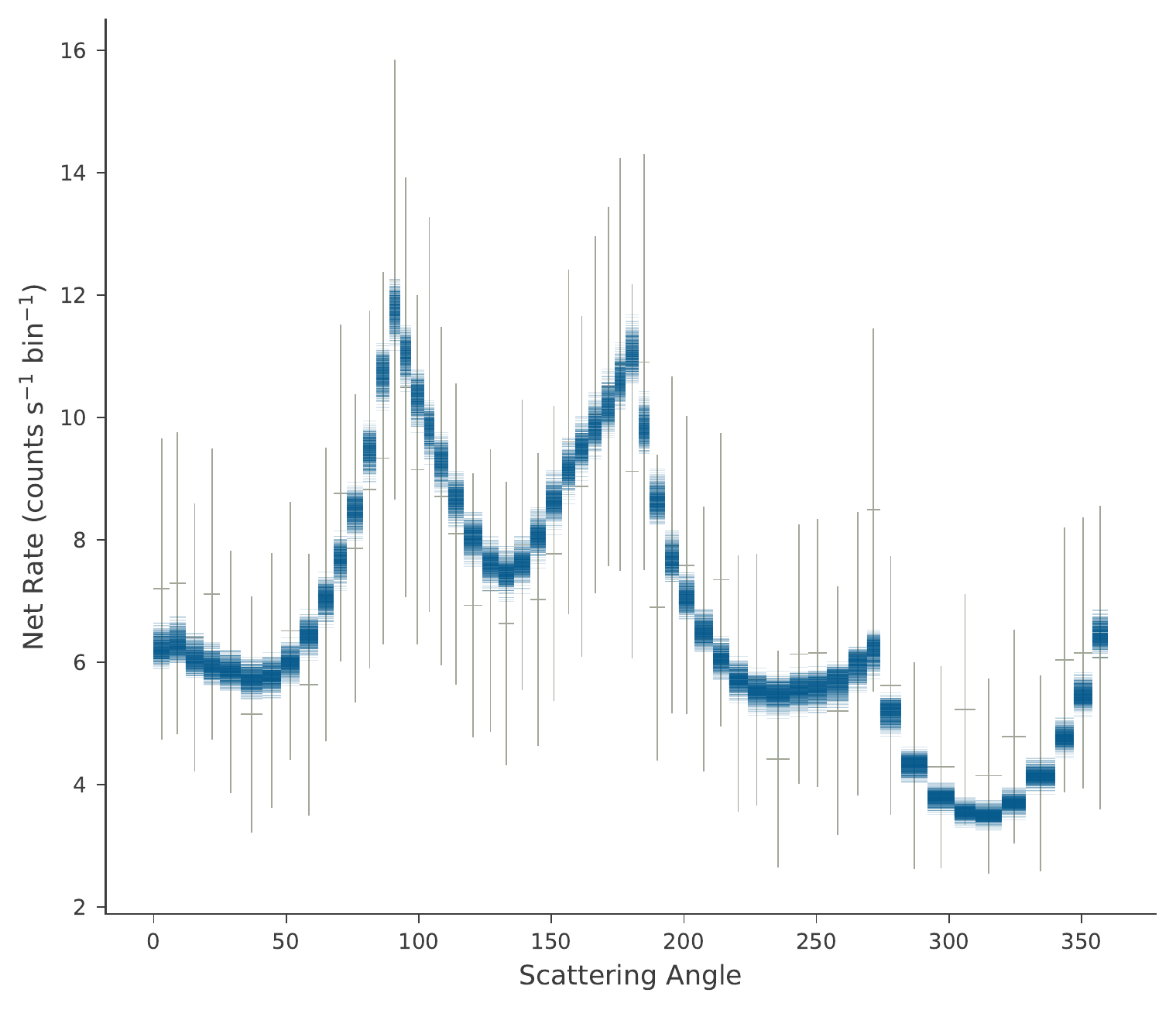}  
  \caption{The measured scattering angle distribution (gray data points with a 6 degree bin size) superimposed by samples from the posterior model predictions (blue) is shown. The errors on the data points are the Poisson errors corrected for the background. }
  \label{fig:170210A_sd}
\end{subfigure}
\caption{The posterior distribution of the polarization parameters (a) together with the scattering angle distribution of GRB 170210A.}
\label{fig:170210A_PD_sd}
\end{figure}

GRB 170210A, a bright long GRB with many overlapping pulses, was detected by both POLAR and \textit{Fermi}-GBM. The latter reported a $T0$ of 2017-02-10 at 02:47:36.58 (UT) \citep{GCN_170210A_Fermi}, which, for convenience will be taken as $T0$ for the analysis presented here as well. A $T_{90}$ of $(47.63\pm2.51)\,\mathrm{s}$ was measured using POLAR data. The light curve using POLAR data, including the signal region (blue) and part of the background region (yellow) can be seen in figure \ref{fig:170210A_lc}. The data from \textit{Fermi}-GBM was used in this analysis. It should be noted that although some significant emission was seen prior to the main onset of the GRB (at $T0$+30 s) these pulses were weak and therefore not taken into account in the analysis here. The spectral fit results showing both the POLAR and \textit{Fermi}-GBM data can be seen in figure \ref{fig:170210A_cs}. The effective area correction (applied to the POLAR data) found in the analysis was $0.95\pm0.02$. The polarization response of POLAR was produced using the location calculated through the IPN \citep{GCN_170210A_IPN}: RA (J2000) = $226.055 ^\circ$, Dec (J2000) = $-65.101^\circ$, a localization error of $2^\circ$ was assumed in the response. This location implies a significant off-axis incoming angle for the GRB of $80.6^\circ$, such an off-axis angle results in a loss of approximately $67\%$ sensitivity at particular polarization angles (at $45^\circ$ and $135^\circ$) compared to the maximum sensitivity found at $0^\circ$ and $90^\circ$, as explained in detail in the section regarding GRB 170207A. The posterior distributions of the spectral and polarization parameters are shown in figure \ref{fig:post_170210A}. Finally, the posterior distribution of the polarization parameters is shown together with measured scattering angle distribution superimposed by samples from the posterior model predictions (blue) in figure \ref{fig:170210A_PD_sd}. The effect of the reduced sensitivity at polarization angles around $135^\circ$ can clearly be observed in the posterior distribution. While very constraining measurements are possible at all other angles, the PD is almost fully unconstrained around this angle. A PD of $11.4\substack{+35.7 \\ -11.4}\%$ is found while a $99\%$ upper limit of $95.2\%$ was found. Although it appears that the posterior distribution is dominated by the lack of sensitivity, the results do clearly exclude high polarization degrees at any polarization angles away from $135^\circ$.

A time-resolved study was performed, however, the polarization results for all studied time bins were found to be similar as for the full GRB. 

\clearpage
\section{170305A}

\begin{figure}[ht]
\begin{subfigure}{.5\textwidth}
  \centering
  % include first image
  \includegraphics[width=.95\linewidth]{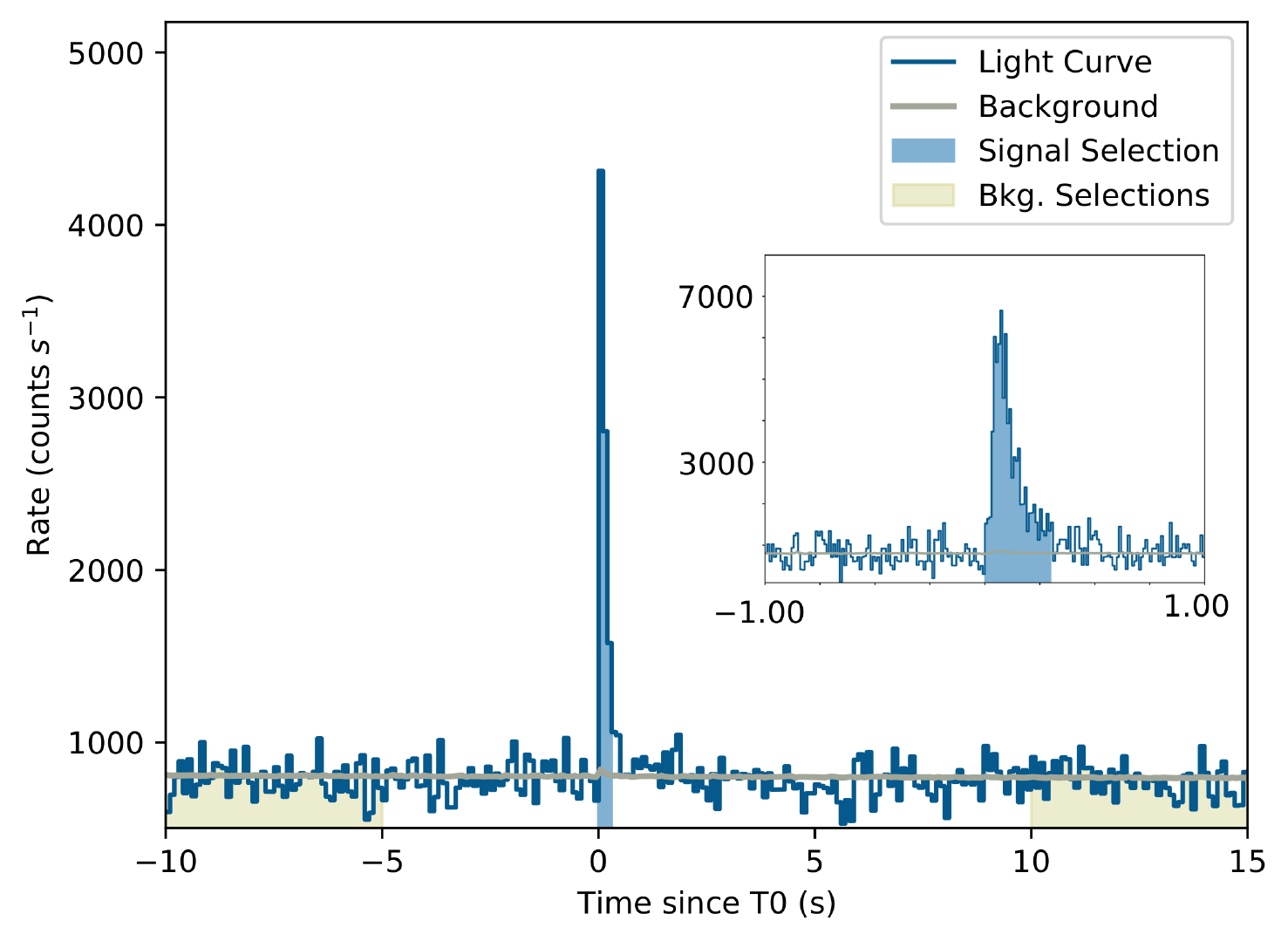}  
  \caption{The light curve of GRB 170305A as measured by POLAR, with an inlay of a zoomed in version around the peak, where $T=0\,\mathrm{s}$ is defined as the $T0$ employed by \textit{Fermi}-GBM in their data products for this GRB \citep{GCN_170305A_Fermi}.}
  \label{fig:170305A_lc}
\end{subfigure}
\newline
\begin{subfigure}{.5\textwidth}
  \centering
  % include second image
  \includegraphics[width=.95\linewidth]{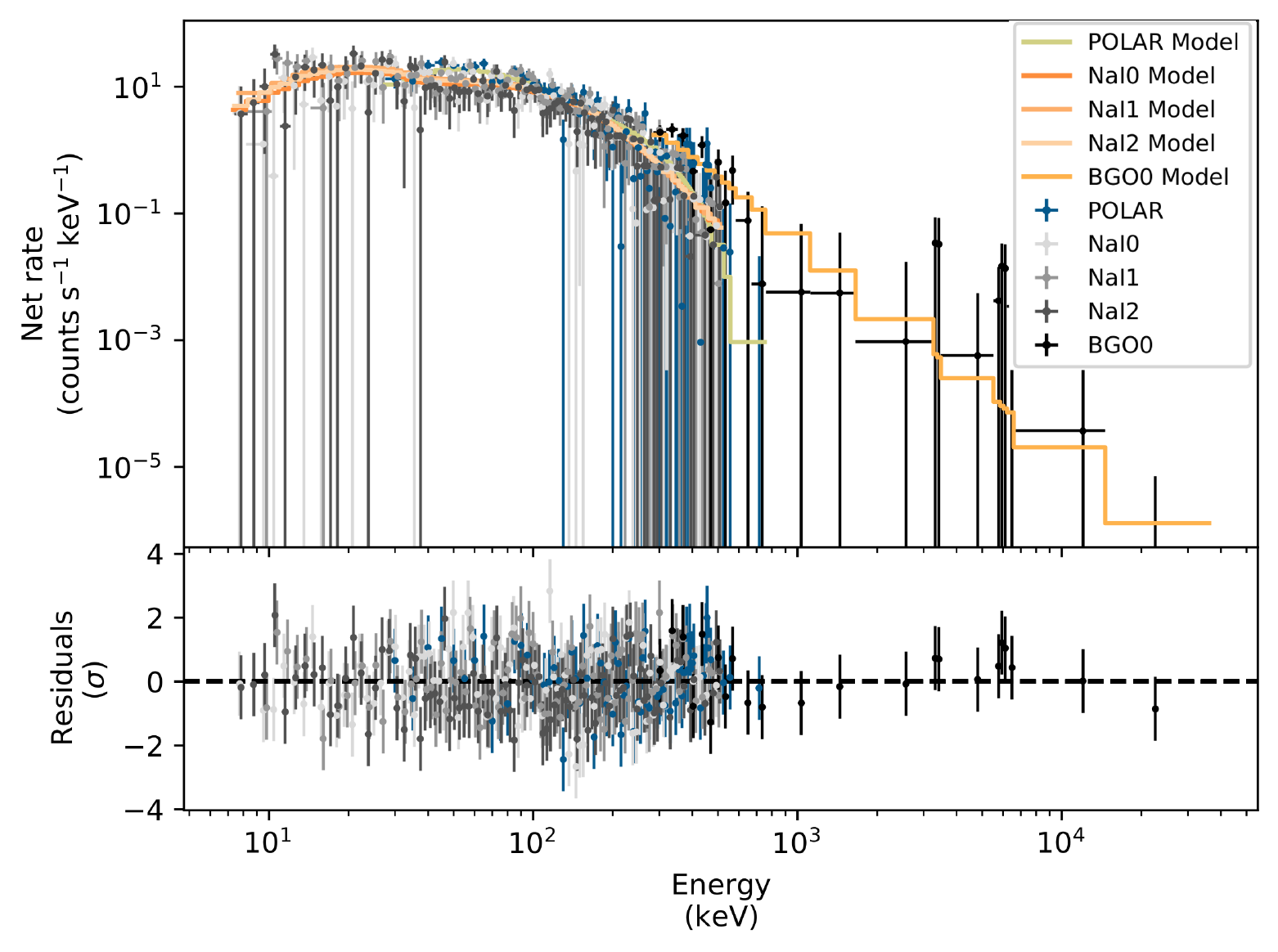}  
  \caption{The joint spectral fit result for 170305A. The number of counts as detected by both POLAR (blue) and the different NaI and BGO detectors of \textit{Fermi}-GBM (gray tints) are shown along with the best fitting spectrum folded through the instrument responses in yellow for POLAR data and in orange tints for the \textit{Fermi}-GBM data. The residuals for both data sets are shown in the bottom of the figure.}
  \label{fig:170305A_cs}
\end{subfigure}
\caption{The light curve as measured by POLAR for GRB 170305A (a) along with the joint spectral fit results of POLAR and \textit{Fermi}-GBM for the signal region indicated in yellow in  figure (a).}
\label{fig:170305A_lc_cs}
\end{figure}

\begin{figure}[!ht]
   \centering
     \resizebox{\hsize}{!}{\includegraphics{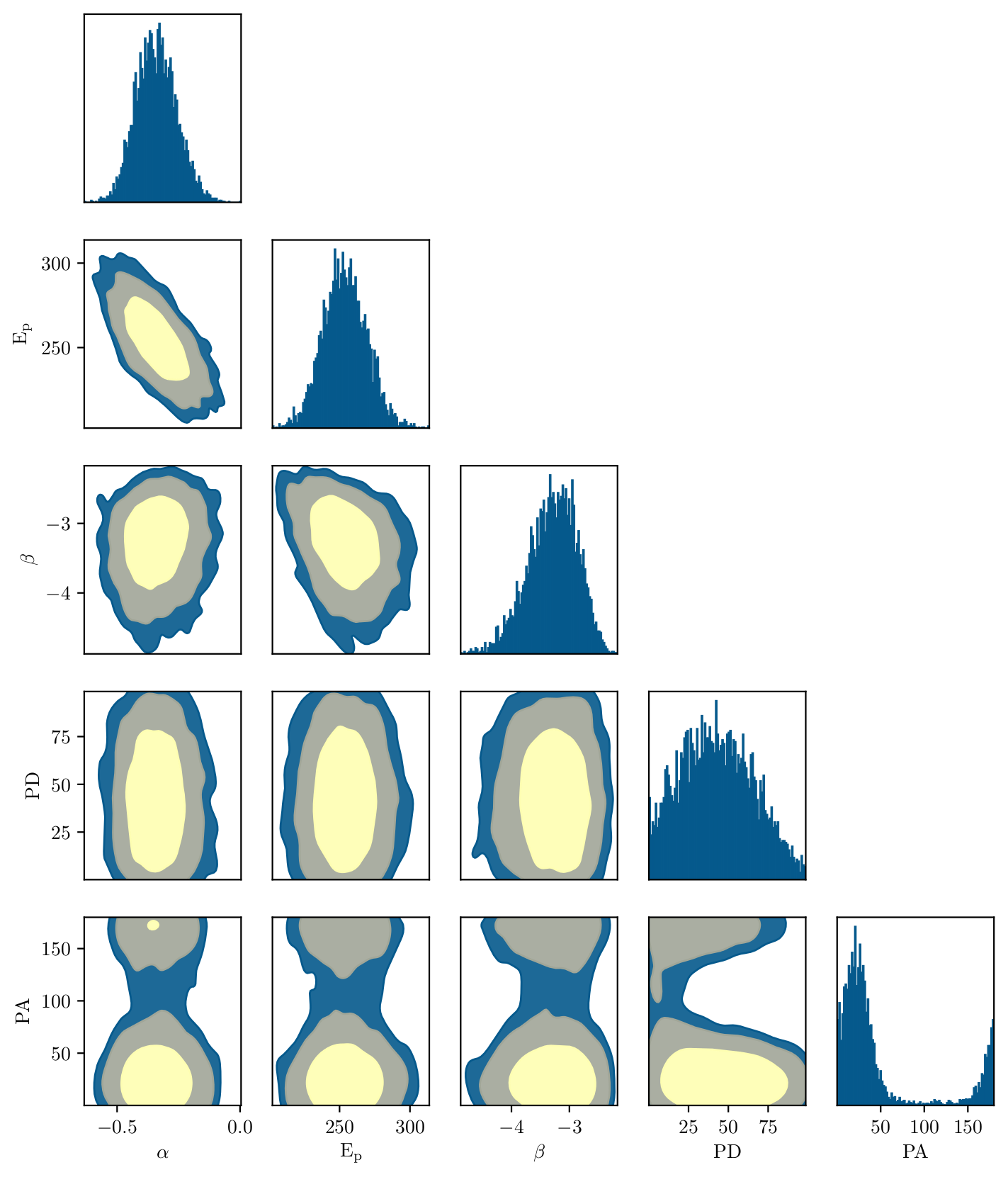}}
   \caption[Spectral and polarization posterior distributions.]
 {The spectral and polarization posterior distributions for GRB 170305A. The 1 and 2 $\sigma$ credibility intervals as well as that corresponding to $99\%$ are indicated. The polarization angle shown here is in the POLAR coordinate system, a rotation in the positive direction of 65 degrees transforms this to the coordinate system as defined by the IAU.}
 \label{fig:post_170305A}
 \end{figure}

\begin{figure}[ht]
\begin{subfigure}{.5\textwidth}
  \centering
  % include first image
  \includegraphics[width=.85\linewidth]{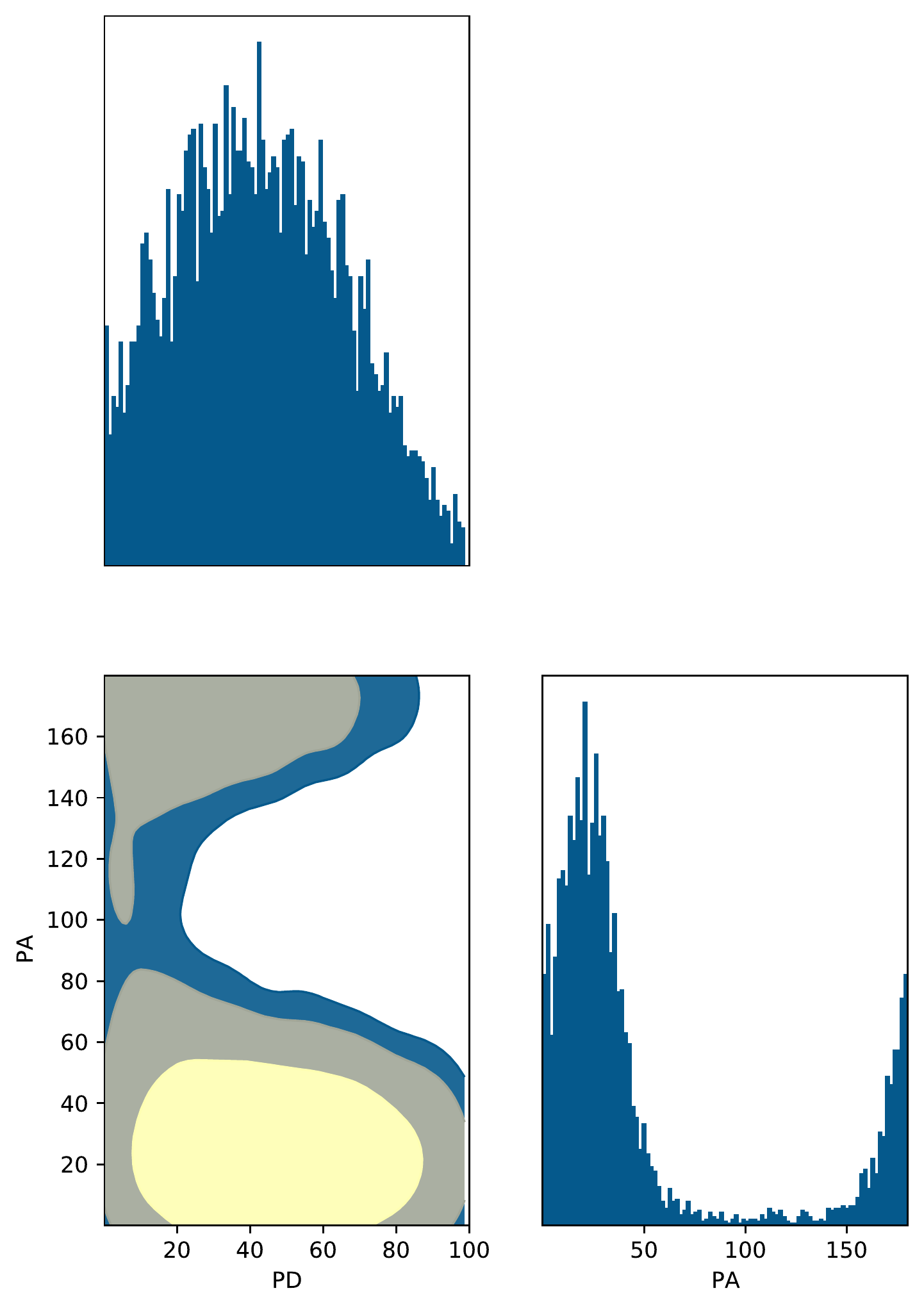}  
  \caption{The polarization posterior distributions for GRB 170305A with the 1 and 2 $\sigma$ credibility intervals as well as that corresponding to $99\%$ credibility. The polarization angle shown here is in the POLAR coordinate system, a rotation in the positive direction of 65 degrees transforms this to the coordinate system as defined by the IAU.}
  \label{fig:170305A_PD}
\end{subfigure}
\newline
\begin{subfigure}{.5\textwidth}
  \centering
  % include second image
  \includegraphics[width=.85\linewidth]{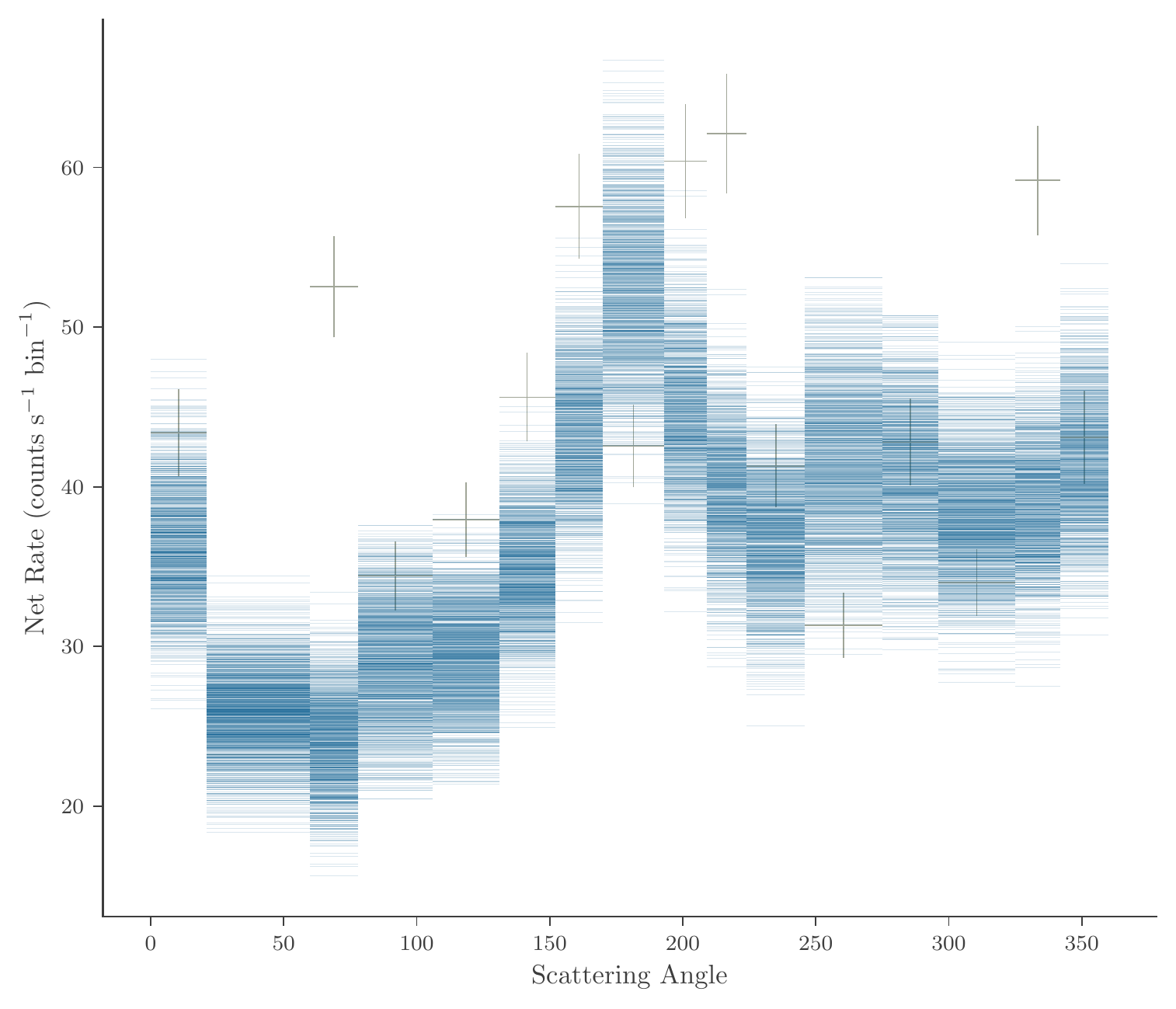}  
  \caption{The measured scattering angle distribution (gray data points with a $22.5$ degree bin size) superimposed by samples from the posterior model predictions (blue) is shown.  The errors on the data points are the Poisson errors corrected for the background. }
  \label{fig:170305A_sd}
\end{subfigure}
\caption{The posterior distribution of the polarization parameters (a) together with the scattering angle distribution of GRB 170305A.}
\label{fig:170305A_PD_sd}
\end{figure}

GRB 170305A, a short GRB, was detected by POLAR and by \textit{Fermi}-GBM. The latter reported a $T0$ of 2017-03-05 at 06:09:06.8 (UT) \citep{GCN_170305A_Fermi}, which, for convenience will be used as the $T0$ for analysis presented here as well. A $T_{90}$ of $(0.45\pm0.01)\,\mathrm{s}$ was measured using POLAR data. The light curve, including the signal region (blue) and the background region (yellow) can be seen in figure \ref{fig:170305A_lc} along with a more detailed inset of the main pulse. Spectral data from \textit{Fermi}-GBM was used in the joint spectral fit along with the POLAR data. The spectral results of the joint fit can be seen in figure \ref{fig:170305A_cs}. The effective area correction (applied to the POLAR data) found in the analysis was $1.06\pm0.07$. The response of POLAR was produced using the location calculated using the IPN \citep{GCN_170305A_IPN}: RA (J2000) = $39.658^\circ$, Dec (J2000) = $9.901^\circ$, corresponding to an off-axis incoming angle of $31.4^\circ$. A localization error of $2^\circ$ was assumed in the response. The posterior distributions of the spectral and polarization parameters are shown in figure \ref{fig:post_170305A}. Finally, the posterior distribution of the polarization parameters is shown together with measured scattering angle distribution superimposed by samples from the posterior model predictions (blue) in figure \ref{fig:170305A_PD_sd}. A PD of $(40.0\pm25.0)\%$ was found along with a $99\%$ credibility upper limit for PD of $97\%$. The rather large uncertainty can be attributed to the relatively low signal over background counts ratio in the POLAR data for this GRB, which had an average flux of $1.6\pm0.2\times10^{-6}\mathrm{erg/cm^2/s}$ as measured over one second. This is just above the set lower limit for our cuts.

 \clearpage
\section{170320A}
 
\begin{figure}[ht]
\begin{subfigure}{.5\textwidth}
  \centering
  % include first image
  \includegraphics[width=.95\linewidth]{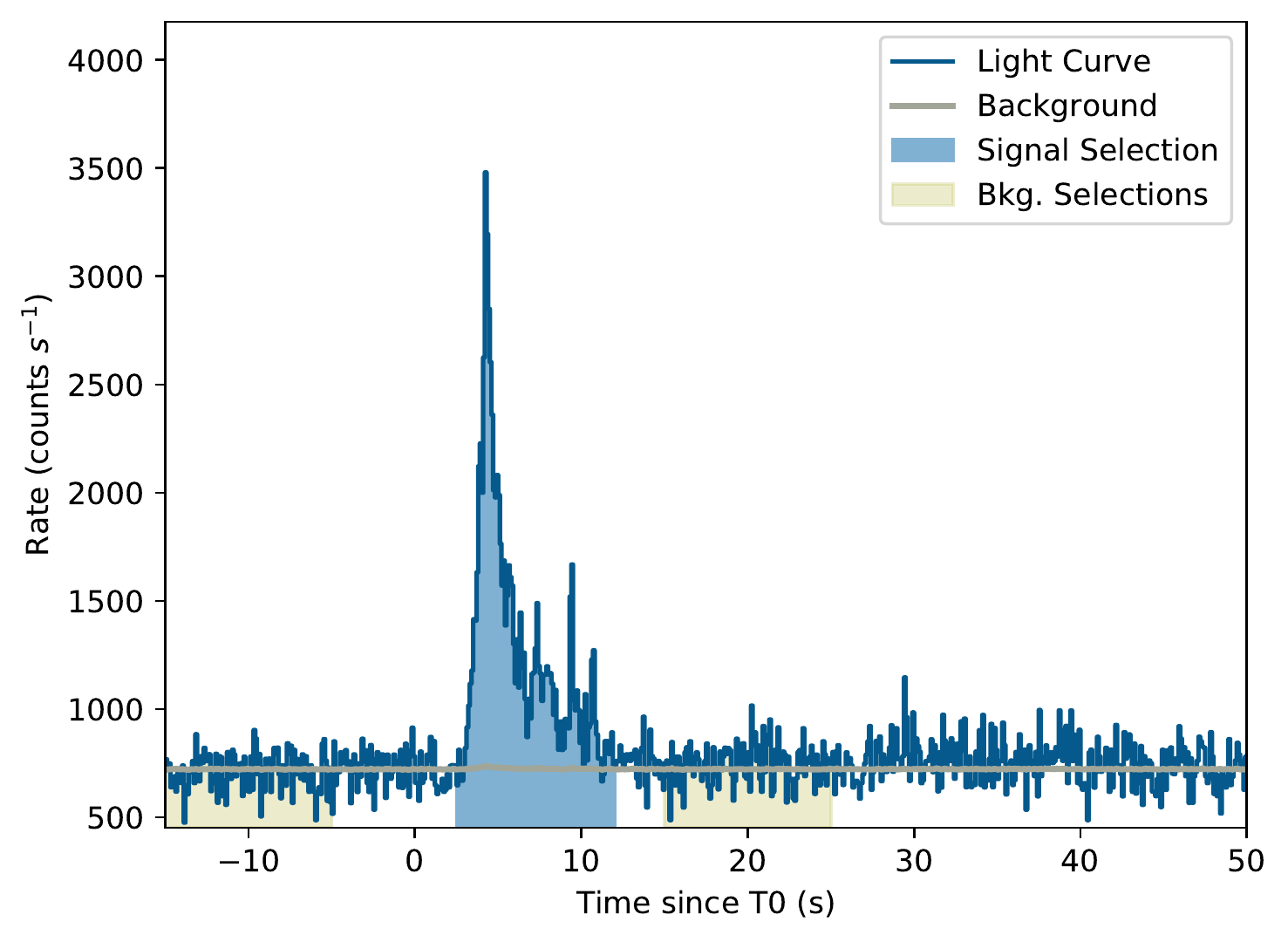}  
  \caption{The light curve of GRB 170320A as measured by POLAR, where $T=0\,\mathrm{s}$ is defined as the $T0$ reported by POLAR in the original GCN \citep{POLAR_GCN_170320A}.}
  \label{fig:170320A_lc}
\end{subfigure}
\newline
\begin{subfigure}{.5\textwidth}
  \centering
  % include second image
  \includegraphics[width=.95\linewidth]{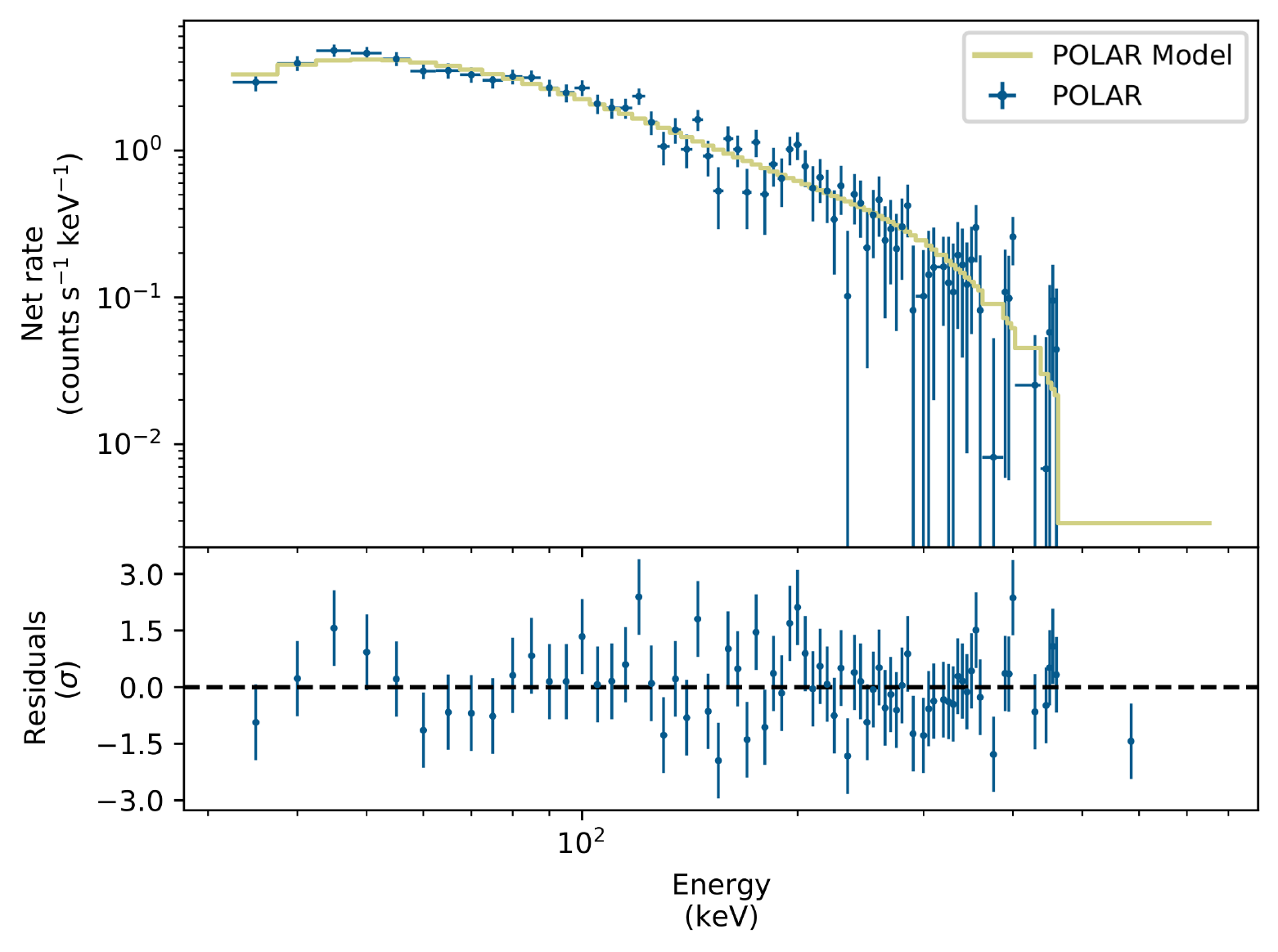}  
  \caption{The joint spectral fit result for 170320A. The number of counts as detected by both POLAR (blue) is shown along with the best fitting spectrum folded through the instrument response in yellow. The residuals are shown in the bottom of the figure.}
  \label{fig:170320A_cs}
\end{subfigure}
\caption{The light curve as measured by POLAR for GRB 170320A (a) along with the joint spectral fit result of POLAR for the signal region indicated in yellow in figure (a).}
\label{fig:170320A_lc_cs}
\end{figure}

\begin{figure}[ht]
\begin{subfigure}{.5\textwidth}
  \centering
  % include first image
  \includegraphics[width=.85\linewidth]{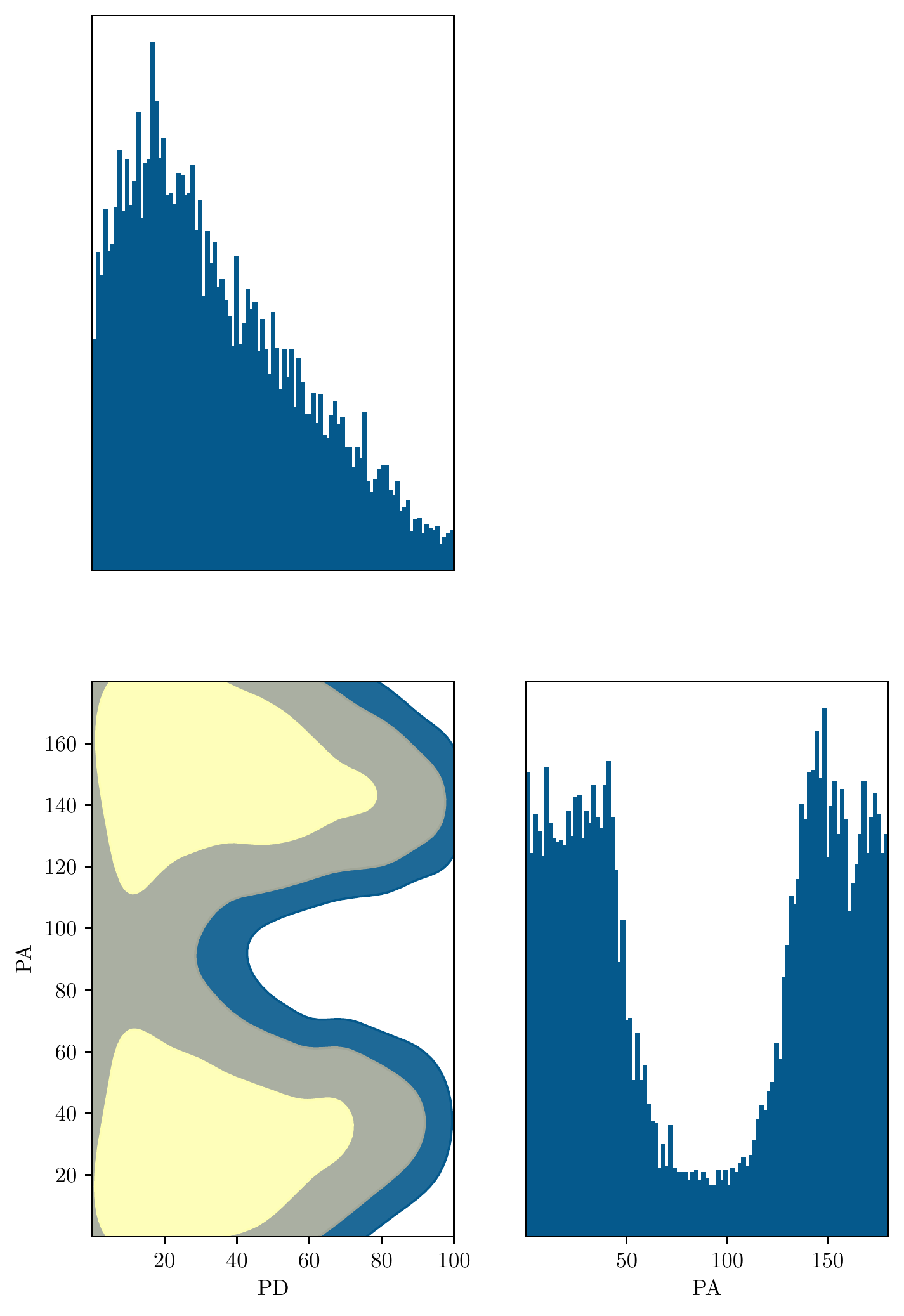}  
  \caption{The polarization posterior distributions for GRB 170320A with the 1 and 2 $\sigma$ credibility intervals as well as that corresponding to $99\%$ credibility. The polarization angle shown here is in the POLAR coordinate system,  a rotation in the positive direction of 56 degrees transforms this to the coordinate system as defined by the IAU.}
  \label{fig:170320A_PD}
\end{subfigure}
\newline
\begin{subfigure}{.5\textwidth}
  \centering
  % include second image
  \includegraphics[width=.85\linewidth]{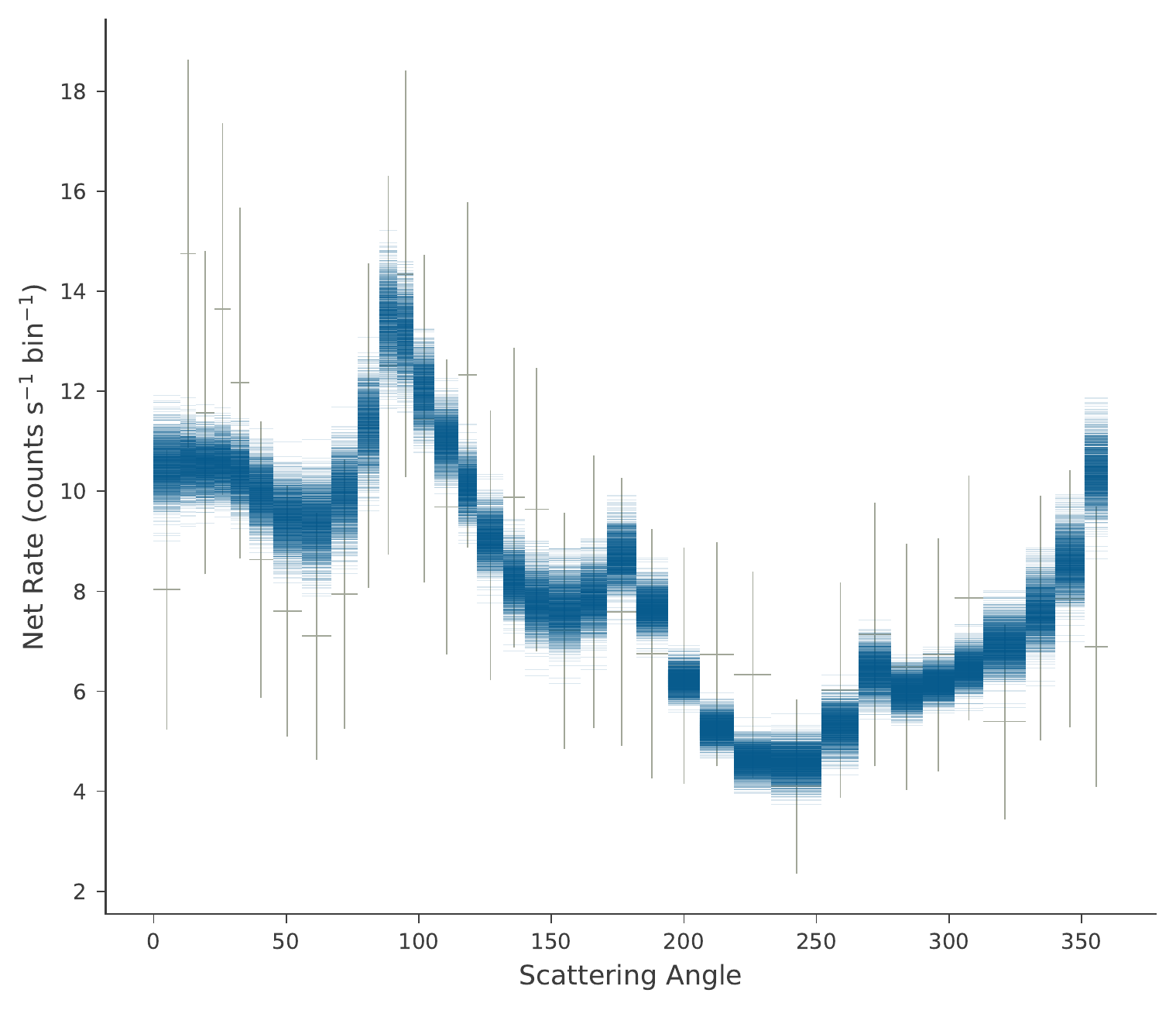}  
  \caption{The measured scattering angle distribution (gray data points with a 10 degree bin size) superimposed by samples from the posterior model predictions (blue) is shown.  The errors on the data points are the Poisson errors corrected for the background. }
  \label{fig:170320A_sd}
\end{subfigure}
\caption{The posterior distribution of the polarization parameters (a) together with the scattering angle distribution of GRB 170320A.}
\label{fig:170320A_PD_sd}
\end{figure}

GRB 170320A was detected by POLAR on 2017-03-20 at 03:42:39.00 (UT) \citep{POLAR_GCN_170320A} which will be defined as $T0$ for the analysis presented here. A $T_{90}$ of $(6.83\pm0.09)\,\mathrm{s}$ was measured using POLAR data. The light curve from POLAR data, including the signal region (blue) and part of the background region (yellow) can be seen in figure \ref{fig:170320A_lc}. As it was not detected by \textit{Fermi}-GBM or \textit{Swift}-BAT no spectral data from other instruments was available to perform a joint fit. Therefore, the spectral and polarization analysis was performed using the POLAR response only. However, the spectral parameters reported by Konus-Wind in \cite{GCN_KONUS_170320A} were used as priors for the spectral fit. It should also be noted that while the main emission took place until $T0+12\,\mathrm{s}$, a soft afterpulse was seen by Konus-Wind \citep{GCN_KONUS_170320A}, and also seen in the POLAR data around $T0+30\,\mathrm{s}$. The time period around the afterpulse was excluded from the the signal and background selection. The response was produced using the location provided by the IPN \citep{GCN_170320A_IPN}: RA (J2000) = $320.074^\circ$, Dec (J2000) = $55.060^\circ$, where an uncertainty of $1^\circ$ on the location was assumed when generating the response. The location corresponds to a large off-axis incoming angle of $84.7^\circ$ which reduces the sensitivity for polarization angles of $45^\circ$ and $135^\circ$ by approximately $85\%$. The spectral results from the joint fit can be seen in figure \ref{fig:170320A_cs}. The posterior distributions of the spectral and polarization parameters are shown in figure \ref{fig:post_170101A}. Finally, the posterior distribution of the polarization parameters is shown together with measured scattering angle distribution superimposed by samples from the posterior model predictions (blue) in figure \ref{fig:170320A_PD_sd}. A PD of $18.0\substack{+28.0 \\ -15.0}\%$ is found. No significant constraints on the polarization can be provided due to the lack of sensitivity for polarization angles of $45^\circ$ and $135^\circ$ although we can exclude high values of PD for the majority of polarization angles.
 
\begin{figure}[!ht]
   \centering
     \resizebox{\hsize}{!}{\includegraphics{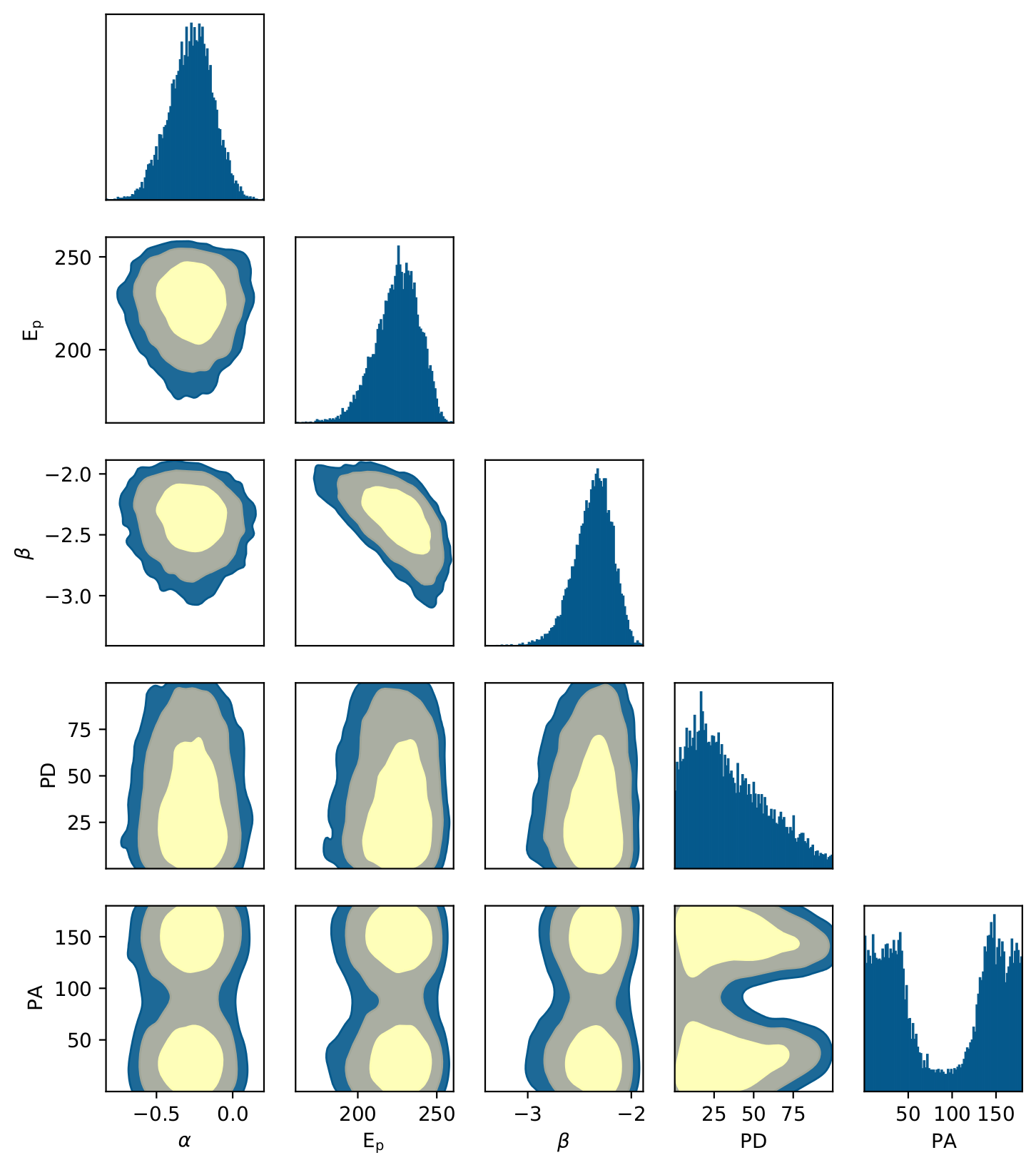}}
   \caption[Spectral and polarization posterior distributions.]
 {The spectral and polarization posterior distributions for GRB 170320A. The 1 and 2 $\sigma$ credibility intervals as well as that corresponding to $99\%$ are indicated. The polarization angle shown here is in the POLAR coordinate system, a rotation in the positive direction of 56 degrees transforms this to the coordinate system as defined by the IAU.}
 \label{fig:post_170320A}
 \end{figure}

\end{document}